\documentclass{aa}  

\usepackage{graphicx}
\usepackage{txfonts}

\newcommand{\iris}{\textit{IRIS}}

\begin{document}

   \title{Modelling of \ion{Mg}{ii} lines in solar prominences}

   \author{P.J. Levens \and N. Labrosse 
          }

   \institute{SUPA, School of Physics \& Astronomy, University of Glasgow, Scotland\\
              \email{Nicolas.Labrosse@glasgow.ac.uk}
             }

   \date{}

  \abstract
   {Observations of the \ion{Mg}{ii} h and k lines in solar prominences with \iris\ reveal a wide range of line shapes from simple non-reversed profiles to typical double-peaked reversed profiles, and  with many other possible complex line shapes. The physical conditions responsible for this variety are not well understood.}
   {Our aim is to understand {how physical conditions inside a prominence slab influence shapes and properties of
        emergent \ion{Mg}{ii} line profiles}.}
   {We compute the spectrum of \ion{Mg}{ii} lines using a one-dimensional non-LTE radiative transfer code {for two large grids of model atmospheres (isothermal isobaric, and with a transition region)}.}
   {The influence of the plasma parameters on the emergent spectrum is discussed in detail. {Our results agree with previous studies. We present several dependencies between observables and prominence parameters which will help with the interpretation of observations. A comparison with known limits of observed line parameters suggests that most observed prominences emitting in \ion{Mg}{ii} h and k lines are cold, low-pressure, and optically thick structures.}
        Our results indicate that there are good correlations between the \ion{Mg}{ii} k line intensities and  the intensities of hydrogen lines, and the emission measure.}
   {One-dimensional non-LTE radiative transfer codes allow us to  understand  the main characteristics of the \ion{Mg}{ii} h and k line profiles in solar prominences, but more advanced codes will be necessary for detailed comparisons.}

   \keywords{Line: profiles -- Radiative transfer -- Sun: filaments, prominences}

   \maketitle

\section{Introduction}
It is necessary to have detailed models that can deal with the full physics of the prominence conditions in order to interpret  observations of the \ion{Mg}{ii} h and k lines in solar prominences. Observations obtained since the launch of the IRIS spacecraft \citep[][{Interface Region Imaging Spectrograph}]{DePontieu2014} have revealed a range of line profile shapes \citep{Schmieder2014,Heinzel2015,Liu2015,Levens2016,Vial2016,Levens2017,2017A&A...606A..30S,2018ApJ...852...79Y,2018ApJ...865..123R,2018A&A...618A..88J} confirming that prominences exhibit a wide range of physical conditions, i.e. different pressures and motions that result in very different line profiles.
Therefore, those models must be equipped to handle the full radiative transfer and statistical equilibrium calculations required to accurately simulate the emission from prominences. 
Many models have been constructed in the past to deal with optically thick emission in prominences \citep[e.g.][]{Heasley1974,Vial1982b,Heinzel1987,Gouttebroze1993,Paletou1993,Gouttebroze1997,Labrosse2001,Heinzel2001b,Gunar2007,Heinzel2014,Heinzel2015}. 
\citet[hereafter HVA]{Heinzel2014} calculated magnesium profiles in prominences using a grid of 27 isothermal isobaric models, and two additional models with a prominence-to-corona transition region (PCTR). {They also considered a set of radiative equilibrium models to quantify temperature variations due to additional radiative losses.}
Radiative transfer modelling in solar prominences was discussed in detail in the reviews by \cite{Patsourakos2002} {and} \cite{Labrosse2010}, and in the books by \cite{Tandberg-Hanssen1974} {and} \cite{2015ASSL..415.....V}.

This work presents an extension to the code called PROM developed by \citet[][hereafter GHV]{Gouttebroze1993}, which in its simplest form is a 1D isothermal isobaric hydrogen prominence model. 
The 1D code was expanded to compute calcium profiles \citep{Gouttebroze1997,Gouttebroze2002} and helium profiles \citep{Labrosse2001}, and to include a prominence-corona transition region \citep[PCTR,][]{Labrosse2004}, as well as 2D cylindrical versions \citep{Gouttebroze2004,Gouttebroze2005,Gouttebroze2006,2009A&A...503..663G} and multi-threaded models \citep{Labrosse2016} which allow for fine-structure modelling. 
The aim of the work presented here is to build an extensive grid of \ion{Mg}{ii} models, based on the PROM code that can be used  to account for the range of prominence conditions observed.

In Section~\ref{s:models} we start by describing {the numerical code,} the atomic data used to construct our atomic model, the incident radiation illuminating the prominence slab determined from \iris\ observations{, and the two grids of prominence atmosphere models used in this work}.
In Section~\ref{sec:results} we present the {results}  for isothermal, isobaric models, {and for} more realistic models that include a transition region between the prominence core and the corona.
We examine correlations between plasma parameters and observables.
Finally, a summary of our results is given in Section~\ref{s:concl} along {with} our conclusions.

\section{Modelling}\label{s:models}
\subsection{Numerical code}
{The numerical code used in this work is based on the code which was developed by \cite{Gouttebroze1997} to compute the lines and continua emitted by H and \ion{Ca}{ii} in solar prominences. Prominences are represented by static one-dimensional plane-parallel slabs positioned vertically above the solar surface and illuminated by the radiation coming from the solar disc. The radiation is taken to be the same on both sides of the prominence slab, resulting in symmetrical boundary conditions. The height of the prominence is that of the line of sight, which can be at any angle with the boundary. In what follows, all computed intensities are for a line of sight normal to the slab. {Non-LTE} calculations are done first for hydrogen to determine electron densities and the internal radiation field, and to compute the emergent radiation in H lines and continua, as in \cite{2000SoPh..196..349G}; the calculations are then repeated for magnesium.}

\subsection{Model \ion{Mg}{ii} atom}
\label{sec:mgii_atom}

The allowed radiative transitions of \ion{Mg}{ii} result in two strong resonance lines and three weaker triplet lines. 
The \ion{Mg}{ii} h and k resonance lines share the \ion{Mg}{ii} ground {term 3s $^2$S} as their lower level and originate from the {3p $^2$P$^o$} excited {term} of \ion{Mg}{ii}.
There are also three UV triplet lines, which are formed from {a multiplet between the 3d $^2$D and the 3p $^2$P$^o$ terms}, so these transitions are important in populating the h and k upper levels. 
The atomic structure of singly ionised magnesium is similar to that of singly ionised calcium. 
Using these similarities, the code developed by \cite{Gouttebroze1997} to compute the \ion{Ca}{ii} lines is used as a basis for this code. 

In our model, ionisations between \ion{Mg}{i} states and the \ion{Mg}{ii} ground level are not considered. 
The effect of \ion{Mg}{i} on the h and k lines is discussed in detail in \citet{Leenaarts2013}, where they find that the inclusion of \ion{Mg}{i} transitions does not affect the h and k line cores at all in the chromosphere. 
Therefore, it is not necessary to include \ion{Mg}{i} for the calculation of \ion{Mg}{ii} transitions in a prominence. 

We use a simplified five-level (\ion{Mg}{ii}) plus one continuum level (\ion{Mg}{iii}) atom, which is sufficient to describe the h and k lines and the corresponding subordinate lines.
{There are five allowed radiative transitions between these levels}: the h and k resonance lines (2803.53~\AA\ and 2796.35~\AA\ respectively), and three triplet lines (2791.60~\AA, 2798.75~\AA, and 2798.82~\AA).
Partial redistribution is used for the \ion{Mg}{ii} resonance lines.
There are also bound-free, continuous radiative transitions from each of the \ion{Mg}{ii} levels, which result in absorption edges at 824.62~\AA, 1168.25~\AA, 1169.50~\AA, 2008.94~\AA, and 2008.97~\AA. 

Most of our atomic data is as in {\ion{Mg}{ii} model
atoms provided with the RH code} \citep{Uitenbroek2001,Pereira2015b}.
The  energy levels are taken from the {NIST Atomic Spectra Database \citep{1980JPCRD...9....1M}}. Photoionisation cross-sections are from \textit{TopBase} \citep{Cunto1993}. {We use  hydrogenic photoionisation cross-sections for the $^2$D term.} Bound-bound collisional {rates are} from \citet{Sigut1995}. Collisional ionisation rates come from 
PANDORA \citep{Avrett1992}. The abundance of magnesium is taken from \citet{Vial1982b}, which gives a fixed value of $3.5 \times 10^{-5}$. 
Einstein coefficients for all five \ion{Mg}{ii} lines are taken from CHIANTI v8.0 \citep{Dere1997,DelZanna2015}, which uses atomic data from \citet{Liang2009} for \ion{Mg}{ii}.

{We also} consider {Stark} broadening and {Van der Waals} broadening for the \ion{Mg}{ii} h and k lines.
Stark broadening, $\Gamma_{\mathrm{elec}}$, is given by Equation~(\ref{eqn:stark}).
Van der Waals broadening, $\Gamma_{\mathrm{VdW}}$, is given by Equation~(\ref{eqn:vdw}):
\begin{equation}
        \label{eqn:stark}
        \Gamma_{\mathrm{elec}} = \Gamma_{QS} N_e \ ,
\end{equation}
\begin{equation}
        \label{eqn:vdw}
        \Gamma_{\mathrm{VdW}} = \Gamma_{VW} T^{0.3} N_\mathrm{H} \ .
\end{equation}
The Stark and Van der Waals coefficients used here, $\Gamma_{QS}$ and $\Gamma_{VW}$ respectively, are taken from \citet{Milkey1974}. 
These take the values $\Gamma_{QS} = 4.8 \times 10^{-7}$ and $\Gamma_{VW} = 6.6 \times 10^{-10}$. 
As suggested in \citet{Milkey1974}, the value for $\Gamma_{VW}$ has been increased by a factor of $10$ compared to the quoted value ($\Gamma_{VW} = 6.6 \times 10^{-11}$), as their method is said to underestimate the Van der Waals coefficient by that much. 
In Equations~\ref{eqn:stark} and \ref{eqn:vdw}, $N_e$ is electron number density, $T$ is temperature, and $N_\mathrm{H}$ is the hydrogen number density. 

\subsection{Incident radiation}
\label{sec:incident_radiation}

An important aspect of the model is the incident radiation on the prominence slab. 
In this work the incident \ion{Mg}{ii} profiles are taken from \iris\ observations. Other \ion{Mg}{ii} codes have used OSO-8 \citep[][{Orbiting Solar Observatory}]{Vial1982b,Paletou1993}, and RASOLBA \citep{Heinzel2014} observations for the incident profiles. 
\iris\ regularly makes observations of the quiet sun at (or near) disc centre with high spectral, spatial, and temporal resolution. 
The study used must contain all five \ion{Mg}{ii} lines of interest in order to have both the resonance lines and the subordinate lines. 

\begin{figure}
        \begin{center}
                \includegraphics[width=\hsize]{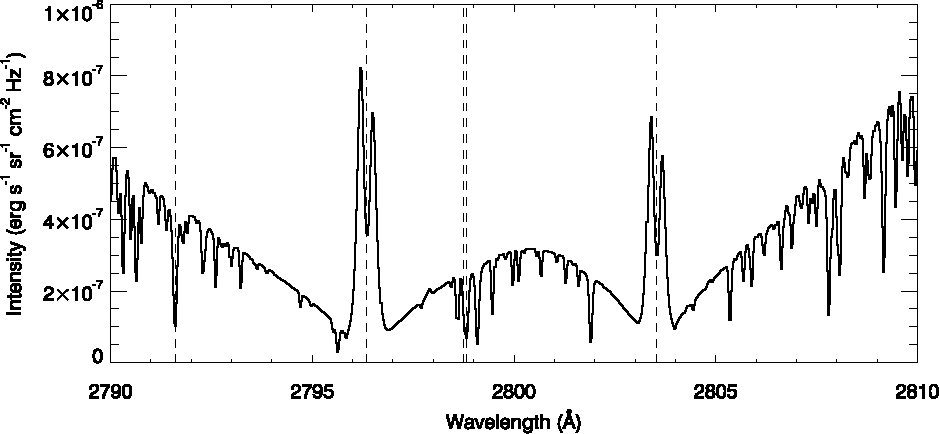}
                \caption{Averaged Sun-centre spectrum from 29 September 2013, used in the calculation of incident profiles in the \ion{Mg}{ii} updates to the PROM code. Dashed lines indicate locations of the five \ion{Mg}{ii} lines of interest. From left to right: 2791.60~\AA, 2796.35~\AA\ (k), 2798.75~\AA, 2798.82~\AA, and 2803.53~\AA\ (h).}
                \label{fig:iris_qs}
        \end{center}
\end{figure}

An observation from 29 September 2013 is used here for the incident profiles, which consisted of one large  raster centred on solar coordinate ($-1$\arcsec,$5$\arcsec). 
This raster is a coarse raster with a total of $64$ $2$\arcsec\ steps, creating a full field of view of $127$\arcsec\ $\times$ $120$\arcsec. 
The raster includes a full CCD readout, meaning that the entire spectral range of \iris\ is available. 
Data is calibrated using standard SSW routines. 

After calibration, the \ion{Mg}{ii} spectra are averaged spatially over the entire  extent of the raster. 
The spatially averaged NUV spectrum around the \ion{Mg}{ii} lines is shown in Figure~\ref{fig:iris_qs}, with positions of the five \ion{Mg}{ii} lines marked with dashed lines. 
From left to right, the lines are 2791.60~\AA\ (subordinate), 2796.35~\AA\ (k), 2798.75~\AA\ (subordinate), 2798.82~\AA\ (subordinate), and 2803.53~\AA\ (h). 

\begin{figure}
        \begin{center}
                \includegraphics[width=\hsize]{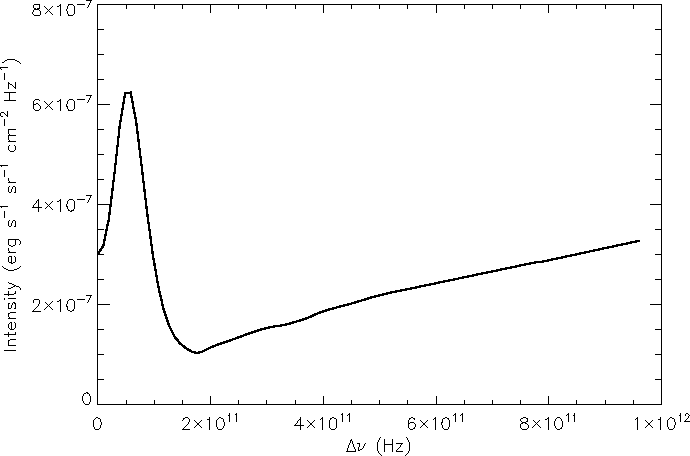}
                \includegraphics[width=\hsize]{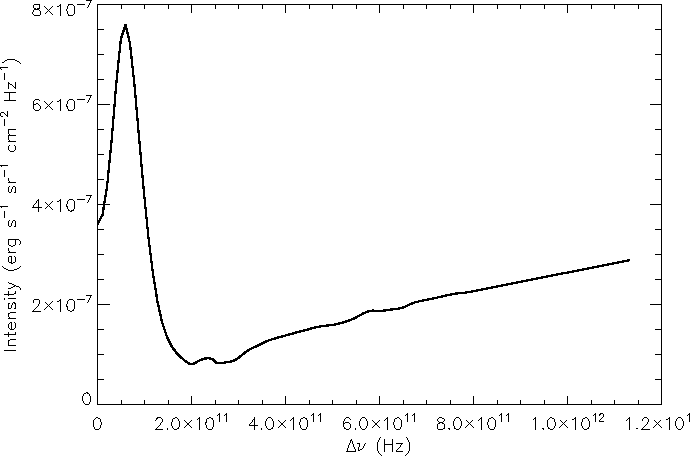}
                \caption{{Incident} half line profiles for \ion{Mg}{ii} h {(\textit{top})} and k {(\textit{bottom})} lines. Line cores ($\Delta \nu \sim 0-2 \times 10^{11}$~Hz) are the average of the blue and red sides of the profile only. Above $\Delta \nu \sim 2 \times 10^{11}$~Hz profiles are averaged and smoothed. Absorption lines in the far wings are removed by interpolating between points in the continuum.}
                \label{fig:final_h_k}
        \end{center}
\end{figure}

Our code requires half profile inputs for each line, which are assumed to be symmetrical about the line core. 
For the h and k lines the profiles are taken from line centre out to $3$~\AA\ in both directions, limited by the shape of the continuum between h and k. 
As can be seen in Figure~\ref{fig:iris_qs} the red and blue peaks of the h and k lines are not symmetrical;   the blue peak is larger in both cases. 
To deal with this asymmetry the half profiles are formed by averaging the blue and red sides of the line together. 
In the wings of these lines, there are a number of absorption lines which are not needed in the calculation of the emergent \ion{Mg}{ii} h and k profiles. 
To help deal with these, smoothing is applied to the wings after averaging the profiles beyond the k$_1$ and h$_1$ positions, and a linear interpolation is performed in the far wings of the profiles in order to remove the remaining absorption features. 
The final incident half profiles for the h and k lines are shown in Figure~\ref{fig:final_h_k}. 

A detailed line profile for the subordinate lines is not required because the incident radiation does not play a large role in the emergent profile.
An approximation is  made such that they can be described by a single representative intensity value at the centroid position of each line. 
This intensity value is taken as the value of the continuum at the line centroid positions -- $4.35 \times 10^{-7}$~erg s$^{-1}$ cm$^{-2}$ sr$^{-1}$ Hz$^{-1}$ for 2791.60~\AA, $2.49 \times 10^{-7}$~erg s$^{-1}$ cm$^{-2}$ sr$^{-1}$ Hz$^{-1}$ for 2798.75~\AA, and $2.54 \times 10^{-7}$~erg s$^{-1}$ cm$^{-2}$ sr$^{-1}$ Hz$^{-1}$ for 2798.82~\AA. 

The photoionisation rates in the \ion{Mg}{ii} continua are calculated using the internal  radiation field resulting from the hydrogen calculations.

\subsection{Grids of prominence model atmospheres}

A basic {non-LTE} 1D model can act as an easy reference for the interpretation of observations. 
This allows a comparison  to be made between observables, such as the line profile characteristics, and the prominence physical parameters, such as temperature, gas pressure, or optical thickness.
{We use two types of prominence model atmospheres, which are described below.}

\subsubsection{Isothermal isobaric models}
To explore the influence of the relevant physical parameters for the formation of the \ion{Mg}{ii} h and k lines in solar prominences, a total of $252$ {isothermal isobaric} models are considered;  the model parameters are shown in Table~\ref{tab:levensgrid}. 
\begin{table}
        \begin{center}
                \caption[Parameters for the extended grid of 1D models.]{Model parameters for the grid of 1D isothermal isobaric models. $T$ denotes temperature, $P$ is gas pressure, and $D$ is slab thickness. $v_T$ and $H$ are microturbulent velocity and prominence height above the solar surface, respectively.}
                \label{tab:levensgrid}
                \begin{tabular}{ l c c }
                        \hline
                        Parameter & Unit & Value \\
                        \hline
                        $T$ & K & $6000$, $8000$, $10000$, \\
                        & & $15000$, $20000$, $25000$, \\
                        & & $30000$, $35000$, $40000$ \\
                        $P$ & dyne~cm$^{-2}$ & $0.01$, $0.02$, $0.05$, $0.1$, \\
                        &  & $0.2$, $0.5$, $1$ \\
                        $D$ & km & $200$, $500$, $1000$, $2000$ \\
                        $v_T$ & km~s$^{-1}$ & $5$ \\
                        $H$ & km & $10000$ \\
                        \hline
                \end{tabular}
        \end{center}
\end{table}
Nine temperatures between $6000$~K and $40000$~K are considered here, as the \ion{Mg}{ii} lines are {formed}  at chromospheric temperatures of $\sim 30000$~K {according to} CHIANTI v8.0 \citep{Dere1997,DelZanna2015}. 
The figure of $30000$~K for the formation of \ion{Mg}{ii} lines comes from calculations considering purely collisional excitation and does not take into account radiative excitation, which is important in prominences. 
Radiative effects can have a large effect on the population of the h and k levels, and the formation of the corresponding resonance lines. In other words, the h and k lines can form at lower plasma temperatures of around $10000$~K \citep{Leenaarts2013,Leenaarts2013b}. 
We use seven gas pressures between $0.01$ and $1$~dyne~cm$^{-2}$.
Slab thicknesses of $200$, $500$, $1000$, and $2000$~km are considered. 
For these isothermal isobaric models the turbulent velocity and prominence height are again kept constant at $v_T = 5$~km~s$^{-1}$ and $H = 10000$~km, respectively.

\subsubsection{PCTR models}
The \ion{Mg}{ii} lines are formed at plasma temperatures around $10000 - 15000$~K, but an isothermal model with that temperature does not best describe the prominence, as we know that there is  lower temperature plasma in the prominence too. 
Therefore, it is more sensible to consider models that have a prominence-to-corona transition region (PCTR) between the cool, dense core and the hot, tenuous corona. 
The PCTR models here follow the temperature and pressure gradients described by \citet{Anzer1999}. 
Equations~(\ref{eqn:pctr_p}) and (\ref{eqn:pctr_t}) show the expressions for the pressure, $P(m)$, and temperature, $T(m)$,  as a function of column mass $m$ across the PCTR: 
\begin{equation}
\label{eqn:pctr_p}
P(m) = 4 P_c \frac{m}{M} \left( 1 - \frac{m}{M} \right) + P_0 \ ,
\end{equation}
\begin{equation}
\label{eqn:pctr_t}
T(m) = T_{\mathrm{cen}} + (T_{\mathrm{tr}} - T_{\mathrm{cen}}) \left( 1 - 4\frac{m}{M} \left( 1 - \frac{m}{M} \right) \right)^\gamma \ .
\end{equation}
The pressure at the outer edge of the PCTR is $P_0$, while at the centre of the slab it is $P_{\mathrm{cen}} = P_c + P_0$. 
The central temperature is $T_{\mathrm{cen}}$, and the temperature at the outer edge of the PCTR is $T_{\mathrm{tr}}$. 
{In Eq.~(\ref{eqn:pctr_t}),} $\gamma$ is a free parameter, with $\gamma \ge 2$. 
When $\gamma$ is small, the PCTR is extended (low temperature gradient), and when $\gamma$ is large, the PCTR is narrow (large temperature gradient). 

For the PCTR models considered here, the $T$ and $P$ values listed in Table~\ref{tab:levensgrid} are taken to be the range of central pressure and temperatures, $P_{\mathrm{cen}}$ and $T_{\mathrm{cen}}$, respectively. 
Column mass values range from around $1 \times 10^{-7}$~g~cm$^{-2}$ to around $1 \times 10^{-3}$~g~cm$^{-2}$. The temperature 
$T_{\mathrm{tr}}$ is set with a value of $1 \times 10^5$~K, with $P_0$ taking a value of $0.01$~dyne~cm$^{-2}$. 
The parameter $\gamma$ is varied and has three values:  $2$, $5$, and $10$. 
The microturbulent velocity is again kept at a constant $5$~km~s$^{-1}$, and the prominence height at $10000$~km. 
A summary of the parameters used in the PCTR models is presented in Table~\ref{tab:pctr}. 
\begin{table}
        \begin{center}
                \caption{Parameters used in the grid of models with a PCTR: $T_{\mathrm{cen}}$ is central temperature, $T_{\mathrm{tr}}$ is temperature at the edge of the slab, $P_{\mathrm{cen}}$ is the pressure at the centre of the slab, $P_0$ is pressure at the edge of the slab, $M$ is column mass, $\gamma$ controls PCTR size, $v_T$ is microturbulent velocity, and $H$ is height above {the} solar surface.}
                \label{tab:pctr}
                \begin{tabular}{ l c c }
                        \hline
                        Parameter & Unit & Value \\
                        \hline
                        $T_{\mathrm{cen}}$ & K & $6000$, $8000$, $10000$, \\
                        & & $15000$, $20000$, $25000$, \\
                        & & $30000$, $35000$, $40000$ \\
                        $T_{\mathrm{tr}}$ & K & $100000$ \\
                        $P_{\mathrm{cen}}$ & dyne~cm$^{-2}$ & $0.01$, $0.02$, $0.05$, $0.1$, \\
                        &  & $0.2$, $0.5$, $1$ \\
                        $P_0$ & dyne~cm$^{-2}$ & $0.01$ \\
                        $M$ & g~cm$^{-2}$ & $1 \times 10^{-7}-1 \times 10^{-3}$ \\
                        $\gamma$ & & $2$, $5$, $10$ \\
                        $v_T$ & km~s$^{-1}$ & $5$ \\
                        $H$ & km & $10000$ \\
                        \hline
                \end{tabular}
        \end{center}
\end{table}
In total there are $252 \times 3 = 756$ possible unique PCTR models. This extended grid covers values that are not necessarily all relevant to typical solar prominence conditions, but will allow us to have a good understanding of the formation mechanisms of \ion{Mg}{ii} lines in these structures.

\section{Results}\label{sec:results}
With spectrometers such as \iris, detailed line profiles of the \ion{Mg}{ii} lines are obtainable, meaning that line features (central reversals, line intensities, etc.) can be measured in prominences with a high spectral resolution. 
This section aims to explore whether any of these observables are related to the prominence parameters (temperature, gas pressure, slab thickness, optical thickness).
Observable characteristics include the k-to-h {line} ratio {(defined as the ratio of frequency-integrated intensities of the two lines)}, or {the} k (or h) line reversal level,  also referred to generally as $I_\mathrm{p}/I_0$ {(i.e. the ratio of specific intensities at the peak, $I_\mathrm{p}$, to those at the line centre, $I_0$)} and specifically for h and k as $I(\text{h}_2)/I({h}_3)$ and $I(\text{k}_2)/I({k}_3)$, respectively.
{In the following, we present results concerning correlations between observables and model parameters, and correlations between \ion{Mg}{ii} and \ion{H}{i} intensities. We start by exploring how observable properties of the \ion{Mg}{ii} h and k lines are related to each other. We focus on the following observables:
\begin{itemize}
        \item frequency-integrated intensities of the \ion{Mg}{ii} h and k lines, and of the \ion{H}{i} Ly$\alpha$ and H$\alpha$ lines;
        \item \ion{Mg}{ii} h and k reversal levels;
        \item emergent \ion{Mg}{ii} h and k line profiles;
        \item \ion{Mg}{ii} k/h line ratio.
\end{itemize} }

\begin{figure}
        \begin{center}
                \includegraphics[width=0.49\hsize,clip=true,trim=1cm 0.5cm 0.9cm 1cm]{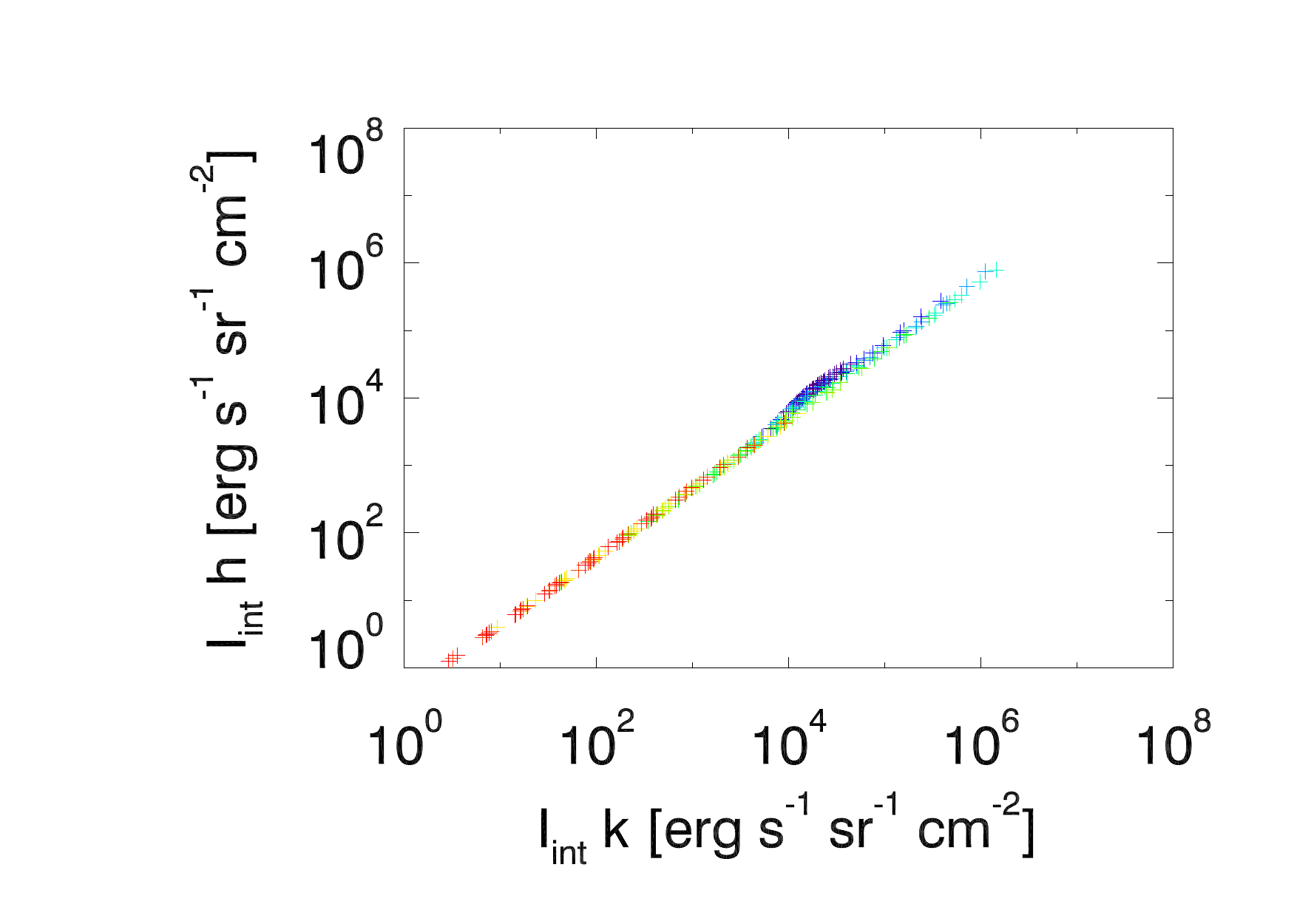}
                \includegraphics[width=0.49\hsize,clip=true,trim=1cm 0.5cm 0.9cm 1cm]{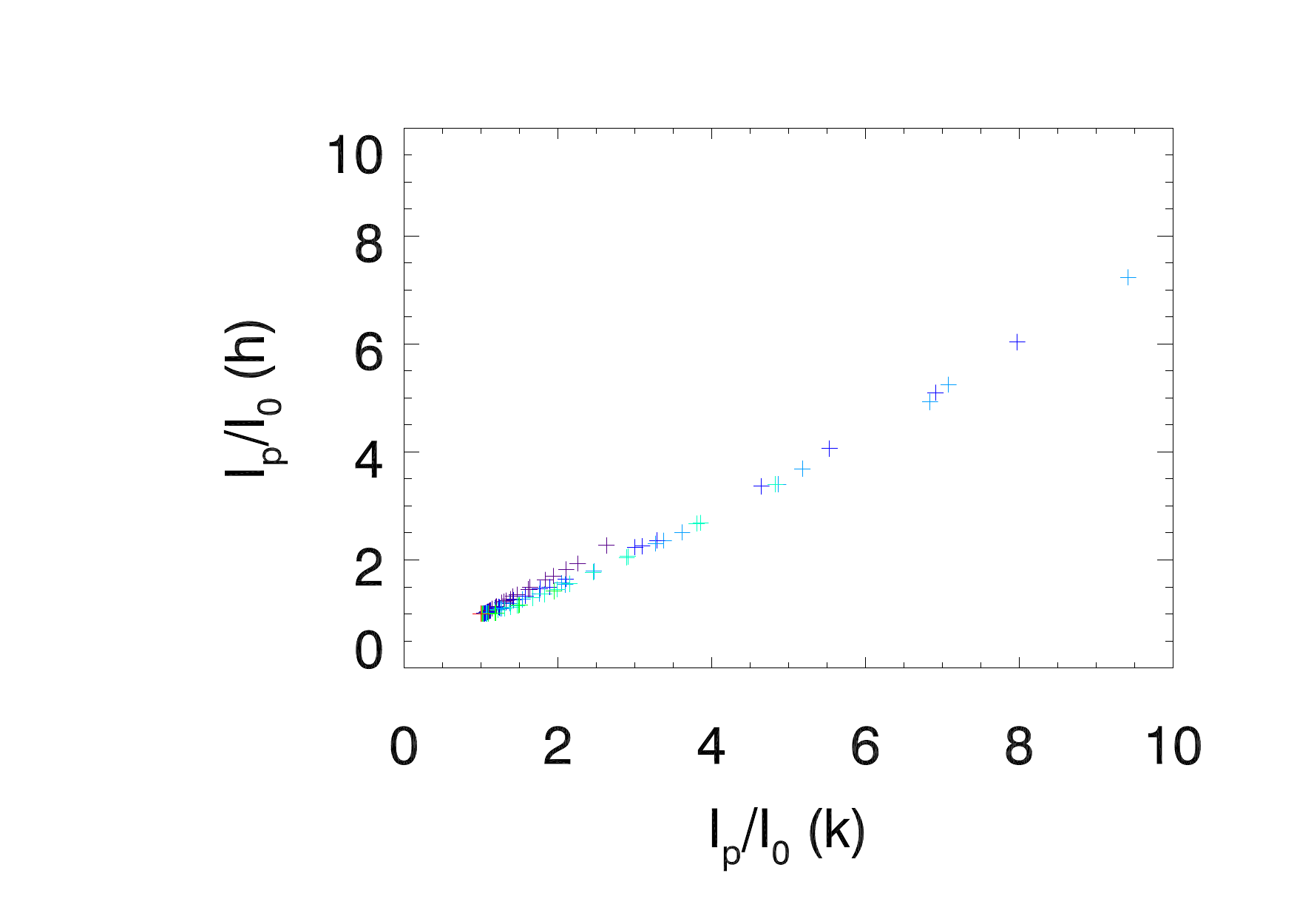}
                \includegraphics[width=0.49\hsize,clip=true,trim=1cm 0.5cm 0.9cm 1cm]{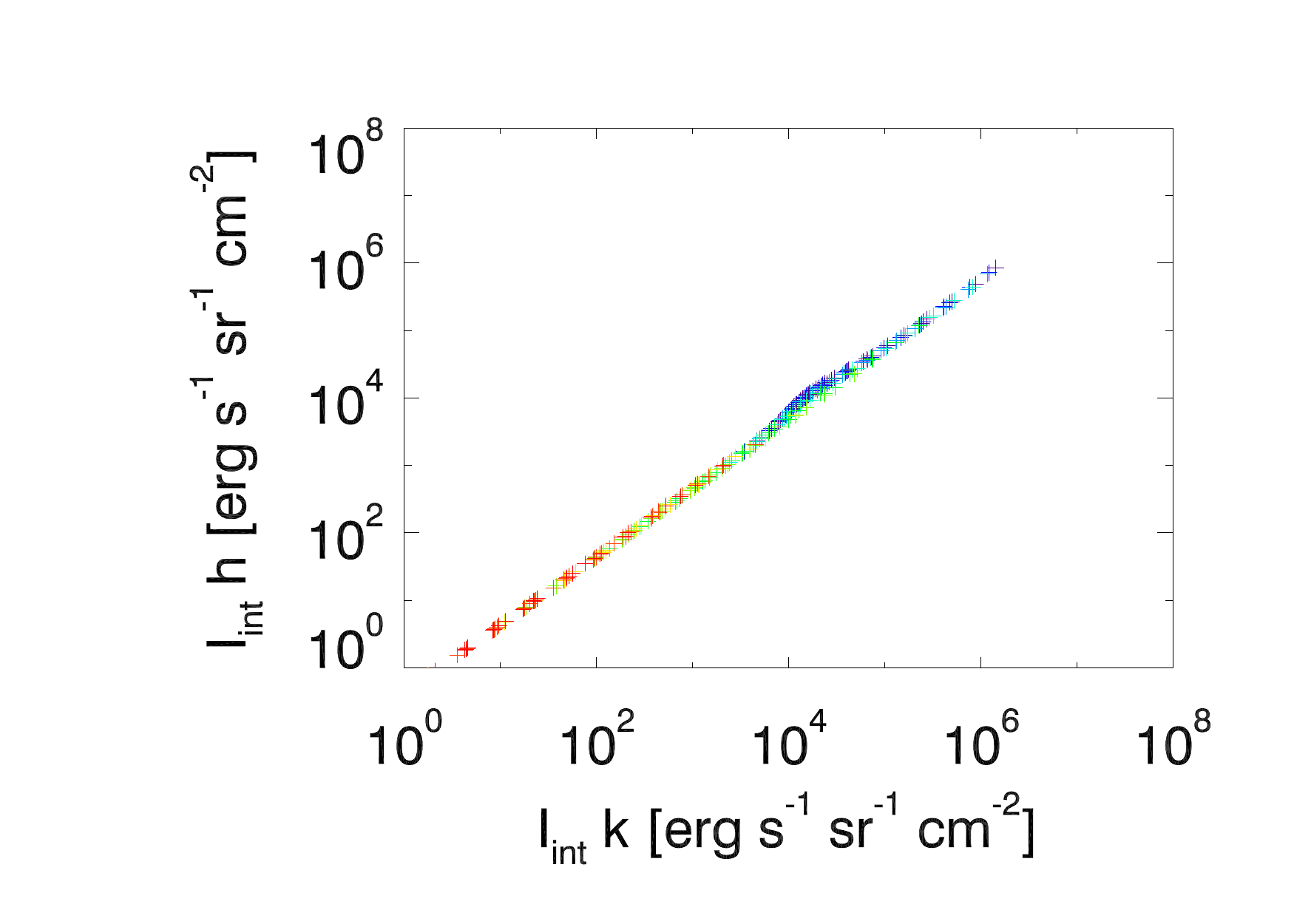}
\includegraphics[width=0.49\hsize,clip=true,trim=1cm 0.5cm 0.9cm 1cm]{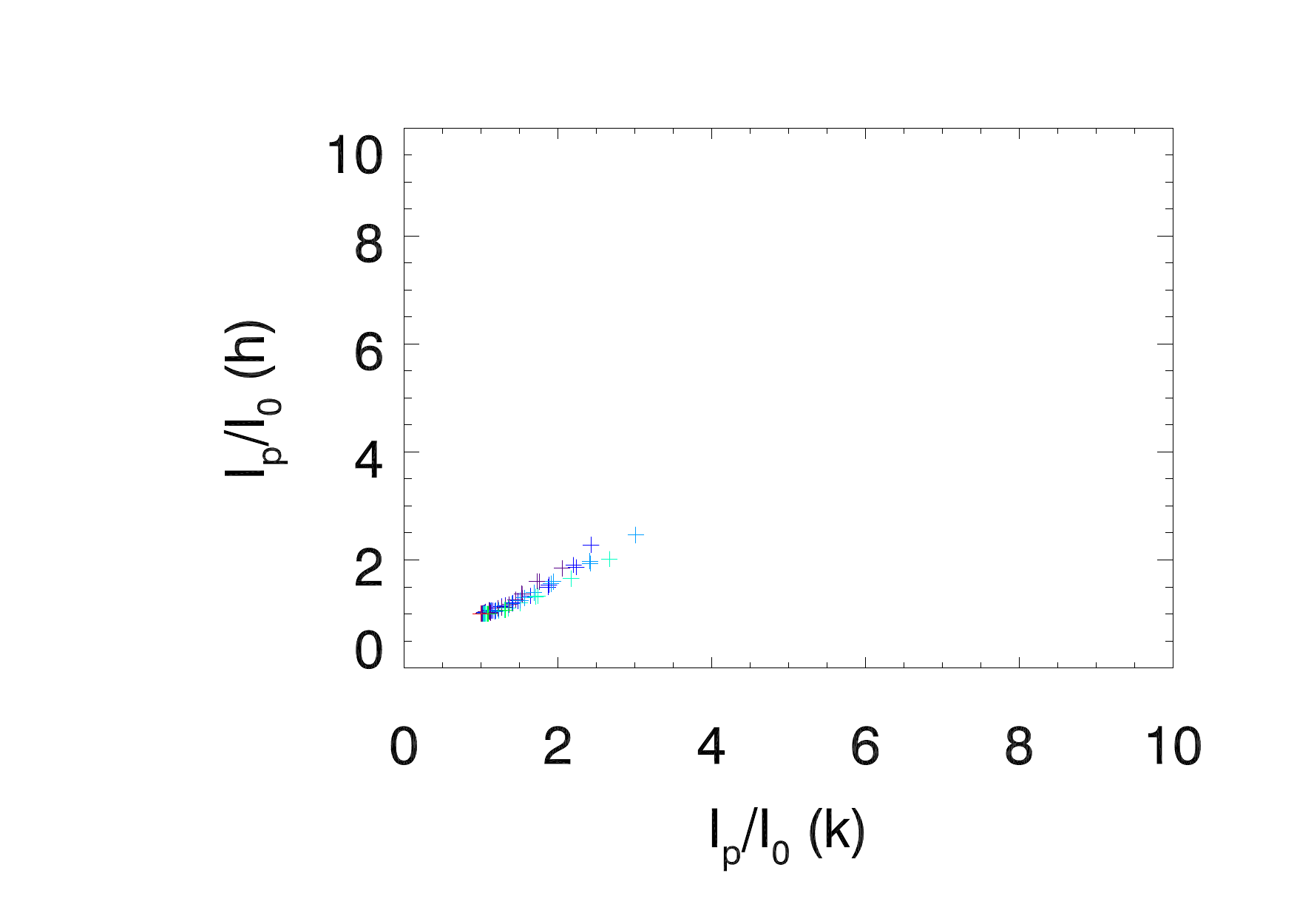}\\
\includegraphics[width=0.49\hsize,clip=true,trim=1cm 0.5cm 0.9cm 1cm]{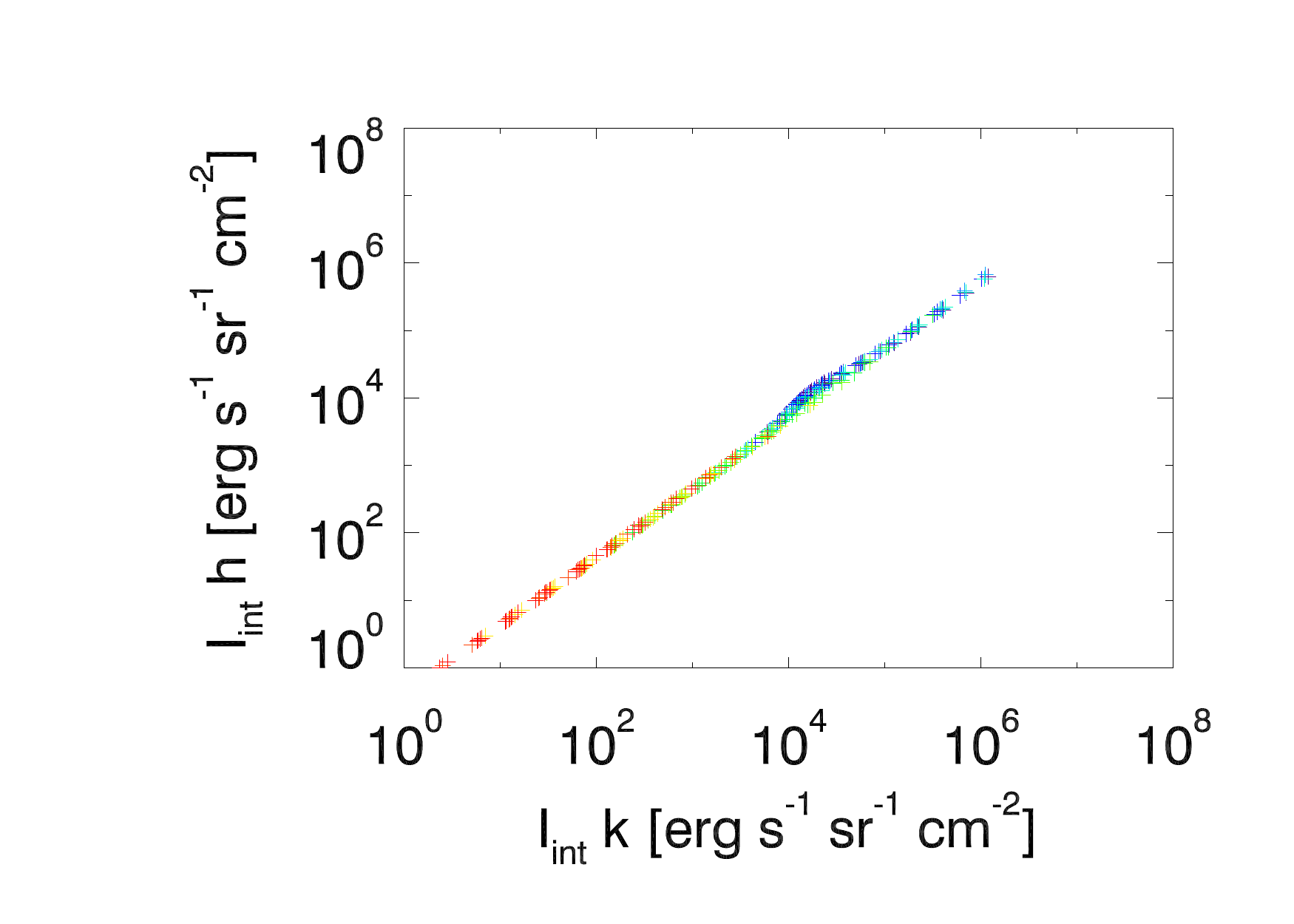}
\includegraphics[width=0.49\hsize,clip=true,trim=1cm 0.5cm 0.9cm 1cm]{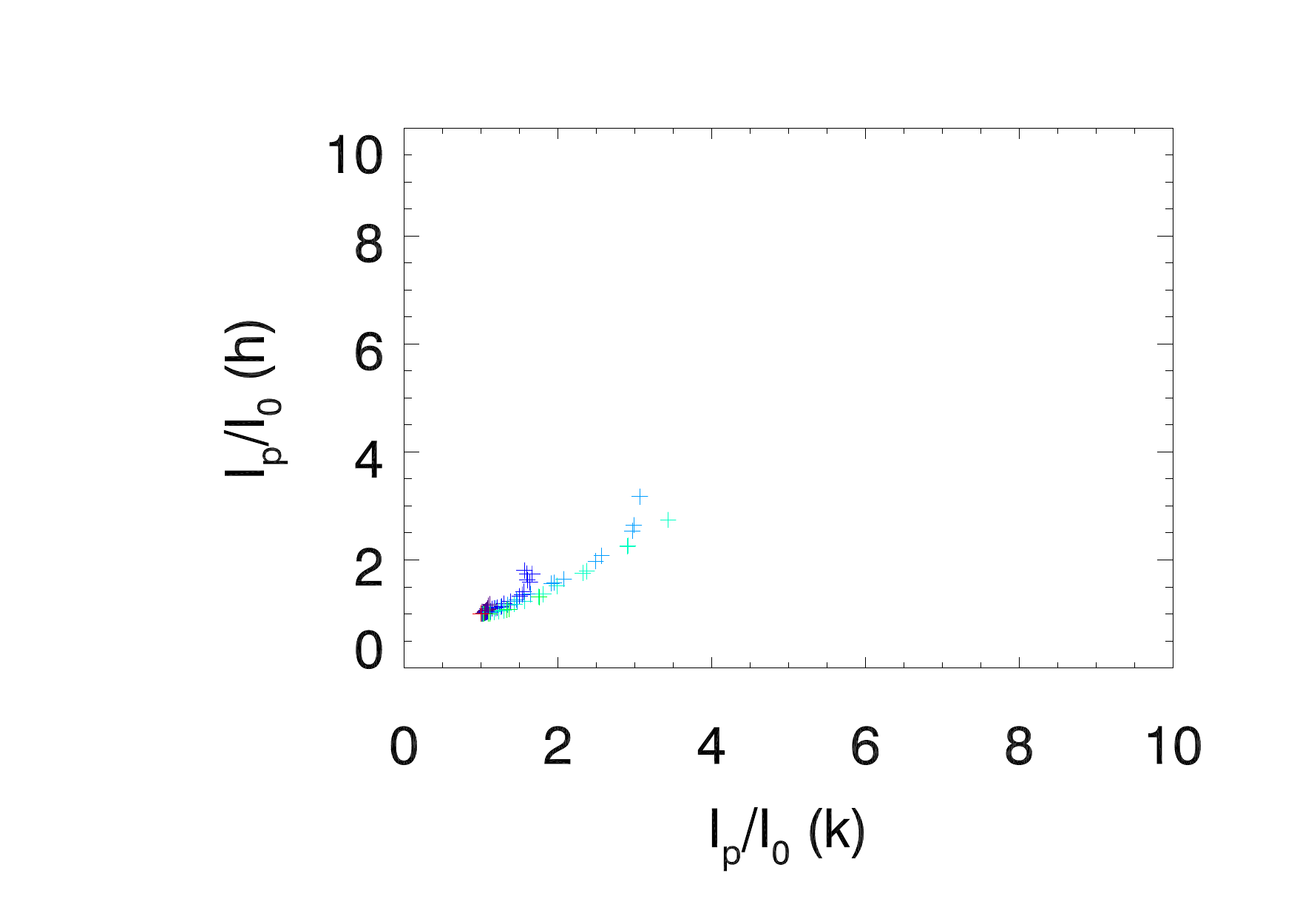}\\
\includegraphics[width=0.49\hsize,clip=true,trim=1cm 0.5cm 0.9cm 1cm]{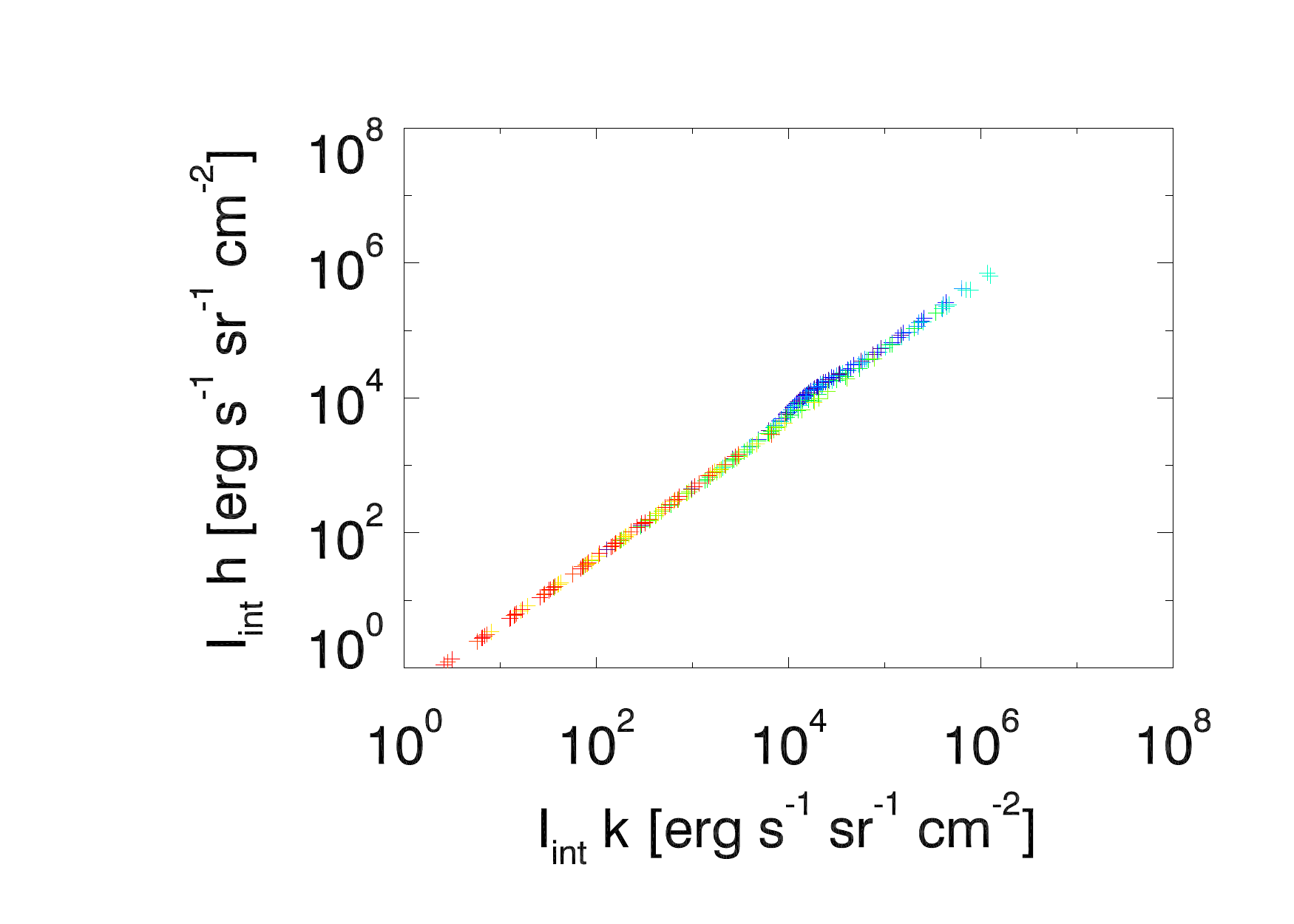}
\includegraphics[width=0.49\hsize,clip=true,trim=1cm 0.5cm 0.9cm 1cm]{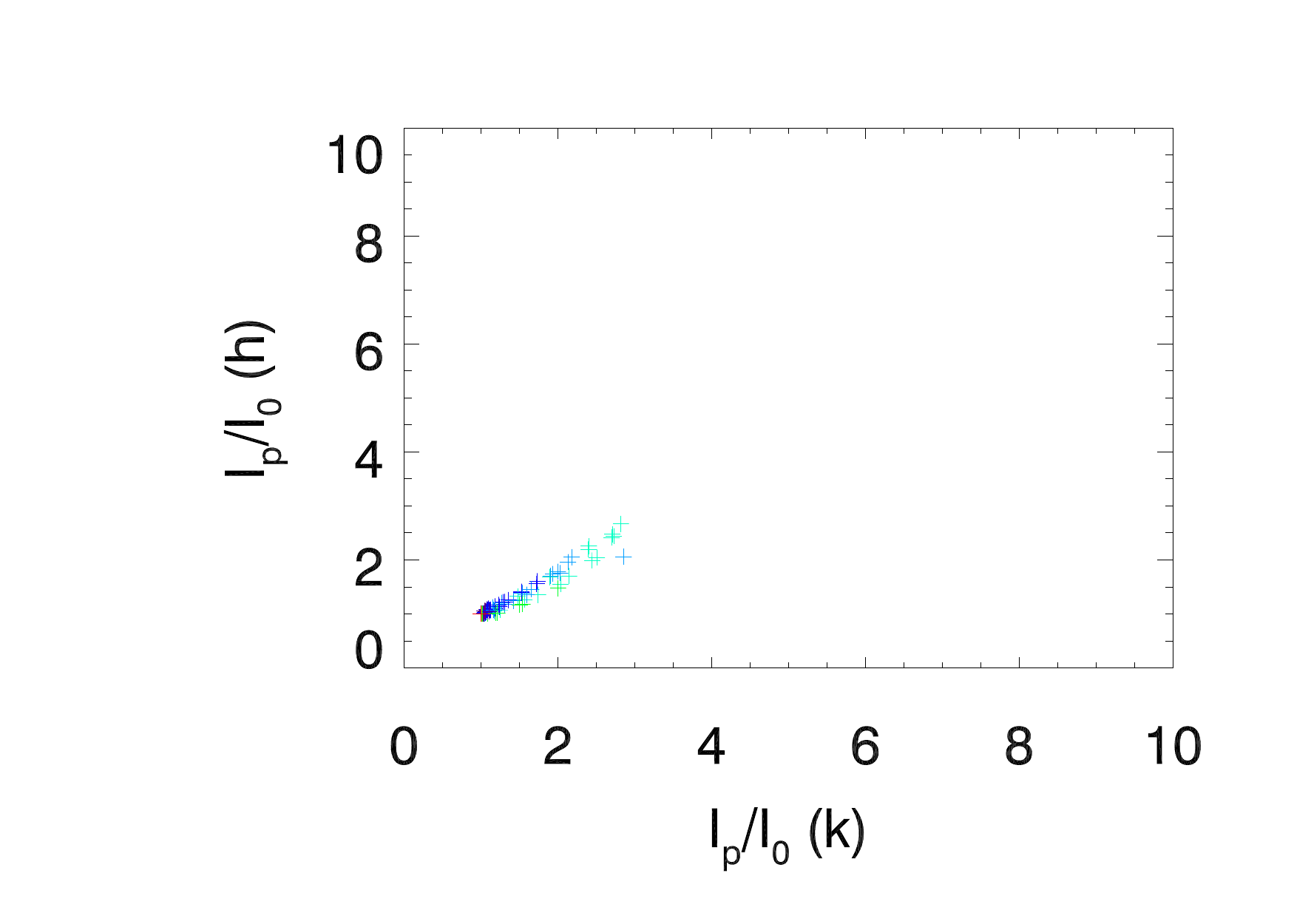}
                \caption{Correlations between observable properties of the \ion{Mg}{ii} h and k lines. \textit{Left:} Correlation between {frequency-}integrated intensities of the \ion{Mg}{ii} h and k lines. \textit{Right:} Correlation between \ion{Mg}{ii} h and k reversal levels. {Top row:} Points are for {all} isothermal isobaric models. {Three bottom rows:} Points are for {all} PCTR models. \textit{Second row:} $\gamma = 2$. \textit{Third row:} $\gamma = 5$. \textit{Bottom:} $\gamma = 10$. {Colour indicates temperature, increasing from purple to red.}}
                \label{fig:h_vs_k_comb}
        \end{center}
\end{figure}

Figure~\ref{fig:h_vs_k_comb} shows relationships between observable parameters for h versus k for all {isothermal, isobaric (top row) and PCTR (three bottom rows for each of the three $\gamma$ values)} models. 
The plots {of the frequency-integrated intensity of \ion{Mg}{ii} h against the frequency-integrated intensity of \ion{Mg}{ii} k} show an almost perfect power-law relationship, with a power-law index of  {$1.01$ in the $\gamma=10$ case, and $1.02$ in all the other cases}, as  expected from the ratio of oscillator strengths of the h and k lines. However there is a small `bump' at around $I_{\mathrm{int}}$ (k) = $10^4$~erg~s$^{-1}$~sr$^{-1}$~cm$^{-2}$ {corresponding to central temperatures lower than 10000~K}. 
This departure from linearity can be explained by considering the optical thickness of the lines{ (see below). The optical thickness is defined as usual, and measures the absorption of radiation across the slab width $$\tau = \int_{0}^{D}\alpha \mathrm{d}s \ , $$ with $D$ the slab width, $\alpha$ the absorption coefficient in a given line, and $\mathrm{d}s$ a small path element.}

The right-hand {panels} of Fig.~\ref{fig:h_vs_k_comb} show the correlation between {reversal levels} of the h and k lines. 
There is a roughly linear relationship between h line reversal {level} and k line reversal {level}. 
Most points are clustered at lower reversal {levels} ($\sim 1 - 2$), with relatively few models showing {a} high reversal {level} in either line. 
{We see} a slightly different spread of values {between the PCTR and} the isothermal isobaric models. 
{In the PCTR case,} there are no models with higher ($> 4$) reversal levels in either line for any $\gamma$ {value}. 
This indicates that the inclusion of a PCTR inhibits the creation of extremely deep central reversals in the line profiles; the presence of a PCTR near the surface of the slab changes the source function in the slab, due to the larger effect of collisional excitation.

\begin{figure}
        \begin{center}
                \includegraphics[width=0.49\hsize,clip=true,trim=1cm 0.7cm 0.9cm 1cm]{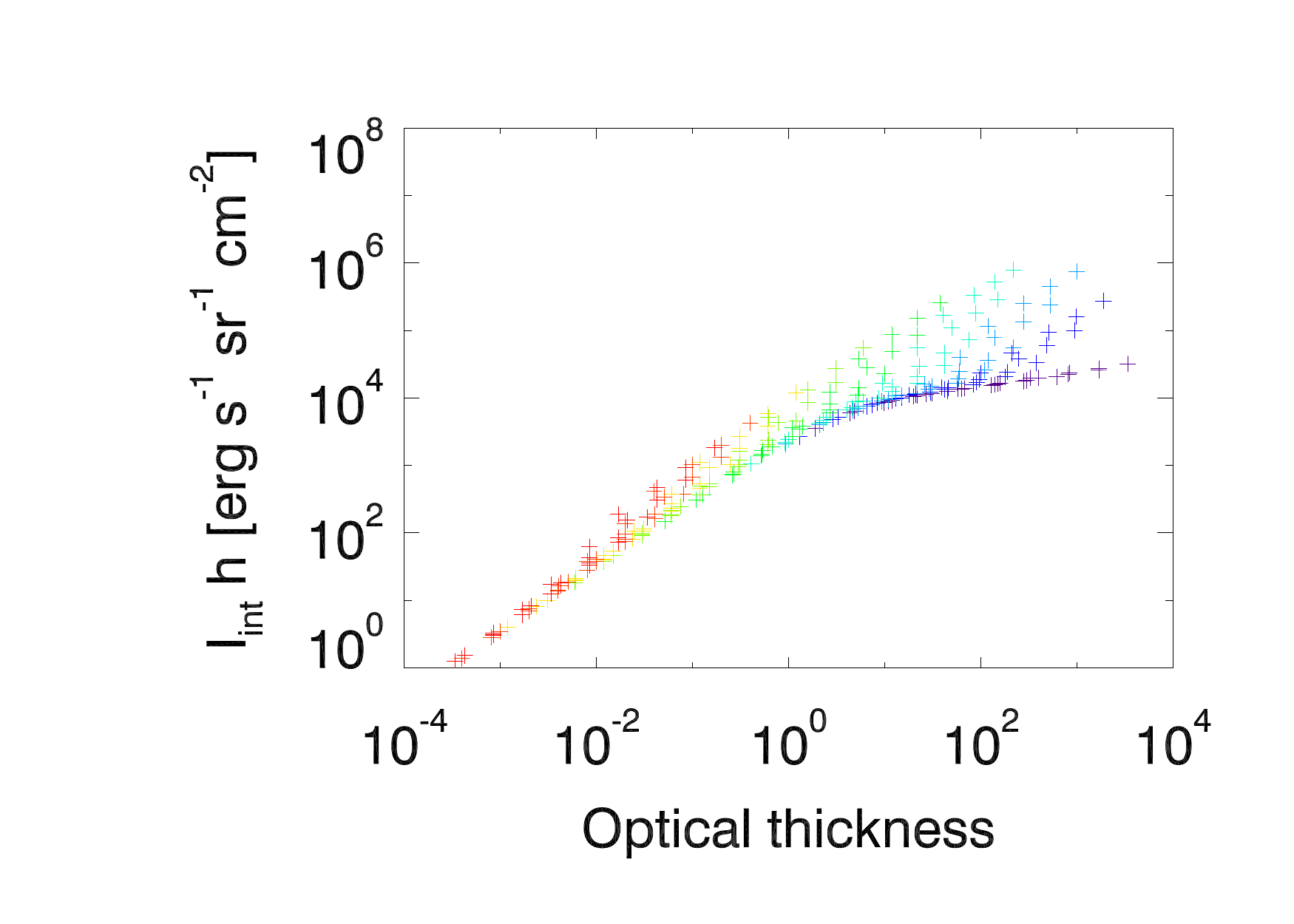}
                \includegraphics[width=0.49\hsize,clip=true,trim=1cm 0.7cm 0.9cm 1cm]{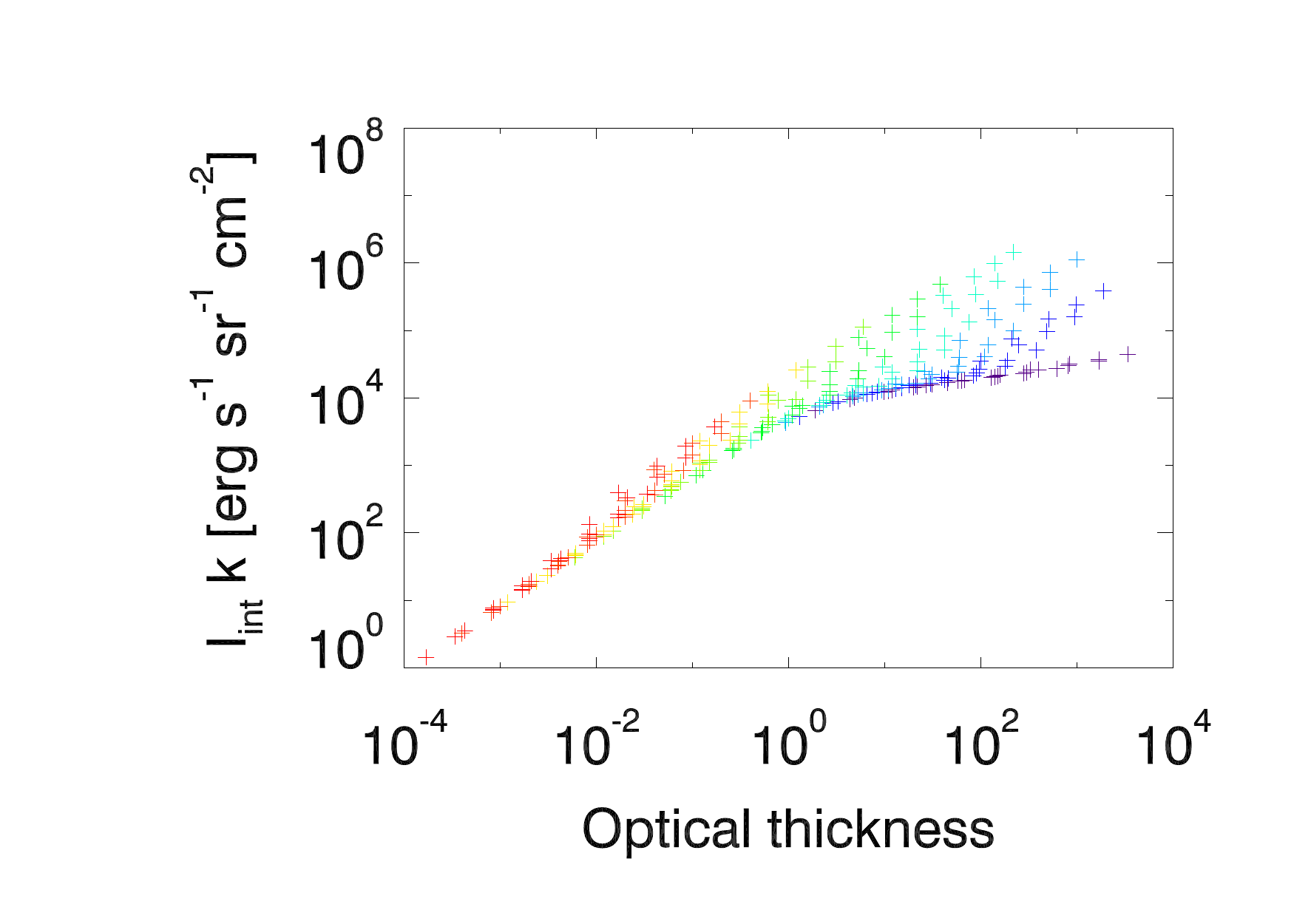}
                \includegraphics[width=0.49\hsize,clip=true,trim=1cm 0.7cm 0.9cm 1cm]{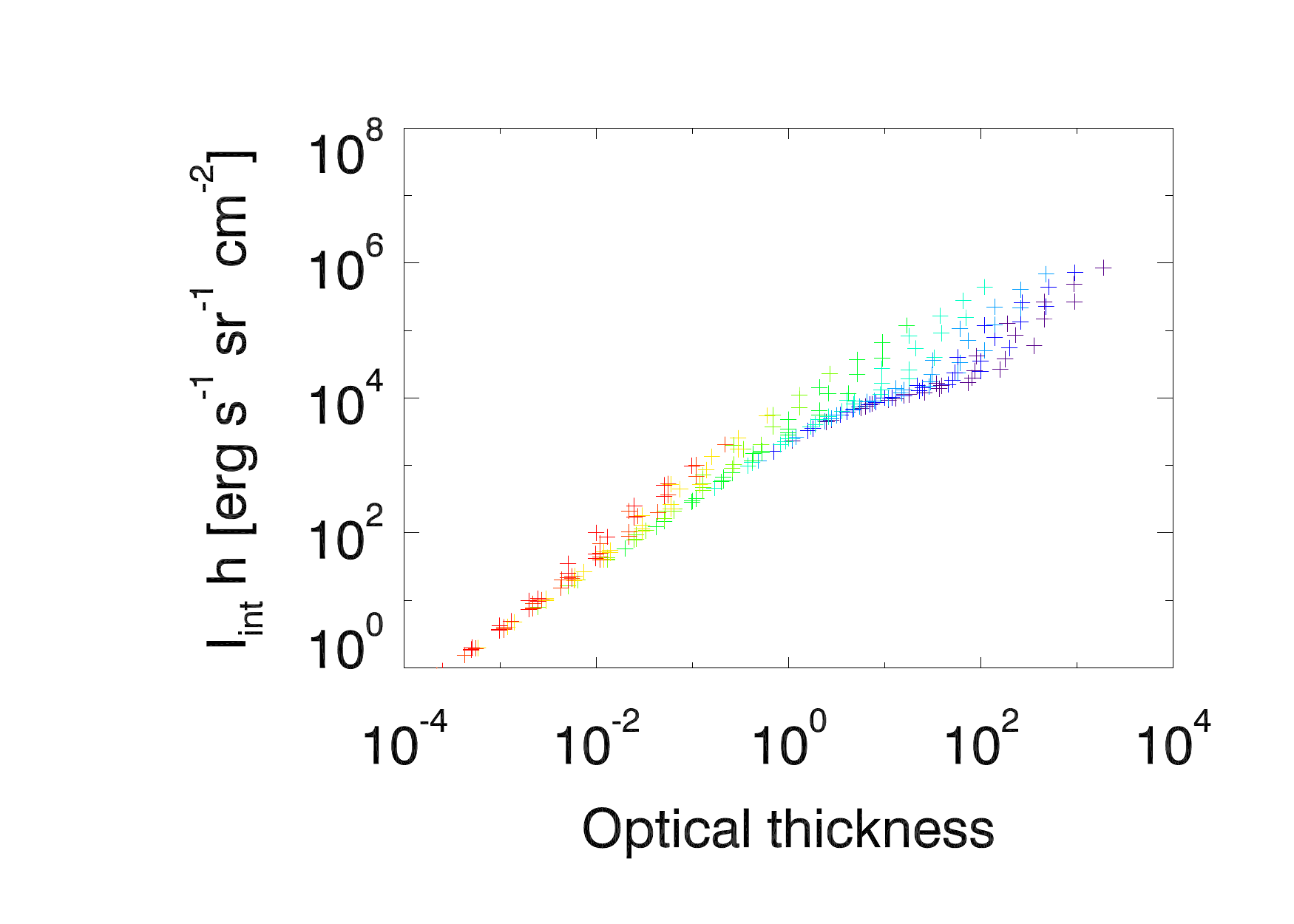}
\includegraphics[width=0.49\hsize,clip=true,trim=1cm 0.7cm 0.9cm 1cm]{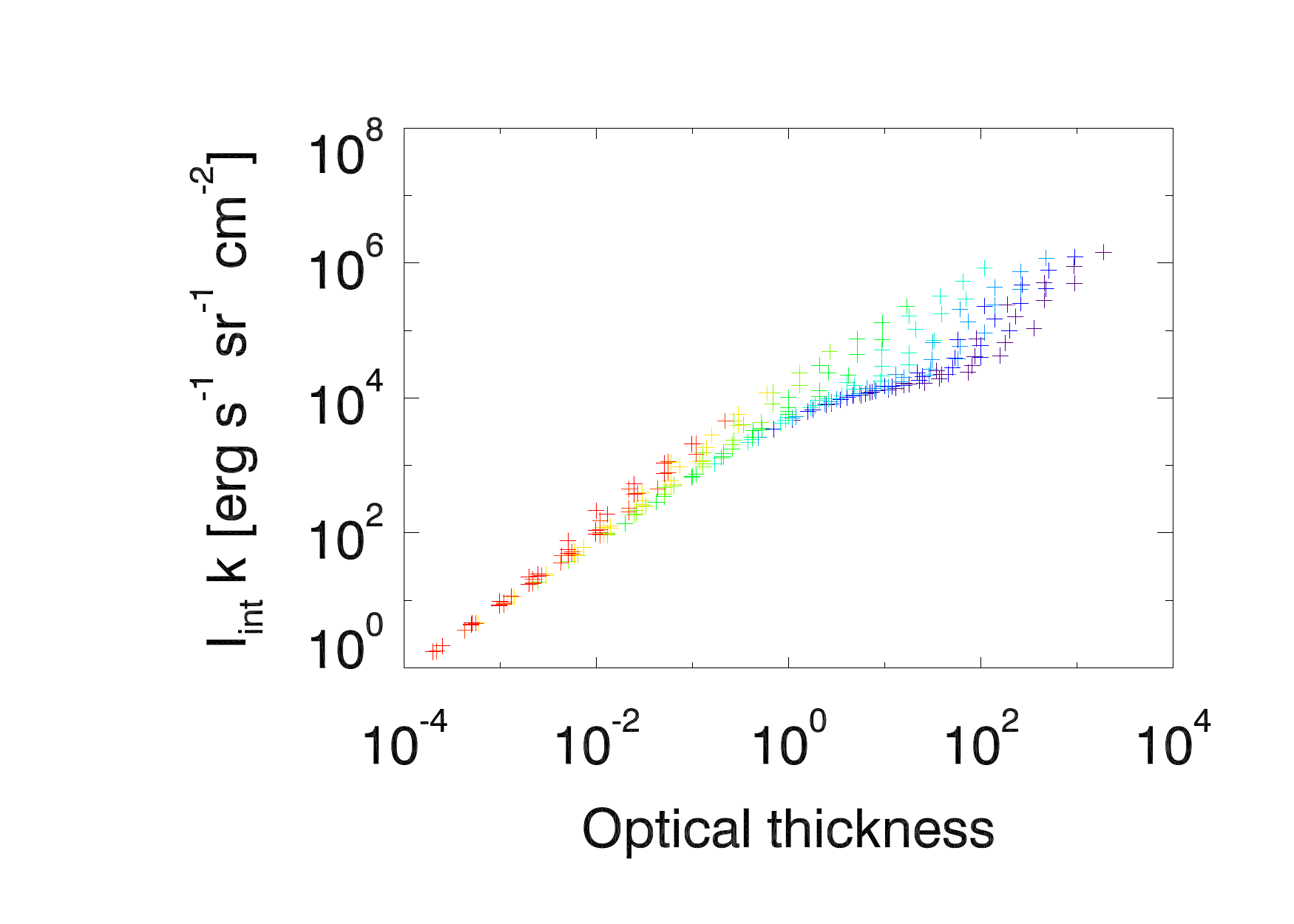}\\
\includegraphics[width=0.49\hsize,clip=true,trim=1cm 0.7cm 0.9cm 1cm]{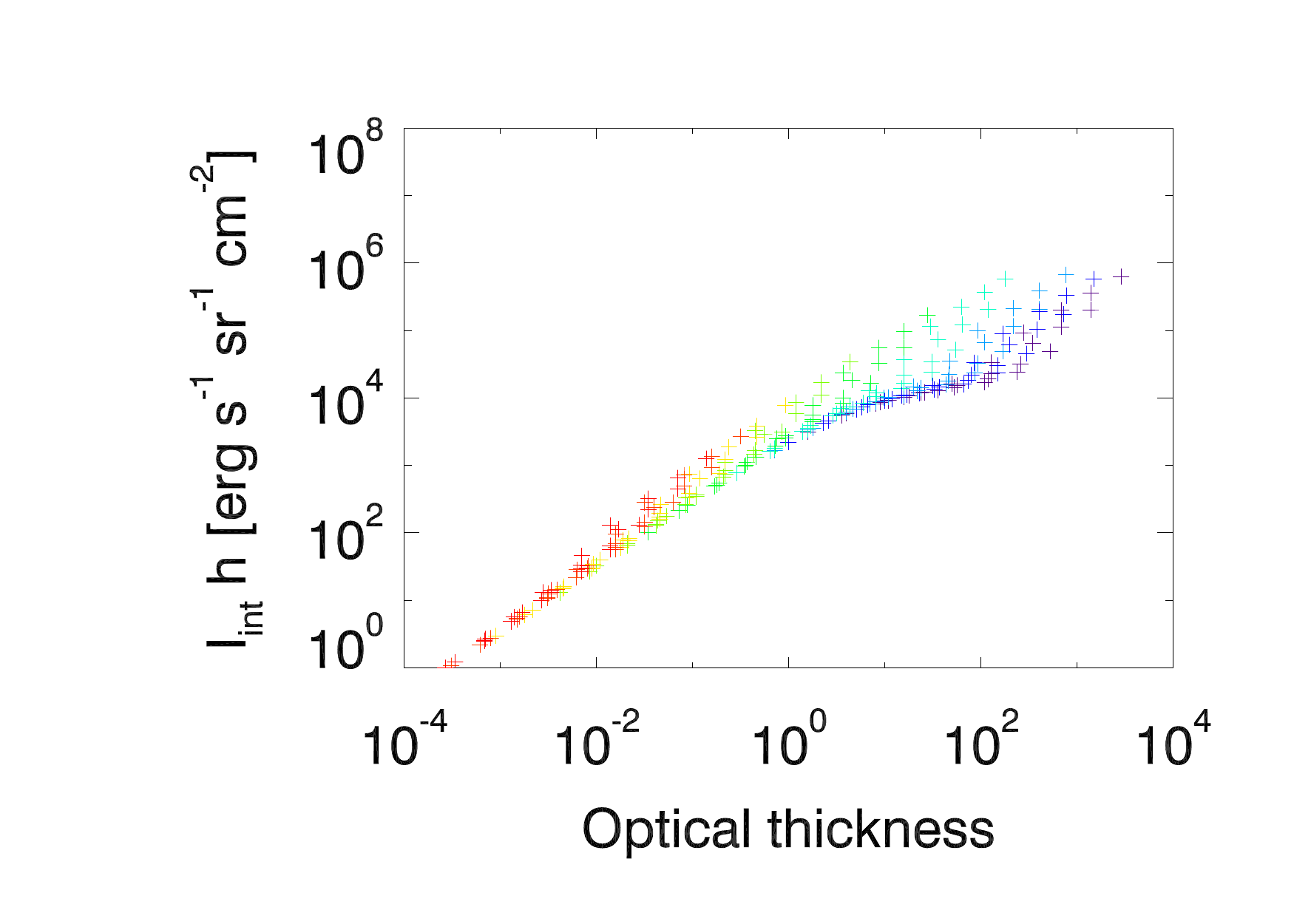}
\includegraphics[width=0.49\hsize,clip=true,trim=1cm 0.7cm 0.9cm 1cm]{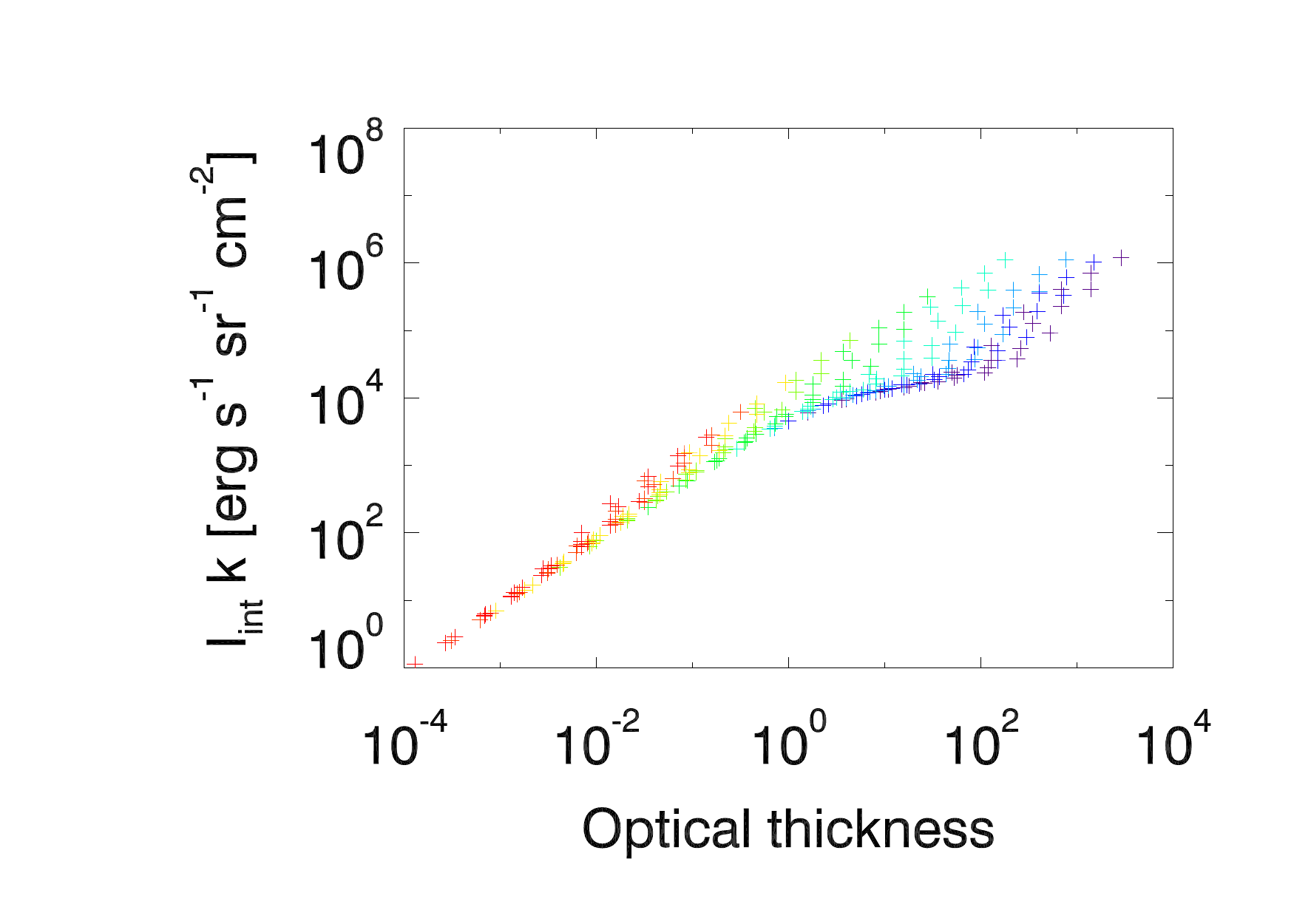}\\
\includegraphics[width=0.49\hsize,clip=true,trim=1cm 0.7cm 0.9cm 1cm]{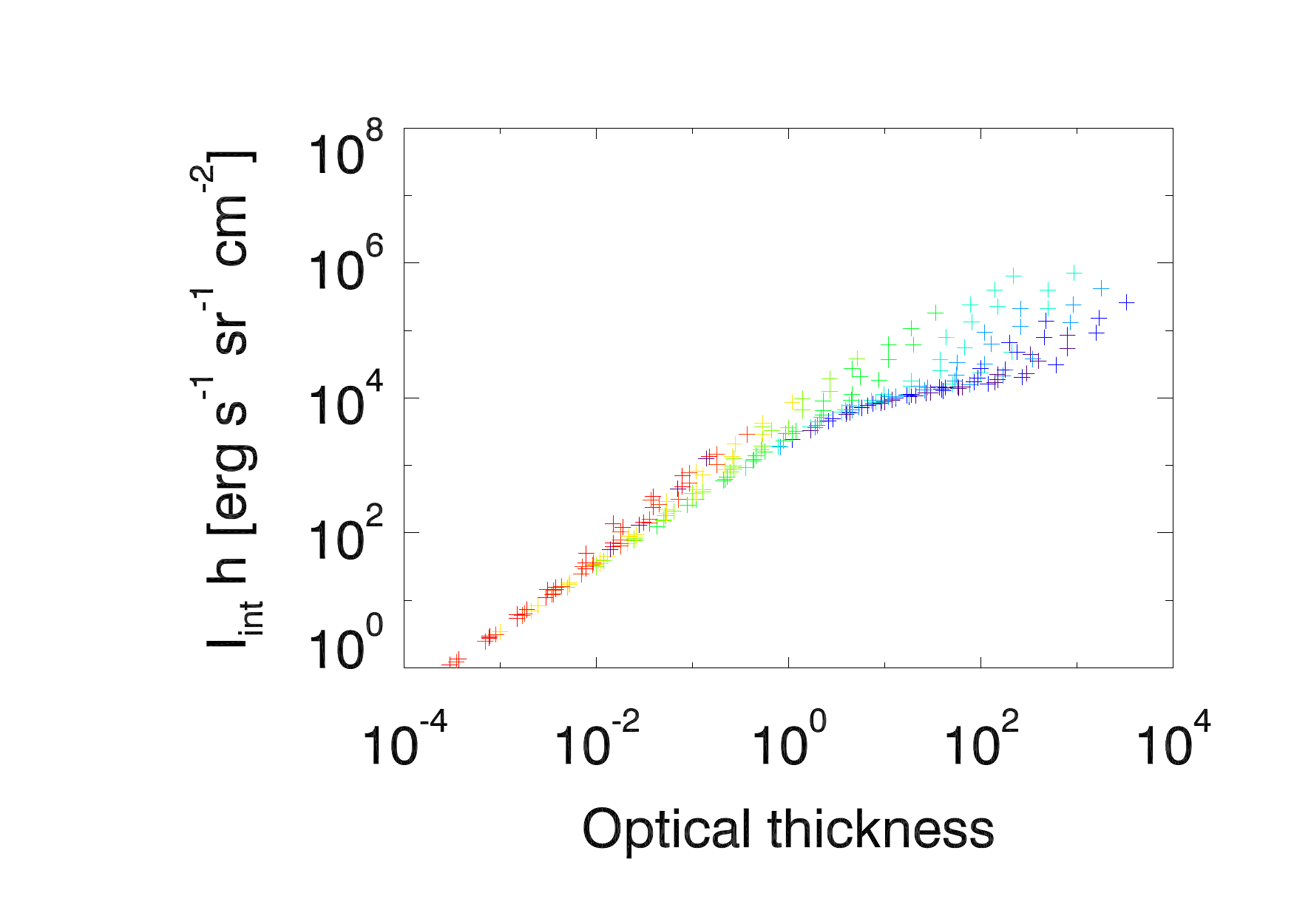}
\includegraphics[width=0.49\hsize,clip=true,trim=1cm 0.7cm 0.9cm 1cm]{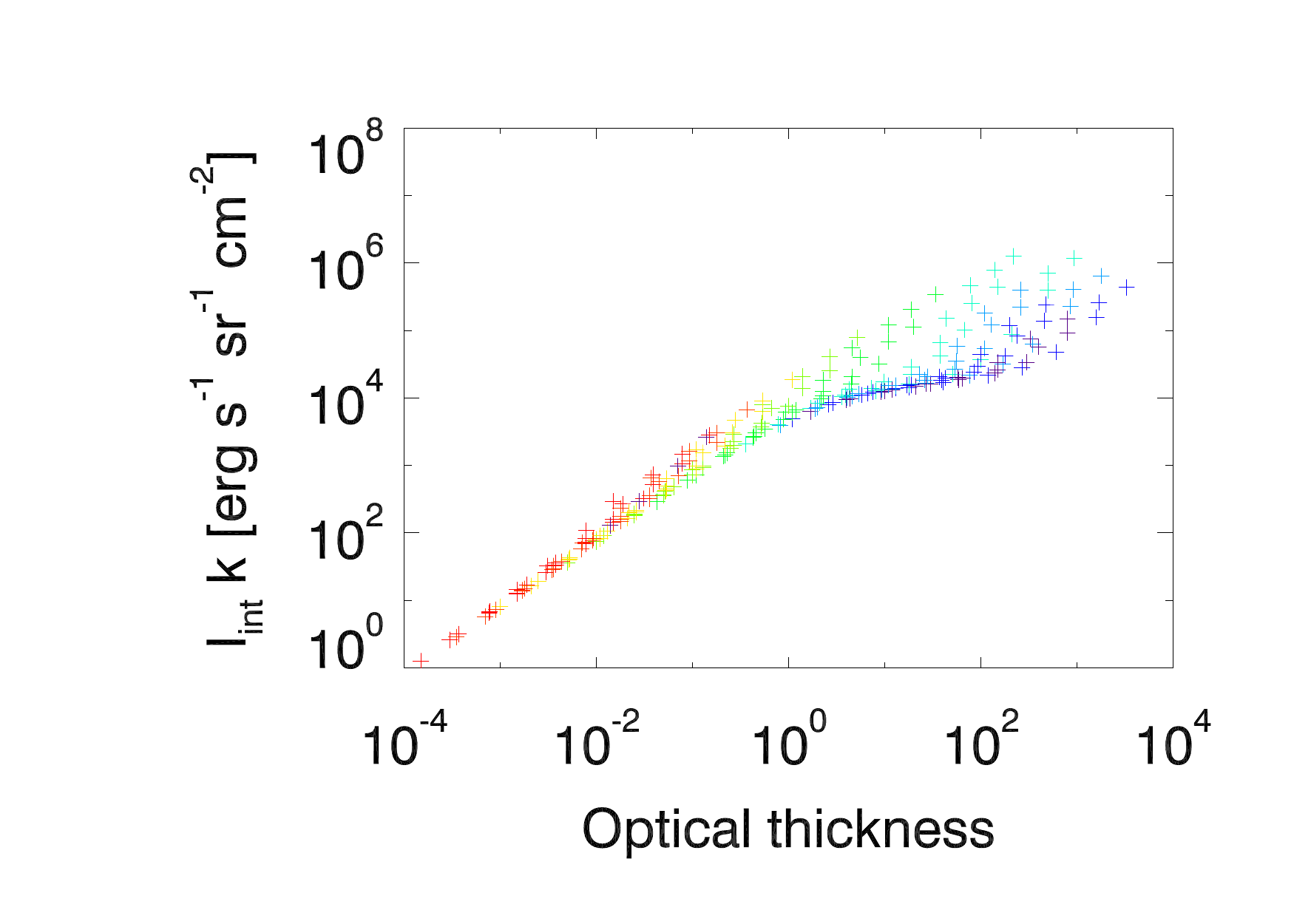}
                \caption{Plots of {frequency-}integrated intensity against optical thickness for the \ion{Mg}{ii} resonance lines. \textit{Left:} \ion{Mg}{ii} h. \textit{Right:} \ion{Mg}{ii} k. {Top row:} Points are for {all} isothermal isobaric models. {Three bottom rows:} Points are for {all} PCTR models. \textit{Second row:} $\gamma = 2$. \textit{Third row:} $\gamma = 5$. \textit{Bottom:} $\gamma = 10$. {Colour indicates temperature, increasing from purple to red.}}
                \label{fig:252_tau_comb}
        \end{center}
\end{figure}

Figure~\ref{fig:252_tau_comb} presents the h  and k  {frequency-}integrated intensities versus the optical thickness of the line for all {isothermal isobaric  and PCTR } models. 
The `bump' in Figure~\ref{fig:h_vs_k_comb} corresponds to the intensity where the h and k lines become optically thick, and the models with the highest optical thickness can be found to have a {frequency-}integrated intensity around $10^4$~erg~s$^{-1}$~sr$^{-1}$~cm$^{-2}$ in both lines. 
A small number of models deviate from this value, having higher intensities for optically thick models, creating a scatter of points in Figure~\ref{fig:252_tau_comb}. 
These mostly continue the approximate power law seen at lower optical thicknesses and {frequency-}integrated intensities.
The optical thickness for the PCTR models is similar to that of the isothermal isobaric values, but at high optical thicknesses there is some variation. 
In the PCTR models, the higher optical thickness values are only found at higher {frequency-}integrated intensities:  the relation is more linear and there is less `flattening' of the scatter plots at $\sim 10^4$~erg~s$^{-1}$~sr$^{-1}$~cm$^{-2}$. 
This has an effect on the resulting line profiles (Fig.~\ref{fig:252_profiles_comb}).

\begin{figure}
        \begin{center}
                \includegraphics[width=0.49\hsize,clip=true,trim=1cm 0.7cm 0.9cm 0cm]{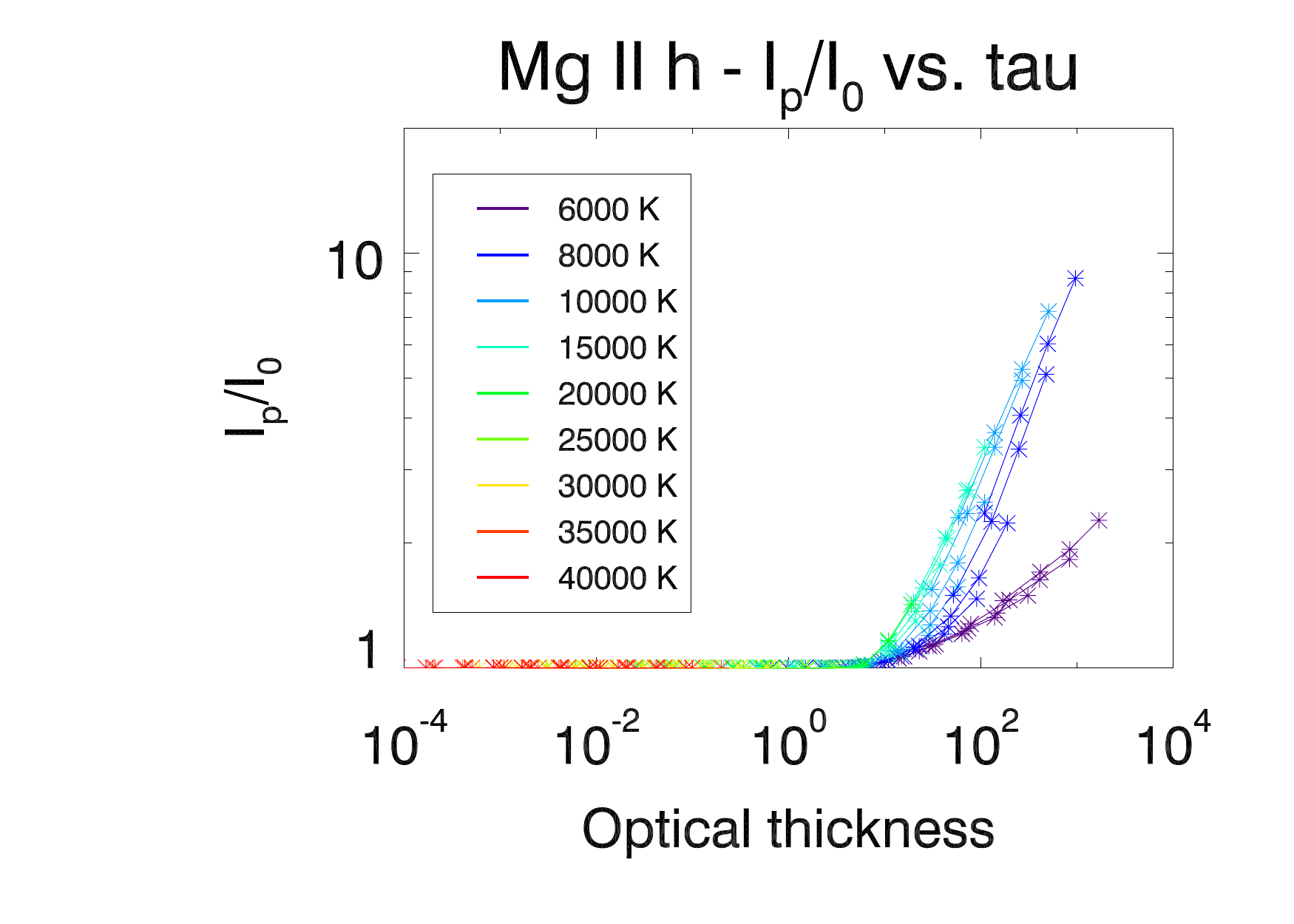}
                \includegraphics[width=0.49\hsize,clip=true,trim=1cm 0.7cm 0.9cm 0cm]{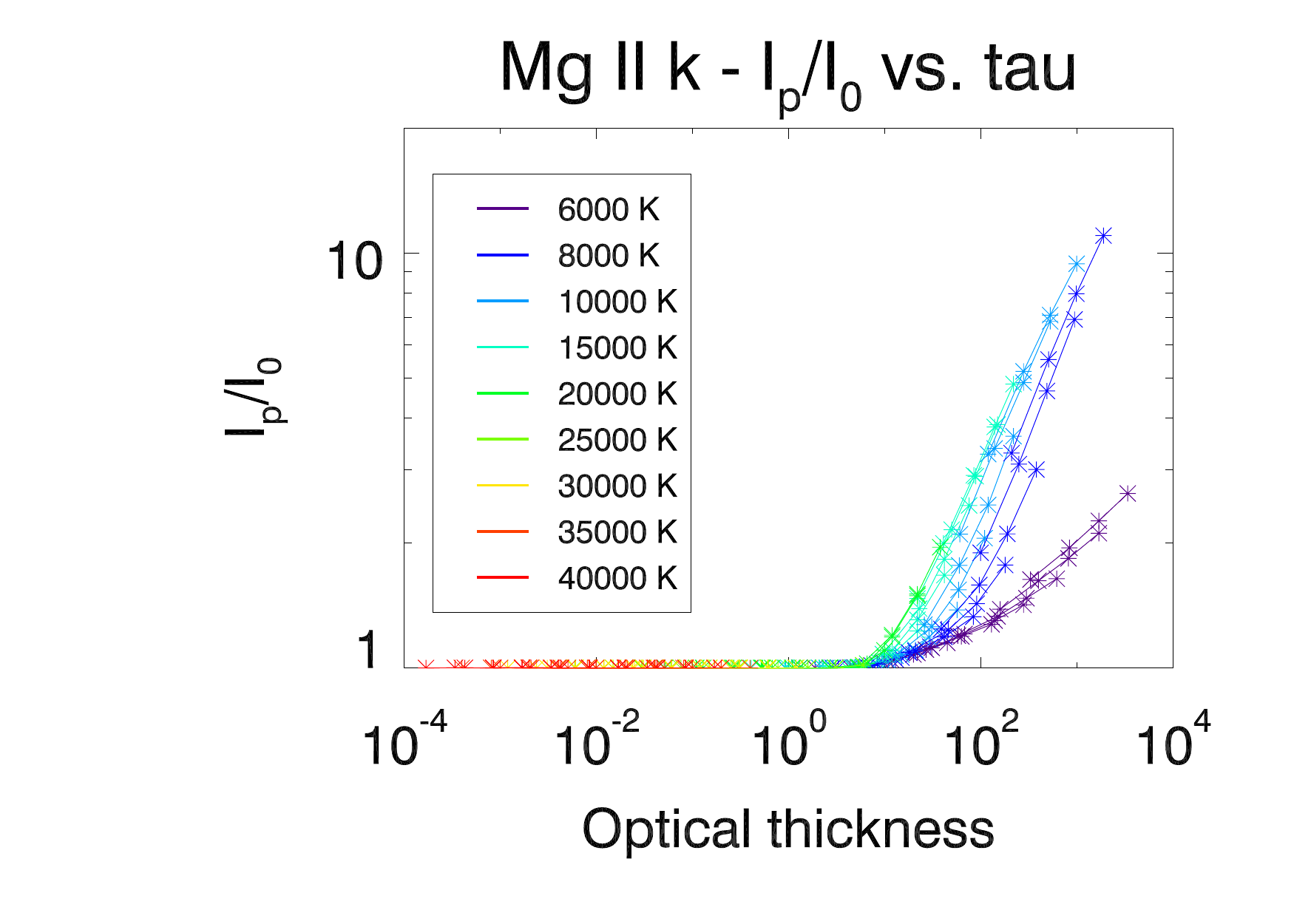}
                \includegraphics[width=0.49\hsize,clip=true,trim=1cm 0.7cm 0.9cm 0cm]{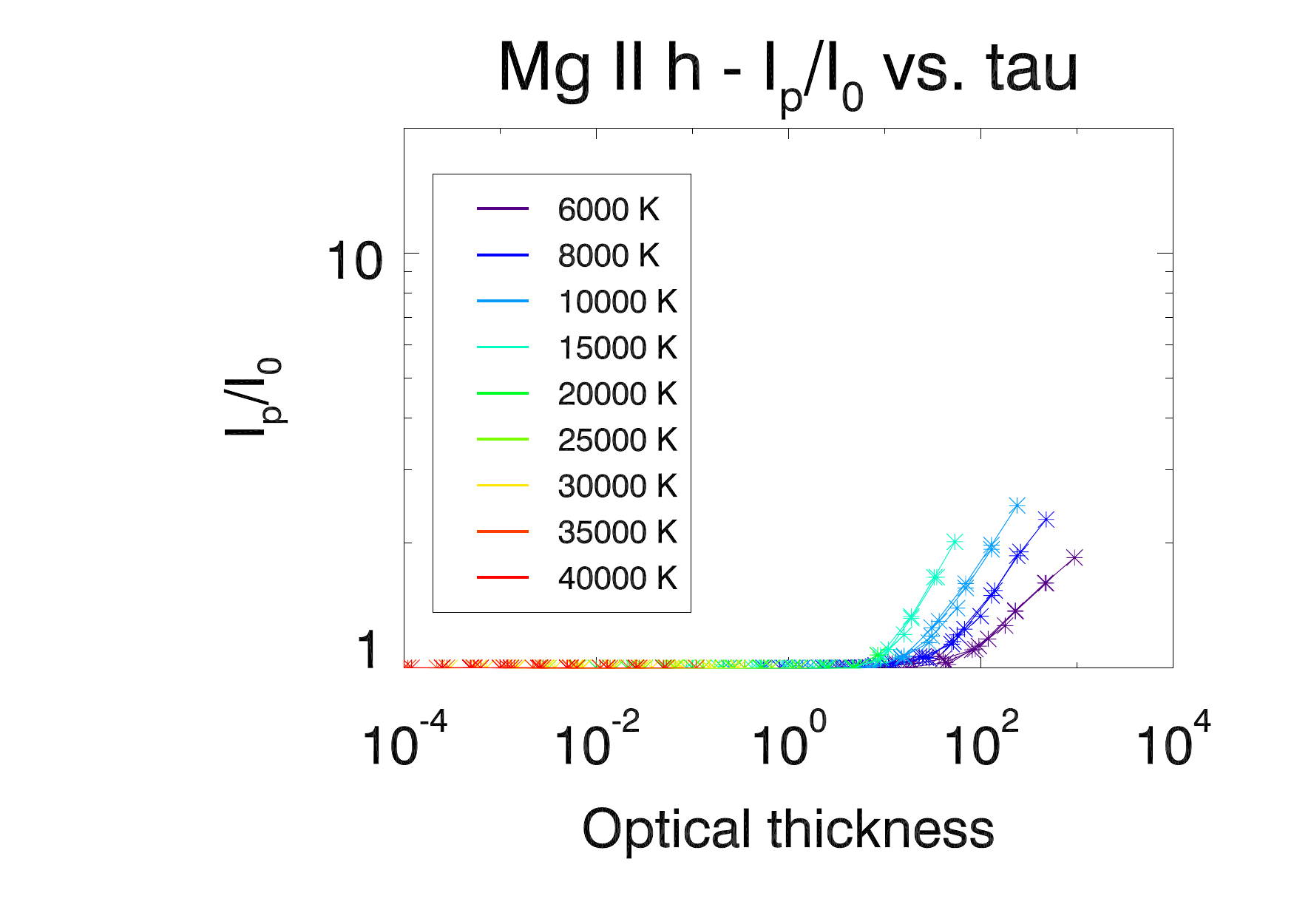}
\includegraphics[width=0.49\hsize,clip=true,trim=1cm 0.7cm 0.9cm 0cm]{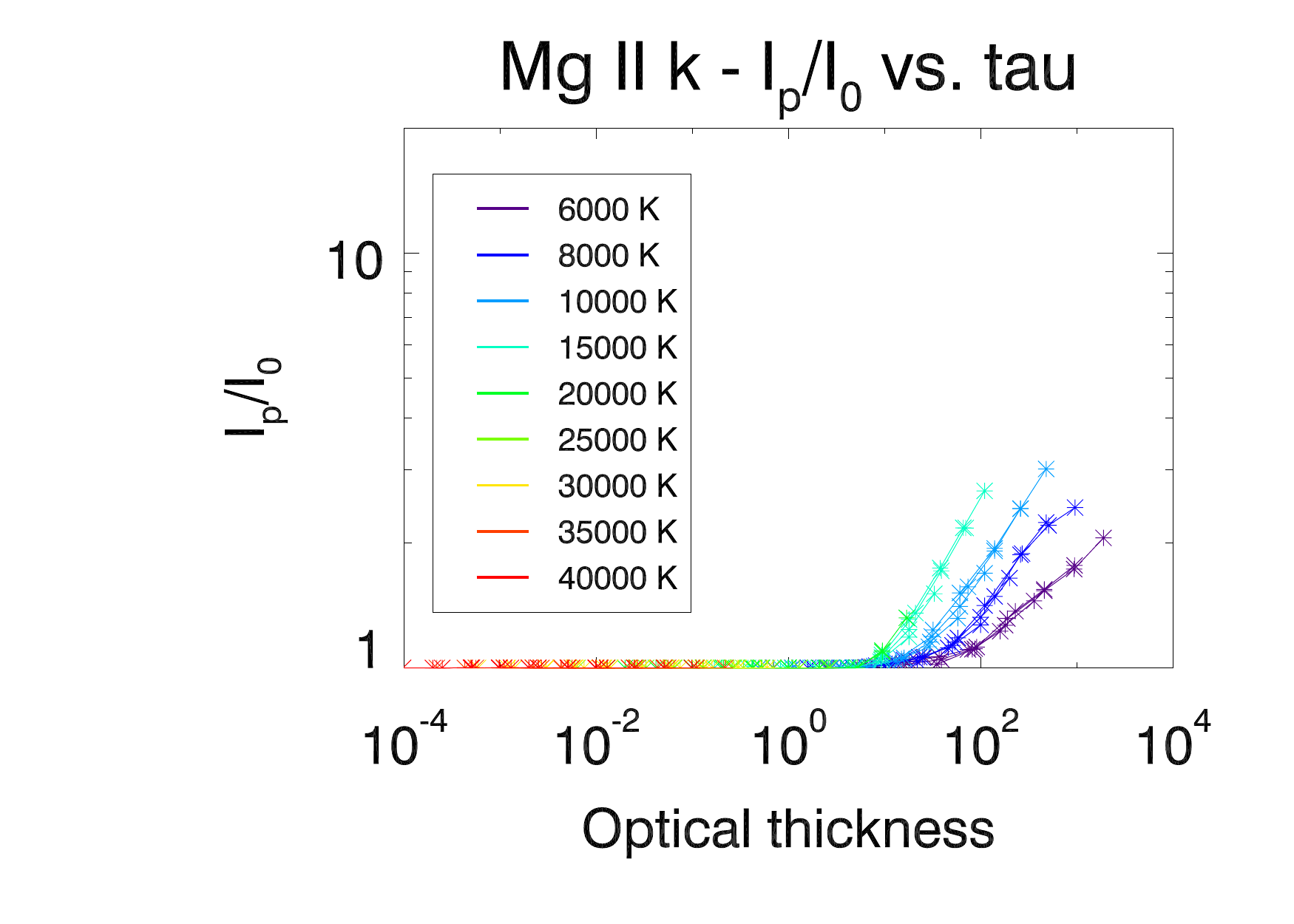}
                \caption{Reversal level against {line} optical thickness of  \ion{Mg}{ii} h (left), and  \ion{Mg}{ii} k (right). {Top row: Isothermal isobaric models. Bottom row: Models with a PCTR and $\gamma = 2$.} Colour indicates temperature, increasing from purple to red. }
                \label{fig:252_rev_tau_comb}
        \end{center}
\end{figure}

Importantly, as shown in Figure~\ref{fig:252_rev_tau_comb}, optical thicknesses higher than around $10$ are where the h and k lines begin to become reversed. {For PCTR models, this occurs for models with $T_{\mathrm{cen}} < 25000$~K. }
Only the PCTR models where $\gamma = 2$ are presented here, as they represent the most extended PCTR case. 
A reversal at line centre results in a reduction of {the frequency-}integrated intensity compared to an equivalent non-reversed profile, which accounts for the flattening of the plots in Figure~\ref{fig:252_tau_comb}. 
Differences in the optical thickness for the h and k lines means there will be different possible relative h and k intensity values depending on the model. 
This then causes the small departure from power-law linearity seen in Figure~\ref{fig:h_vs_k_comb}. 
{Figure~\ref{fig:252_rev_tau_comb} illustrates again that the}
inclusion of a PCTR inhibits the creation of deep central reversals.

\subsection{Correlations between observables and model parameters}

\begin{figure*}
        \begin{center}
                \includegraphics[width=0.16\hsize,clip=true,trim=1cm 0.7cm 0.9cm 0cm]{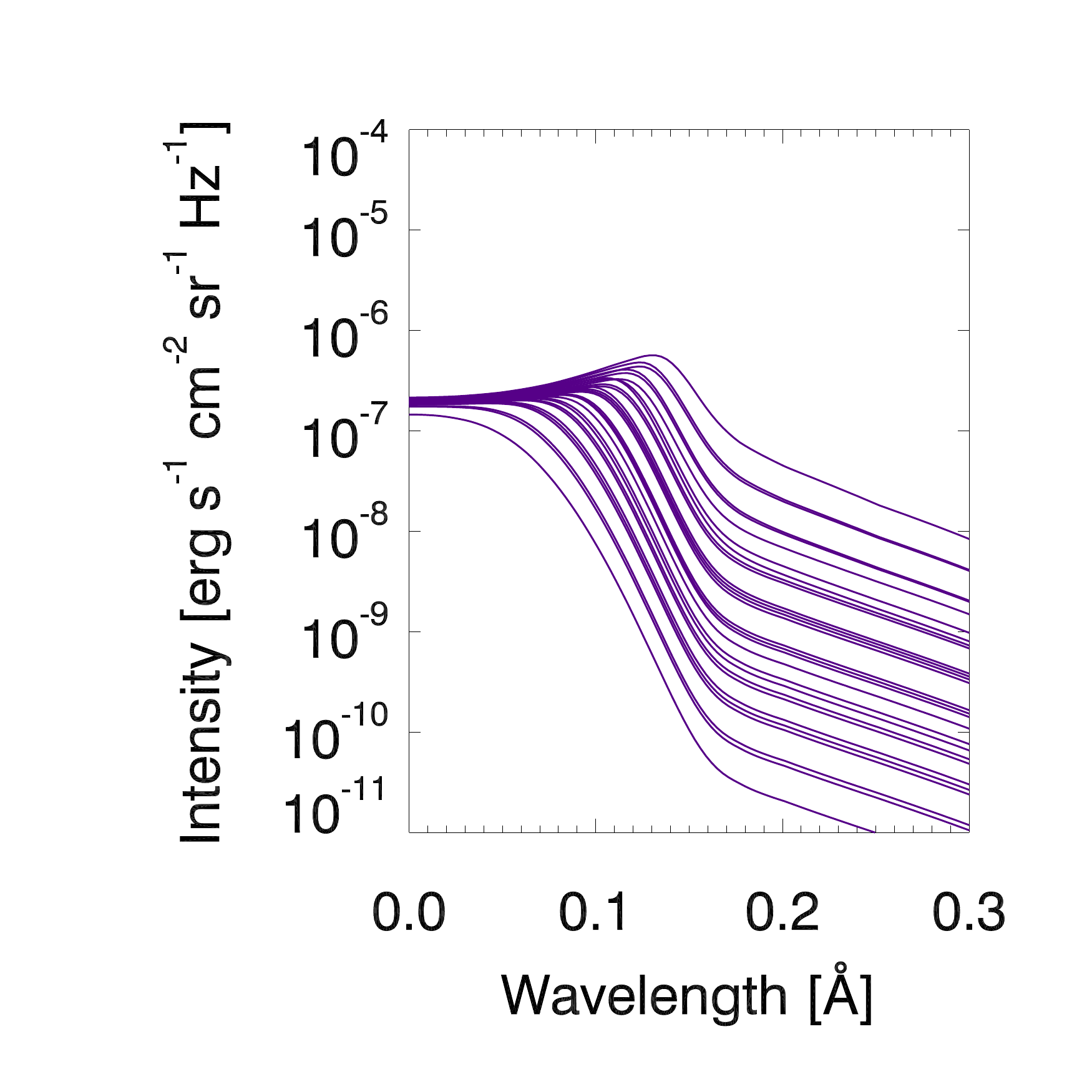}
                \includegraphics[width=0.16\hsize,clip=true,trim=1cm 0.7cm 0.9cm 0cm]{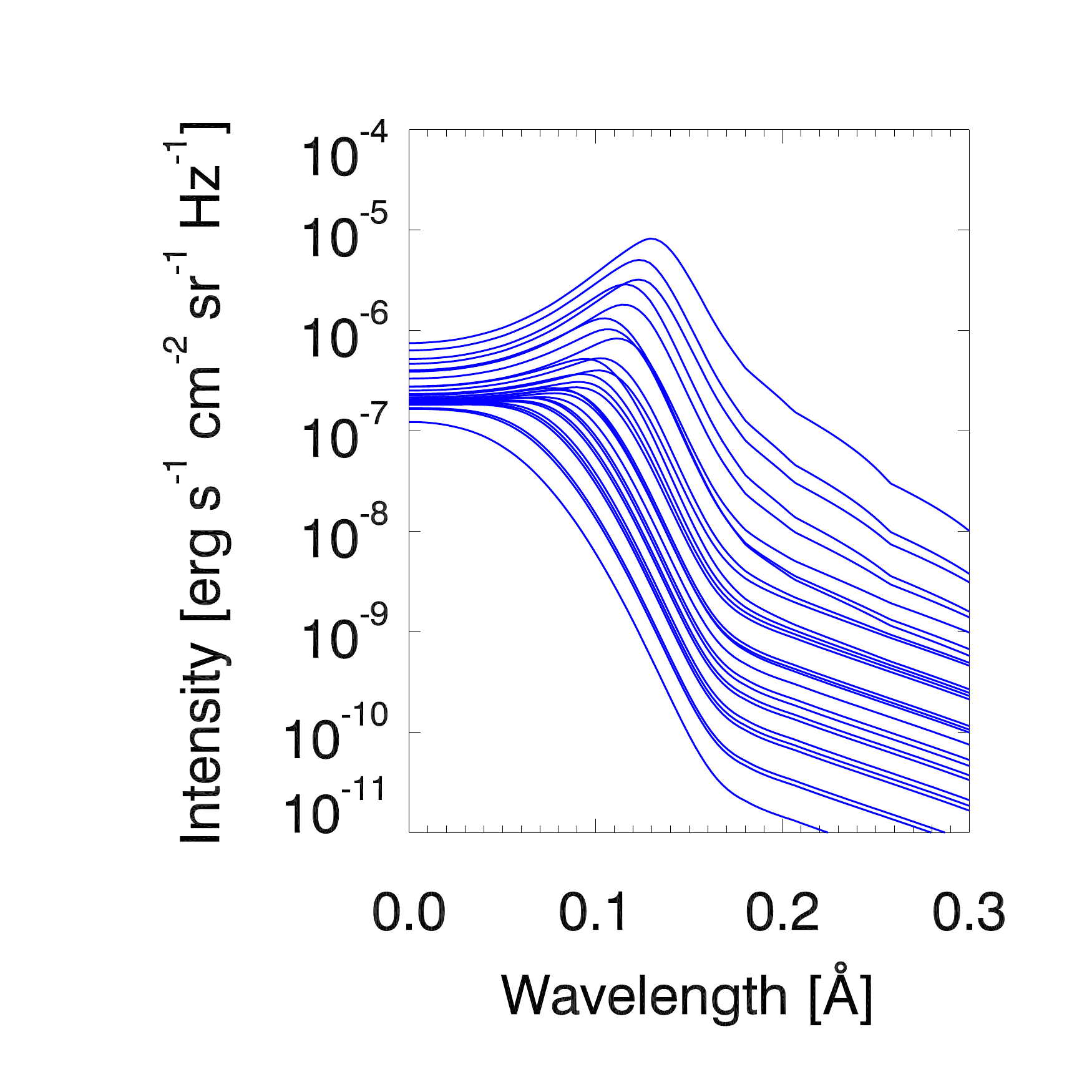}
                \includegraphics[width=0.16\hsize,clip=true,trim=1cm 0.7cm 0.9cm 0cm]{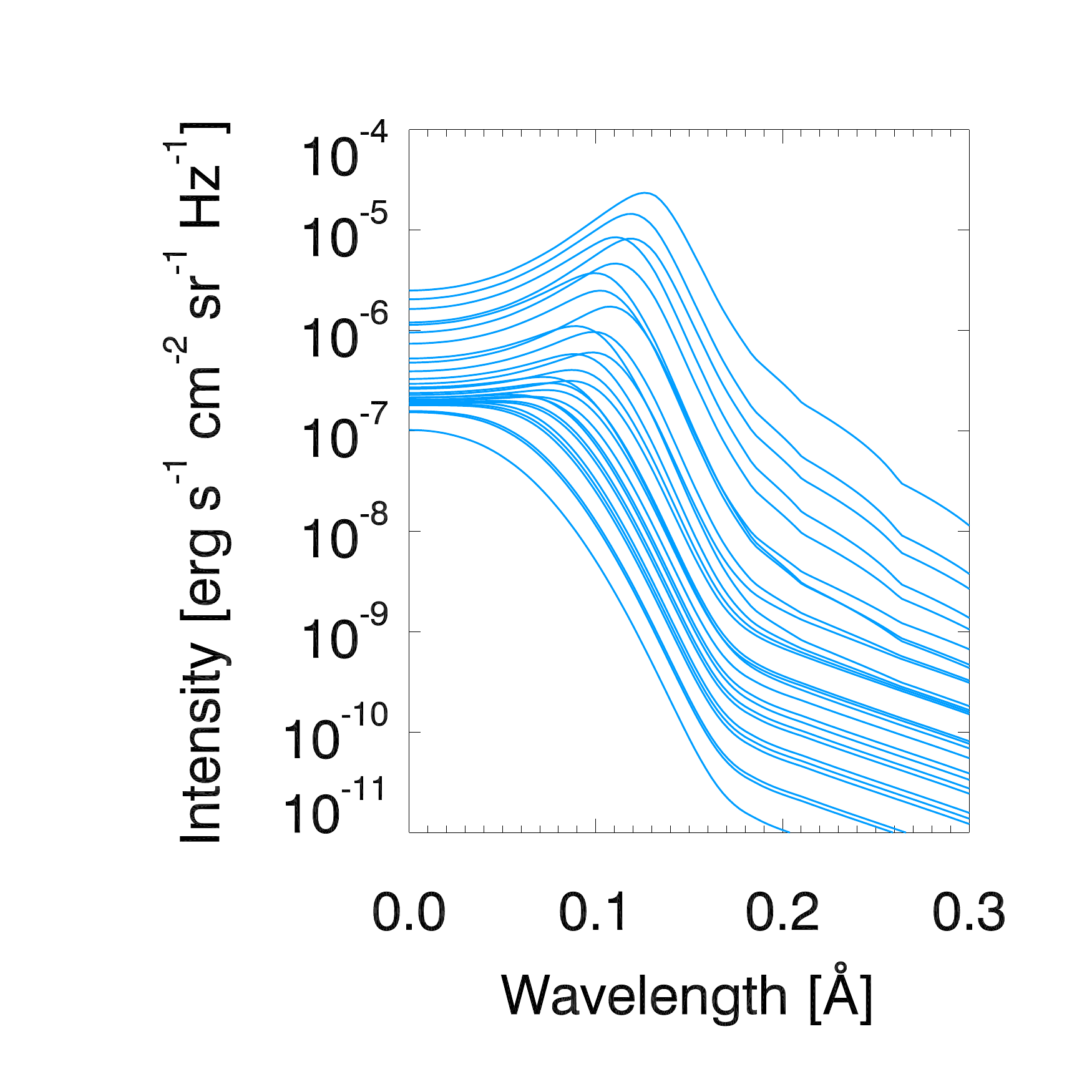}
                \includegraphics[width=0.16\hsize,clip=true,trim=1cm 0.7cm 0.9cm 0cm]{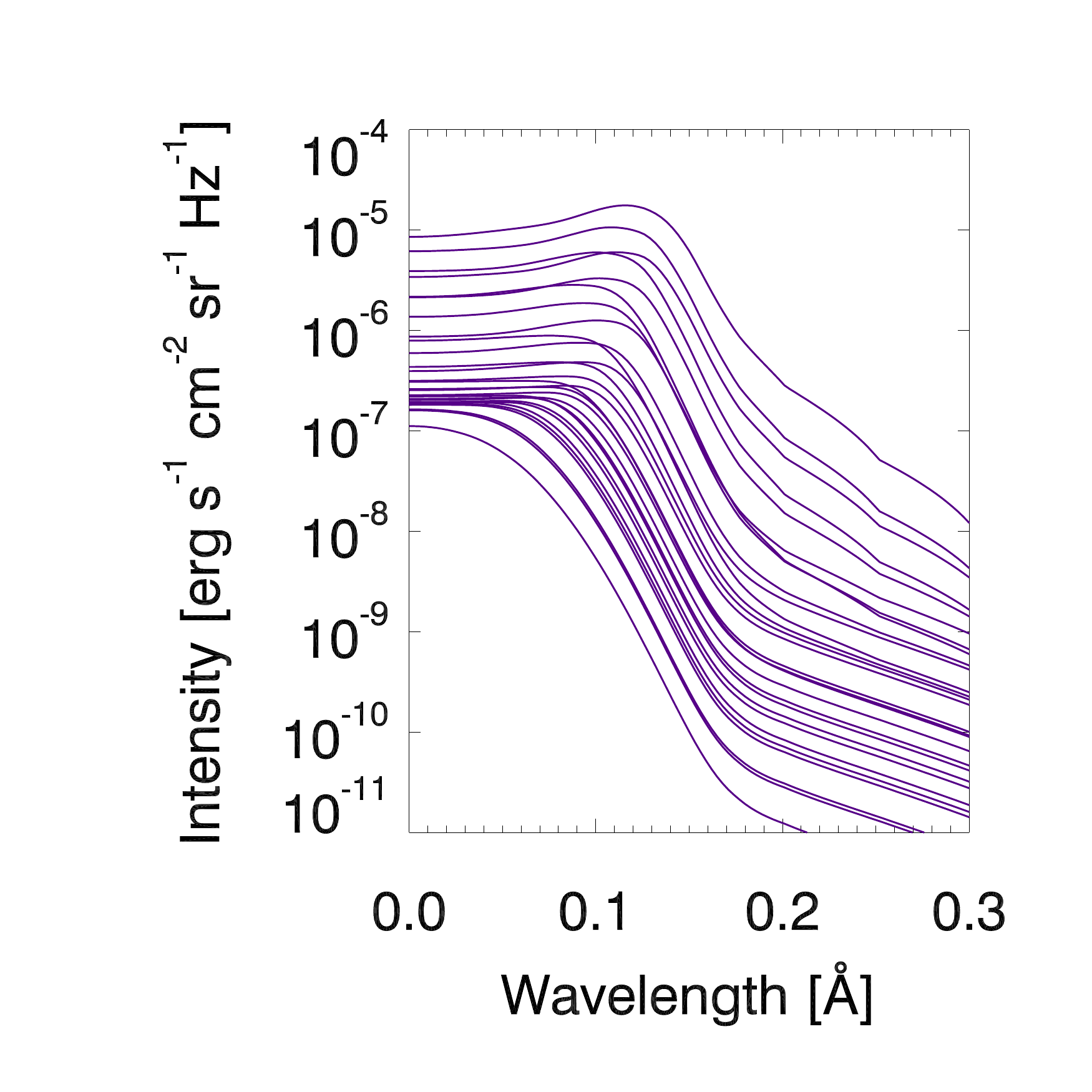}
\includegraphics[width=0.16\hsize,clip=true,trim=1cm 0.7cm 0.9cm 0cm]{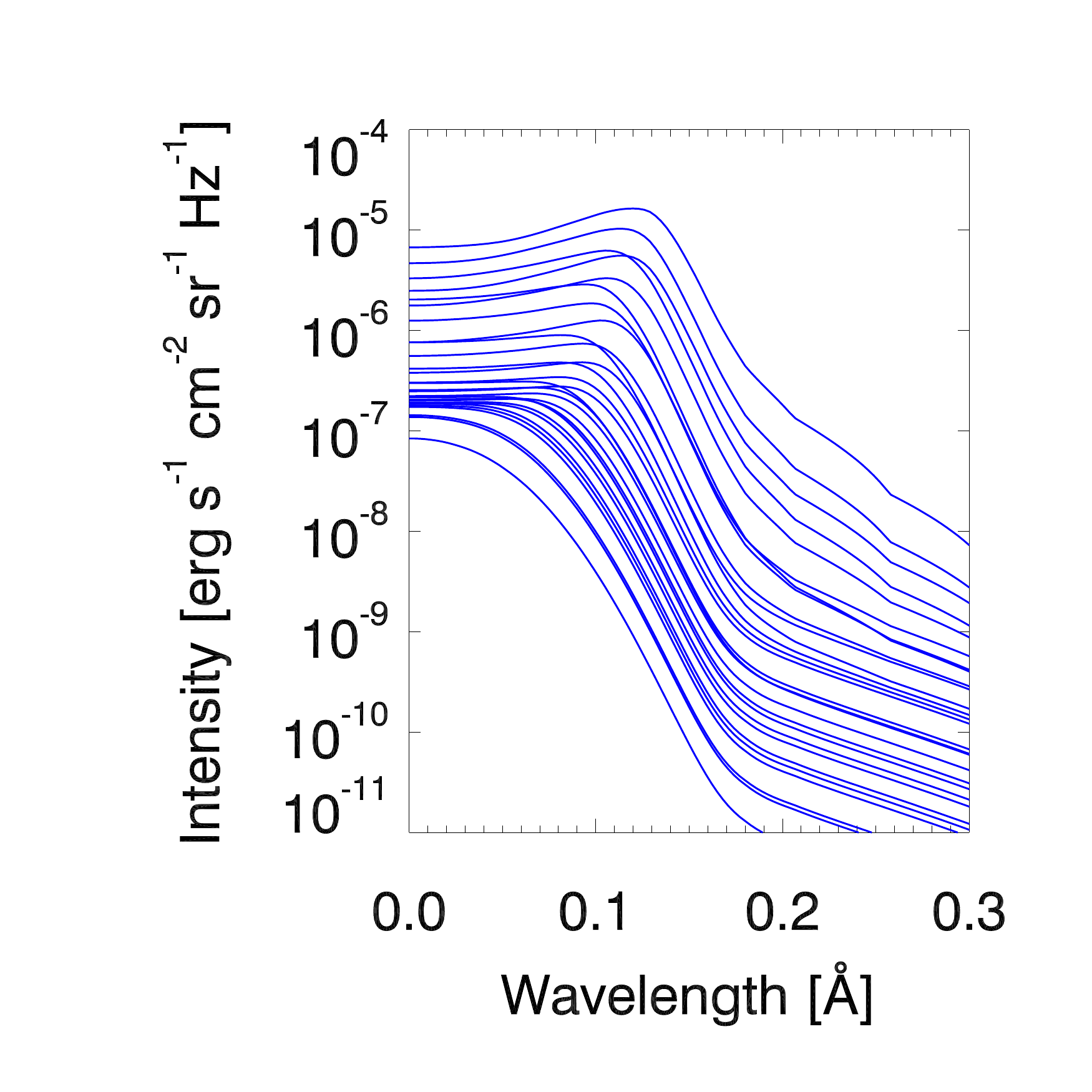}
\includegraphics[width=0.16\hsize,clip=true,trim=1cm 0.7cm 0.9cm 0cm]{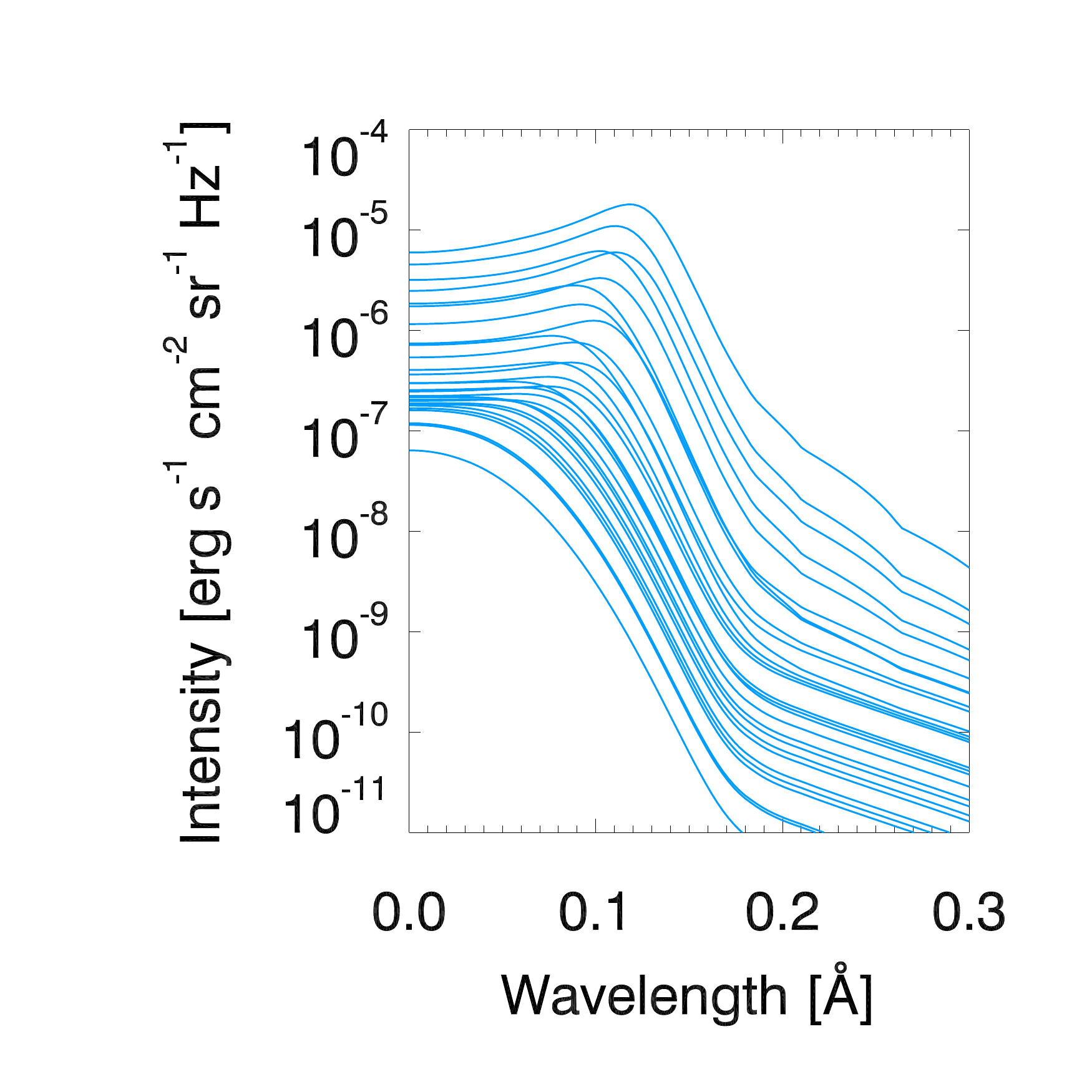}\\
                \includegraphics[width=0.16\hsize,clip=true,trim=1cm 0.7cm 0.9cm 0cm]{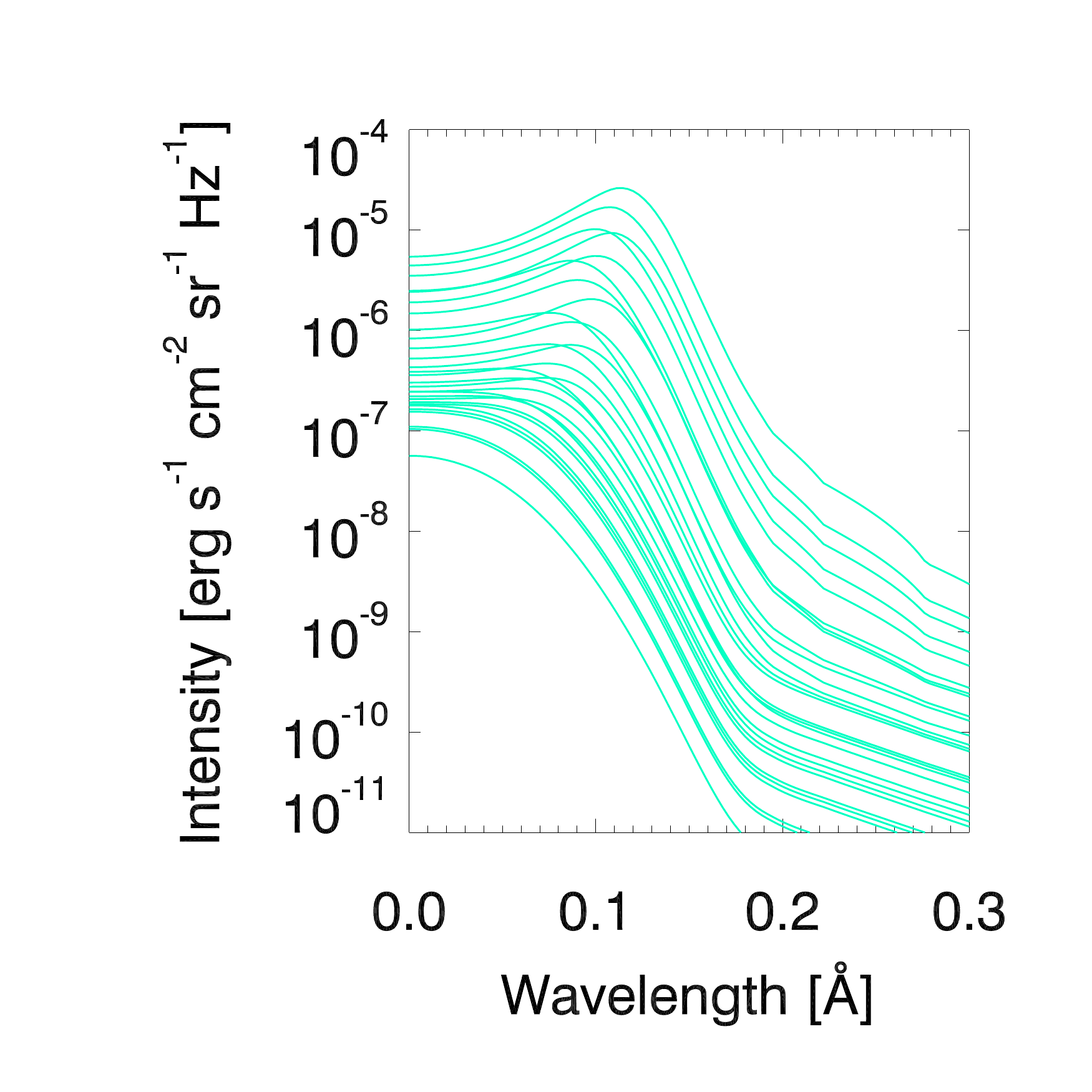}
                \includegraphics[width=0.16\hsize,clip=true,trim=1cm 0.7cm 0.9cm 0cm]{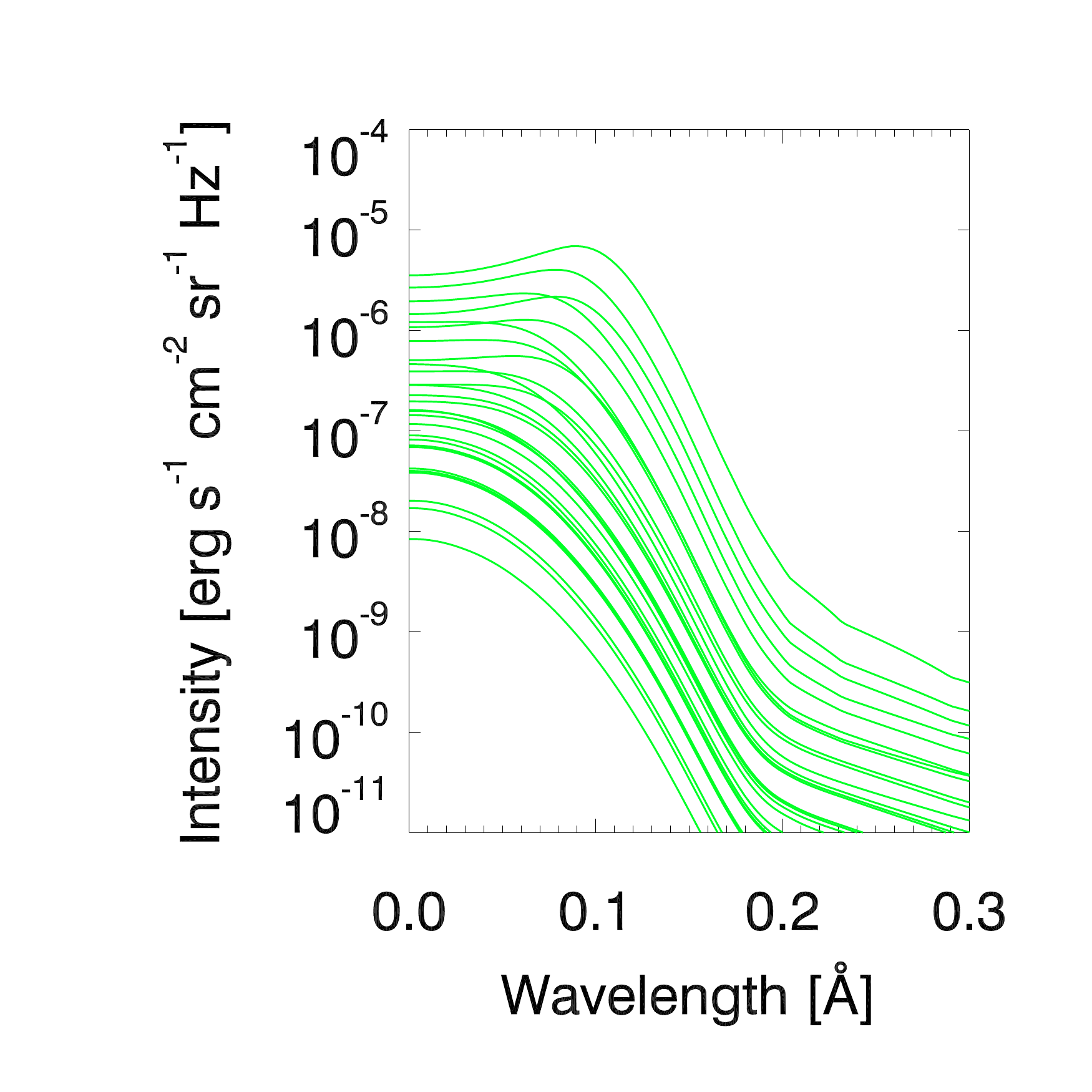}
                \includegraphics[width=0.16\hsize,clip=true,trim=1cm 0.7cm 0.9cm 0cm]{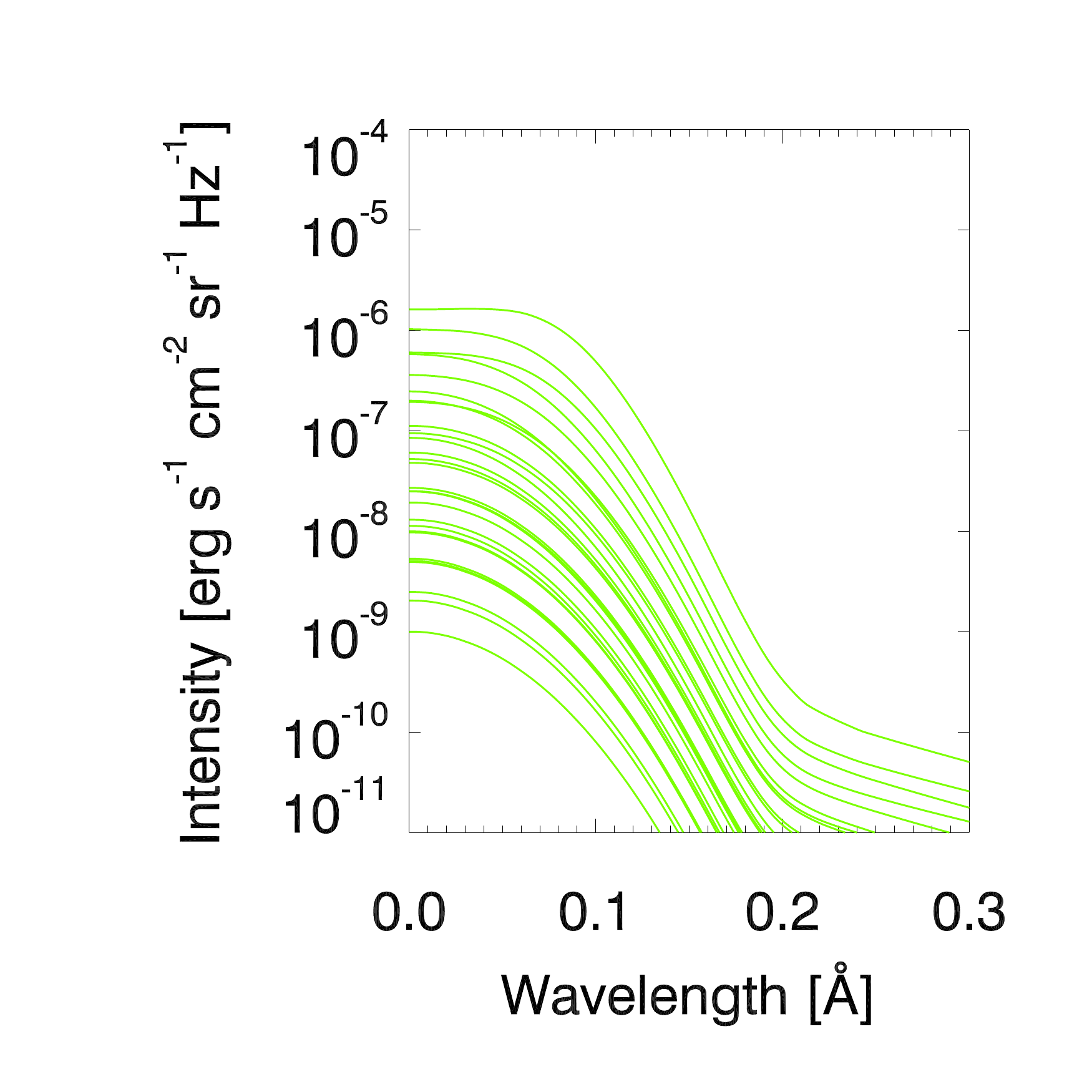}
                \includegraphics[width=0.16\hsize,clip=true,trim=1cm 0.7cm 0.9cm 0cm]{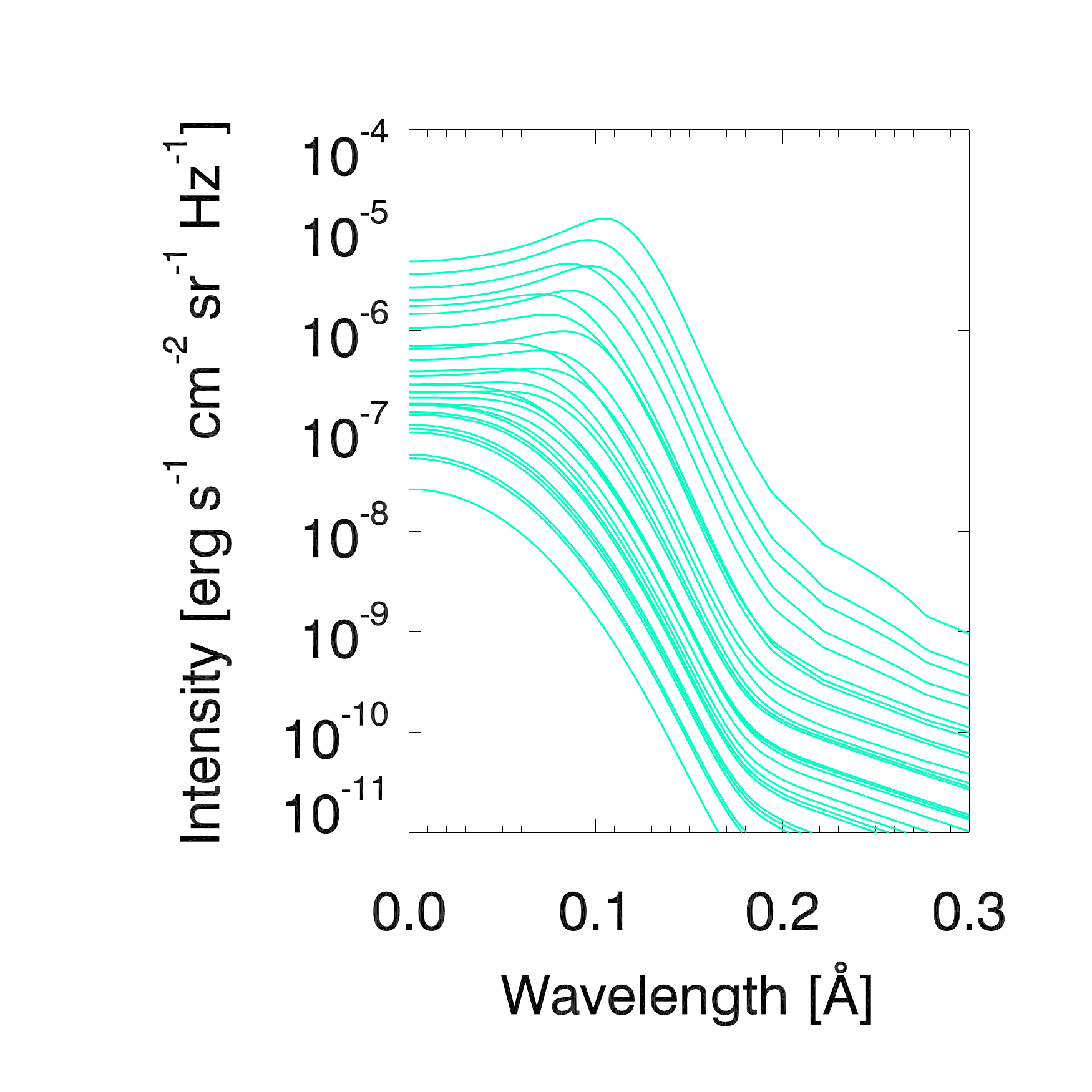}
\includegraphics[width=0.16\hsize,clip=true,trim=1cm 0.7cm 0.9cm 0cm]{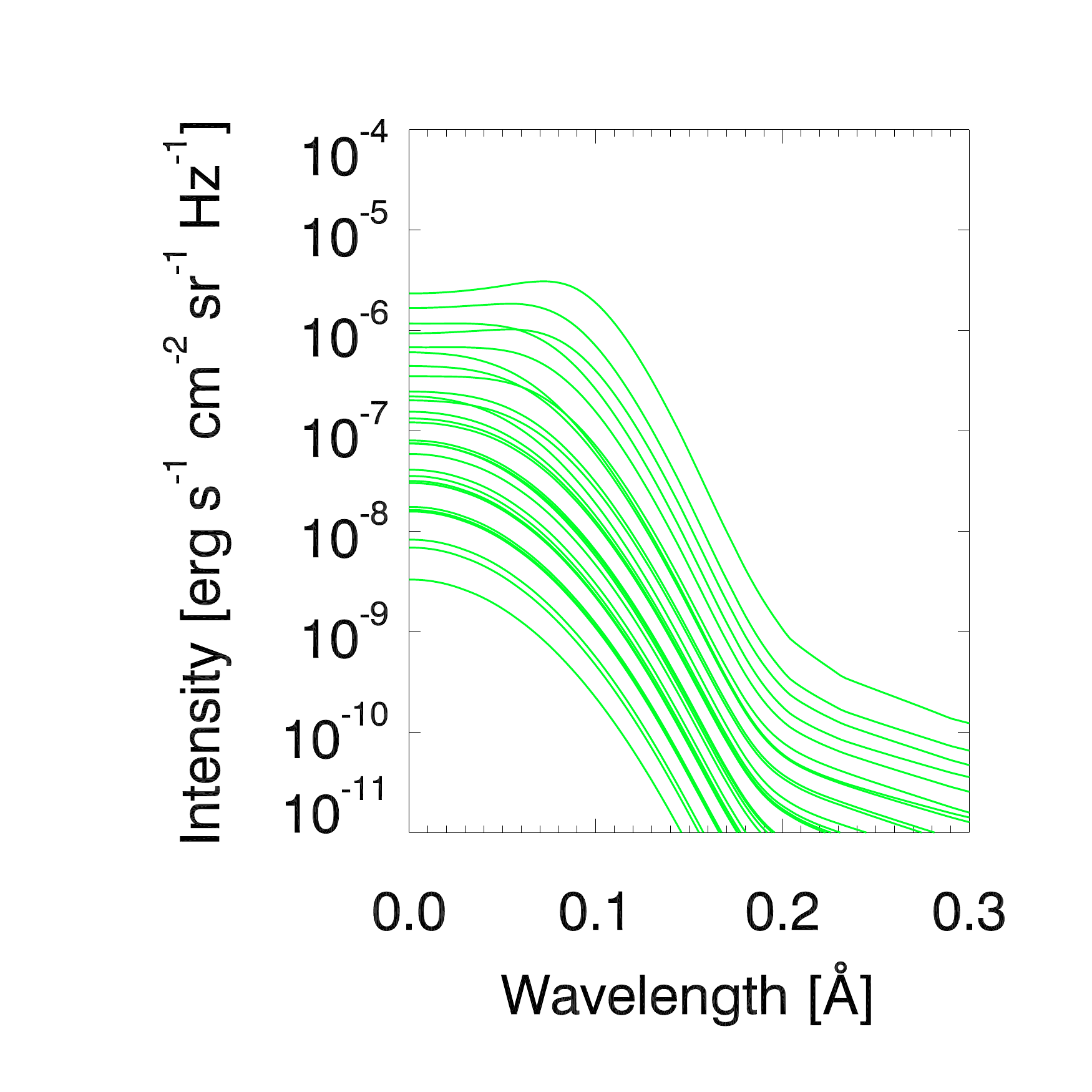}
\includegraphics[width=0.16\hsize,clip=true,trim=1cm 0.7cm 0.9cm 0cm]{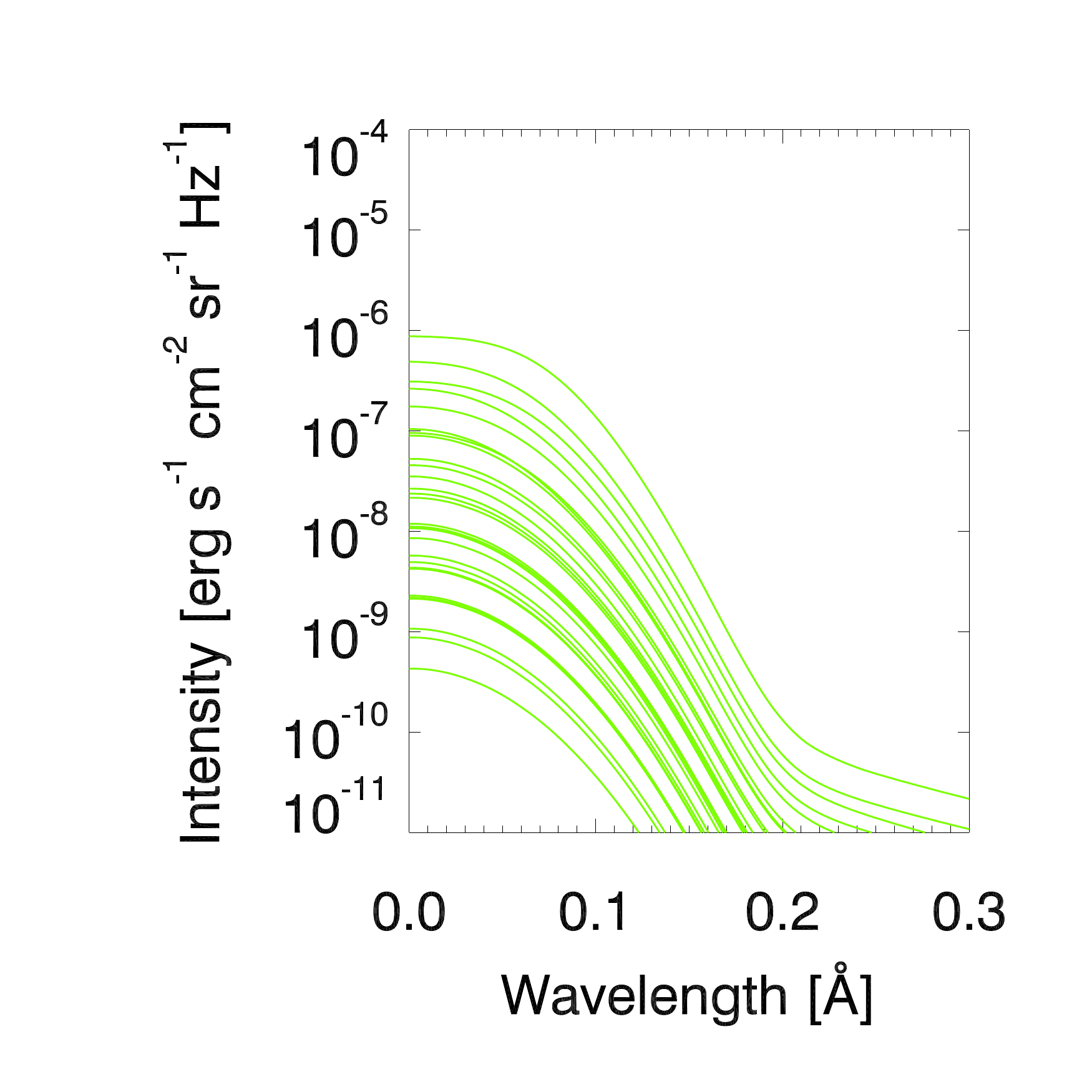}\\
                \includegraphics[width=0.16\hsize,clip=true,trim=1cm 0.7cm 0.9cm 0cm]{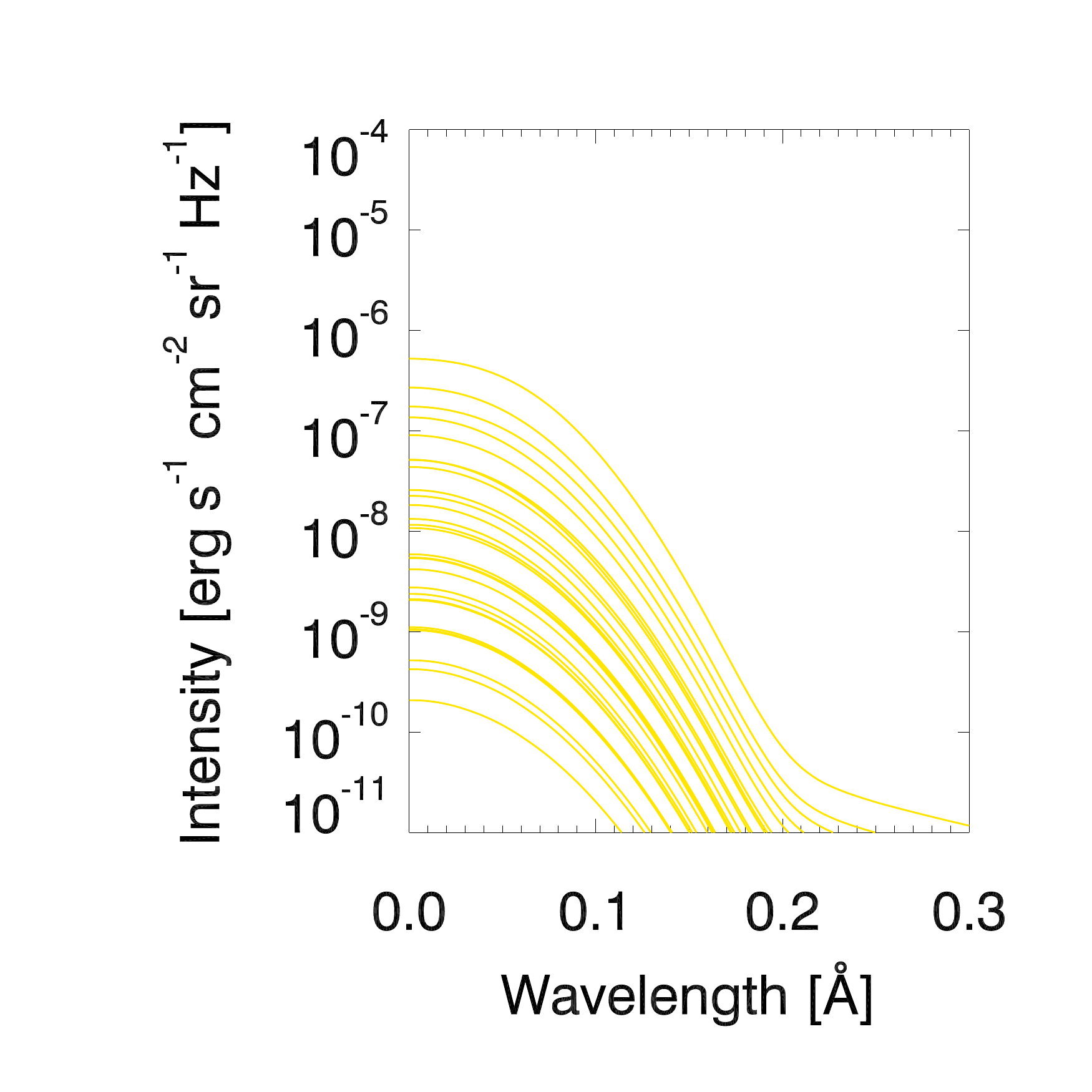}
                \includegraphics[width=0.16\hsize,clip=true,trim=1cm 0.7cm 0.9cm 0cm]{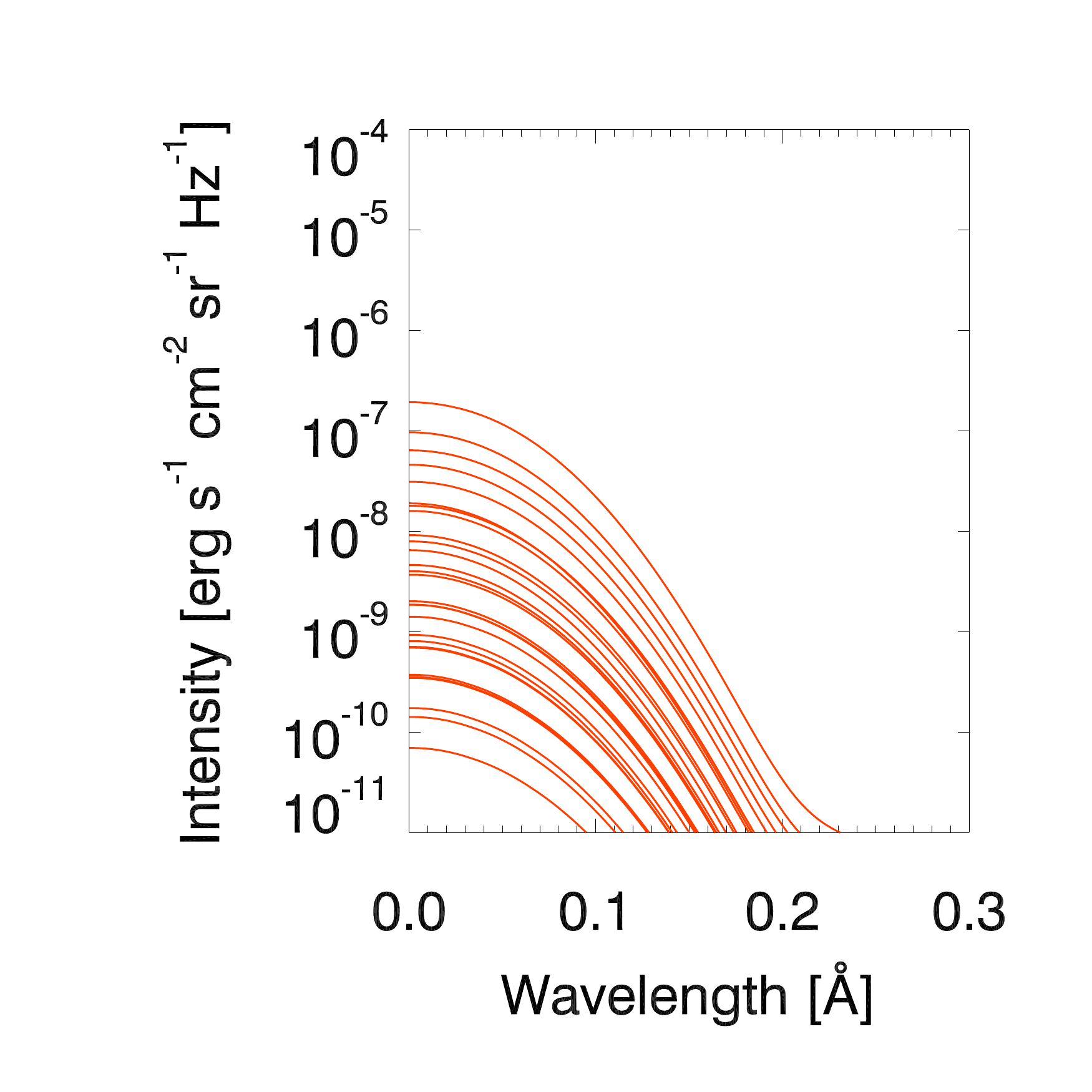}
                \includegraphics[width=0.16\hsize,clip=true,trim=1cm 0.7cm 0.9cm 0cm]{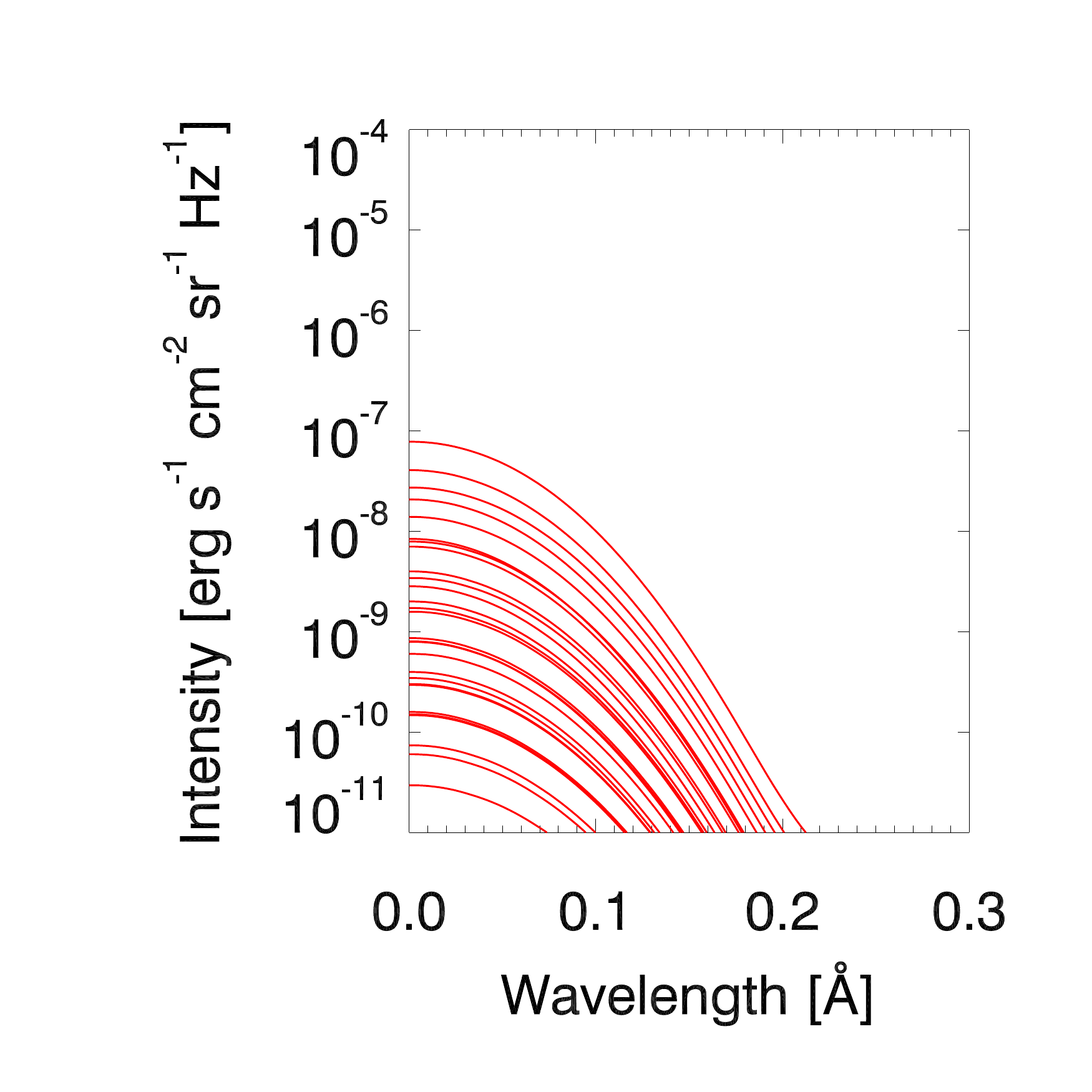}
                \includegraphics[width=0.16\hsize,clip=true,trim=1cm 0.7cm 0.9cm 0cm]{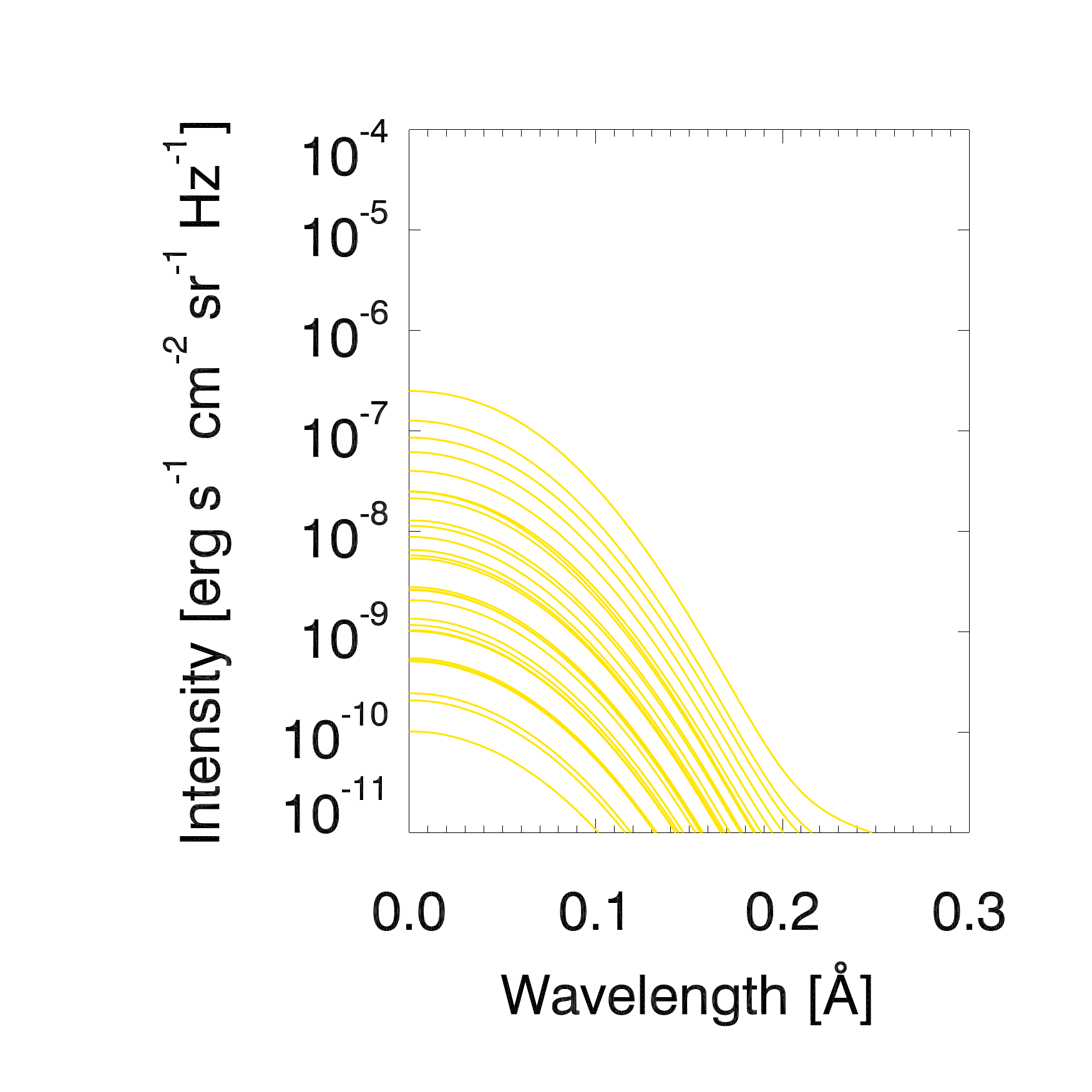}
                \includegraphics[width=0.16\hsize,clip=true,trim=1cm 0.7cm 0.9cm 0cm]{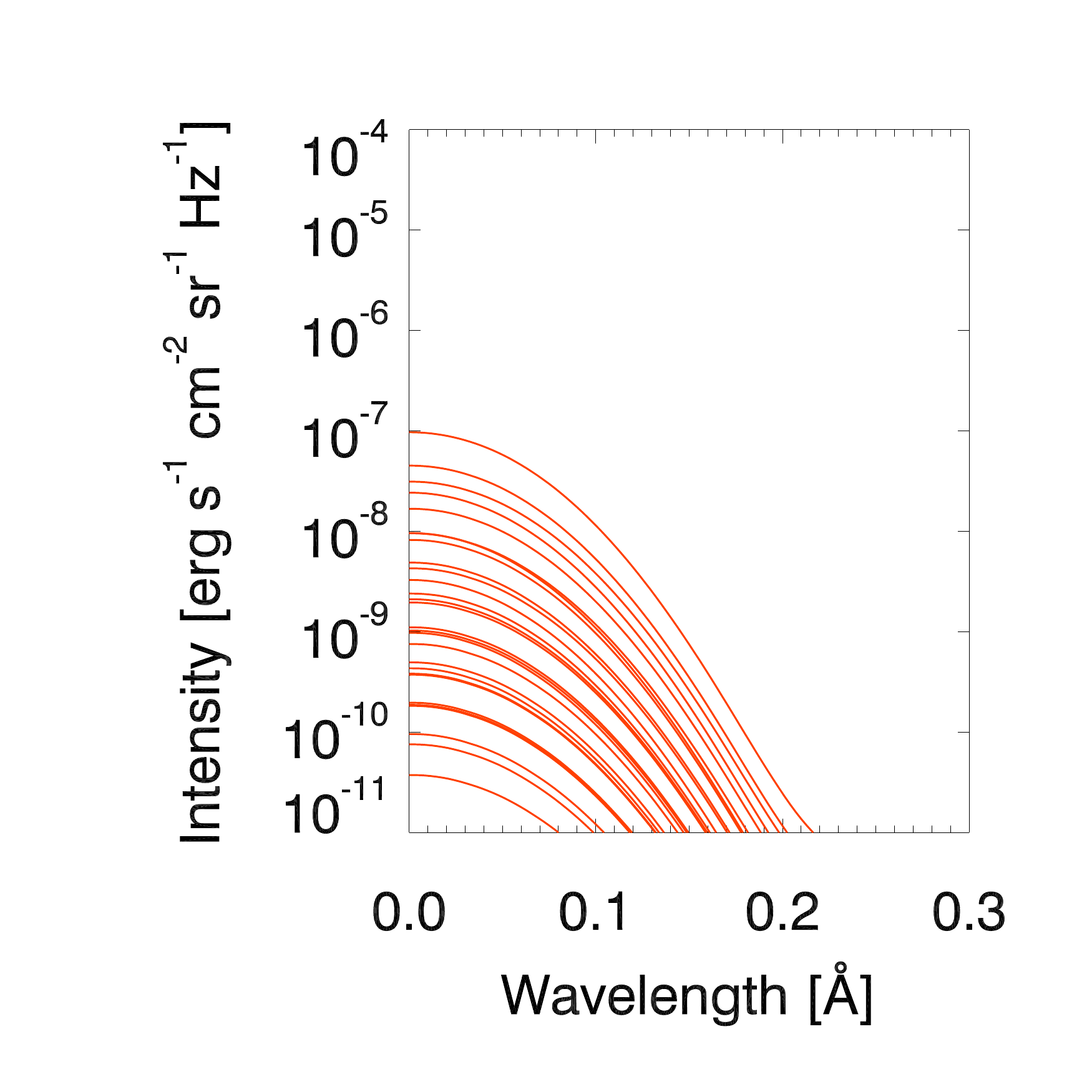}
                \includegraphics[width=0.16\hsize,clip=true,trim=1cm 0.7cm 0.9cm 0cm]{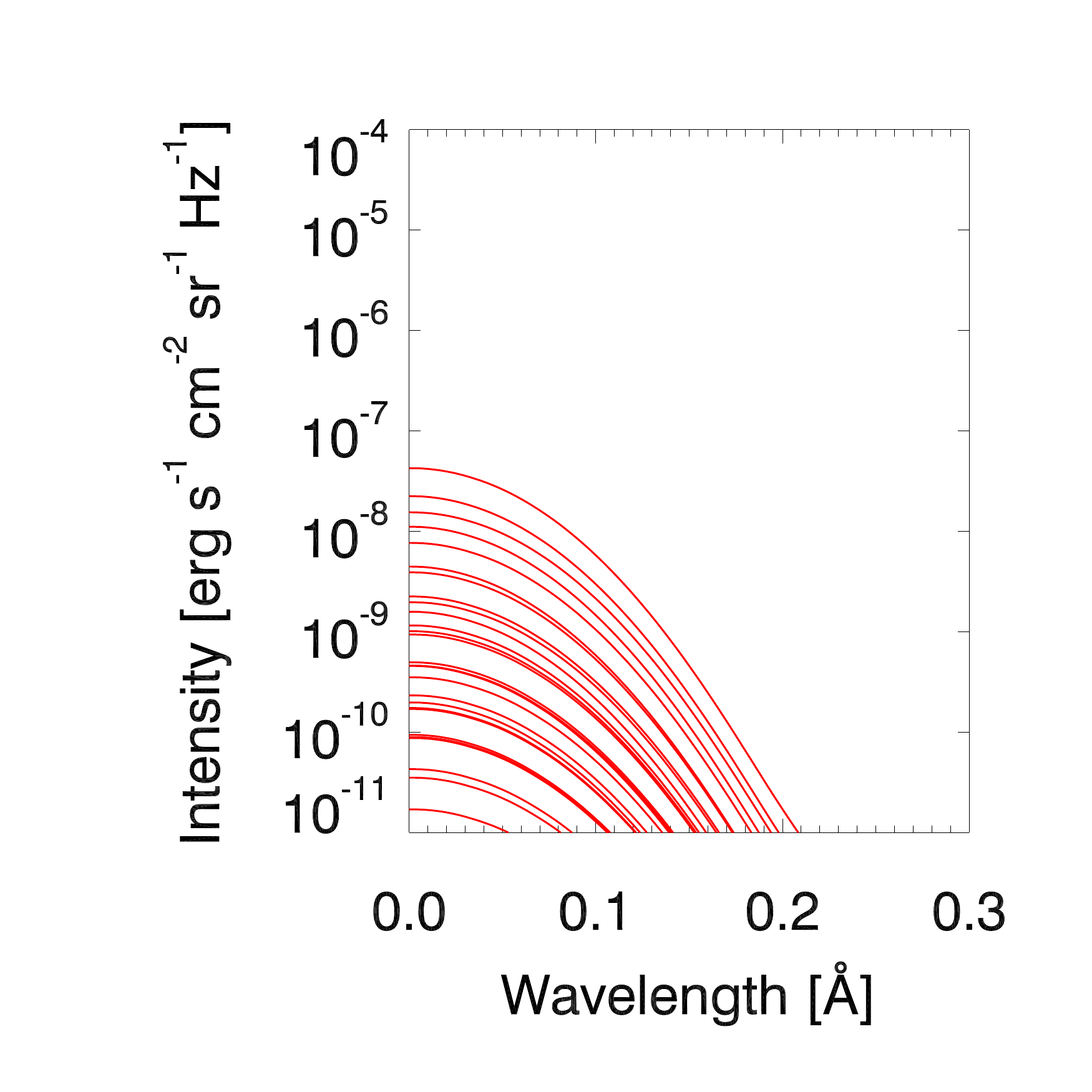}
                \caption{Emergent  half-line \ion{Mg}{ii} k profiles. {The first three columns show isothermal isobaric models; the last three columns show PCTR models with $\gamma =2$.} Each panel is colour-coded to show a different temperature: $6000$~K (purple), $8000$~K (dark blue), $10000$~K (light blue), $15000$~K (teal), $20000$~K (green), $25000$~K (lime green), $30000$~K (yellow), $35000$~K (orange), and $40000$~K (red). Intensities generally increase for higher pressure and slab thickness.}
                \label{fig:252_profiles_comb}
        \end{center}
\end{figure*}

Figure~\ref{fig:252_profiles_comb} shows the emergent \ion{Mg}{ii} k profiles for the {isothermal isobaric and PCTR} models, with each panel showing all {pressure} $P$ and {slab thickness} $D$ combinations for a single temperature. 
{Line profiles for PCTR models} are only shown for $\gamma =2$. 
Notably, {in the isothermal isobaric case,} lower temperature models ($\leq 20000$~K) show a mix of reversed and non-reversed profiles, whereas higher temperature models ($> 20000$~K) do not show any line reversals. 
It {will} therefore be  worth exploring these high-temperature models in more detail to gauge their relevance to observations.
Line centre intensities for models with low central temperature ($T_{\mathrm{cen}} = 6000, 8000, 10000$~K) have a much larger spread of values for the PCTR models than the isothermal models. 
This occurs because  there is plasma with much higher temperatures than the stated core temperature in the slab, which is due to the presence of the PCTR. 
A value of $\gamma = 2$ means that there will be more high-temperature plasma nearer the slab centre. 
As $\gamma$ becomes larger the PCTR becomes narrower, and the solutions for high $\gamma$ should converge to the isothermal case \citep{Labrosse2004}. 
As a consequence, the central reversals seen in the h and k profiles should become deeper as $\gamma$ increases. 
The effects of this are not seen, however, for the three values of $\gamma$ considered here, as seen in the right panels of Figure~\ref{fig:h_vs_k_comb}. Values of $\gamma$ much higher than 10 (maximum value adopted here) are required to see this effect \citep{Labrosse2004,Heinzel2014}.

\begin{figure}
        \begin{center}
                \includegraphics[width=0.49\hsize,clip=true,trim=1cm 0cm 0.95cm 0cm]{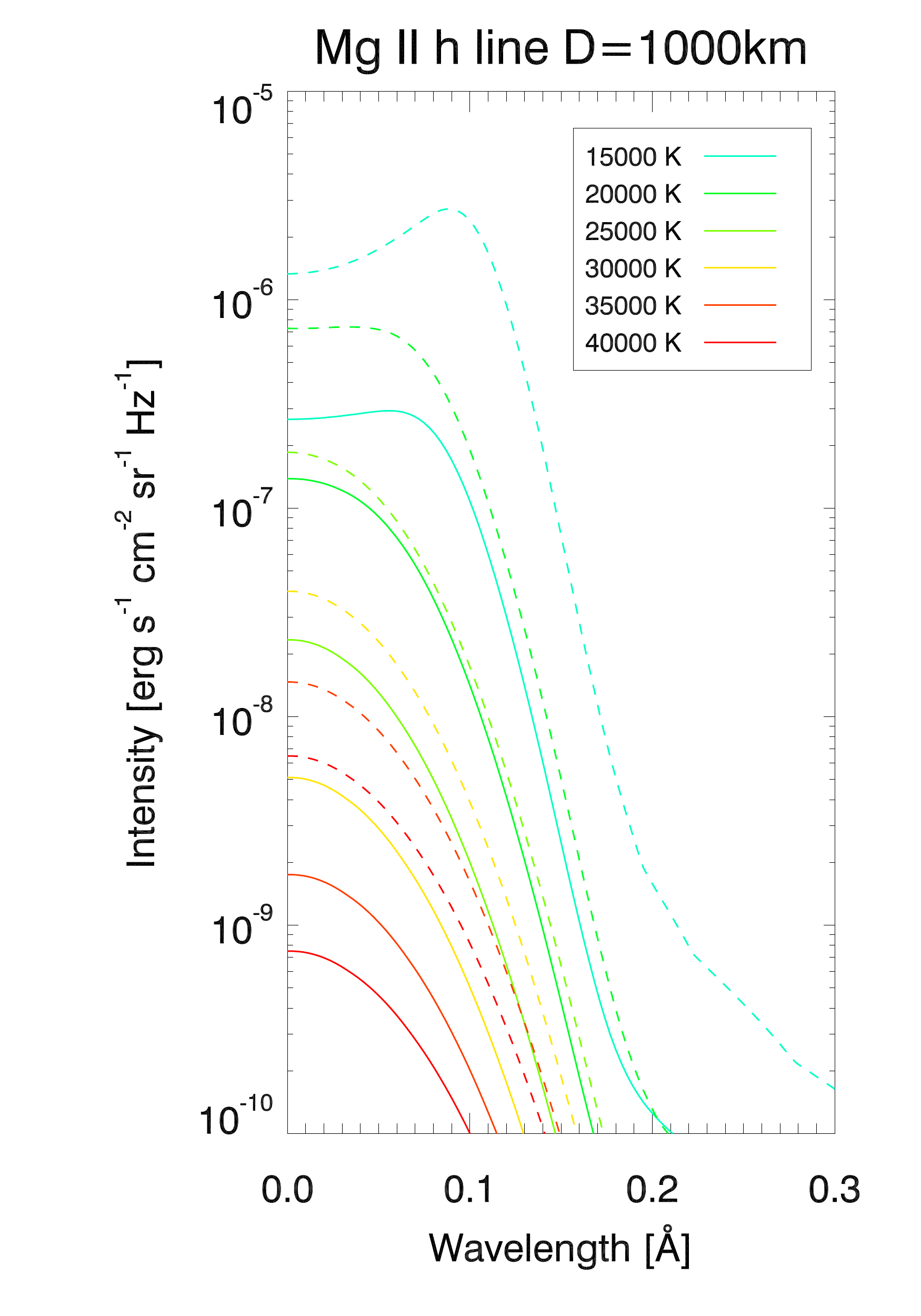}
                \includegraphics[width=0.49\hsize,clip=true,trim=1cm 0cm 0.95cm 0cm]{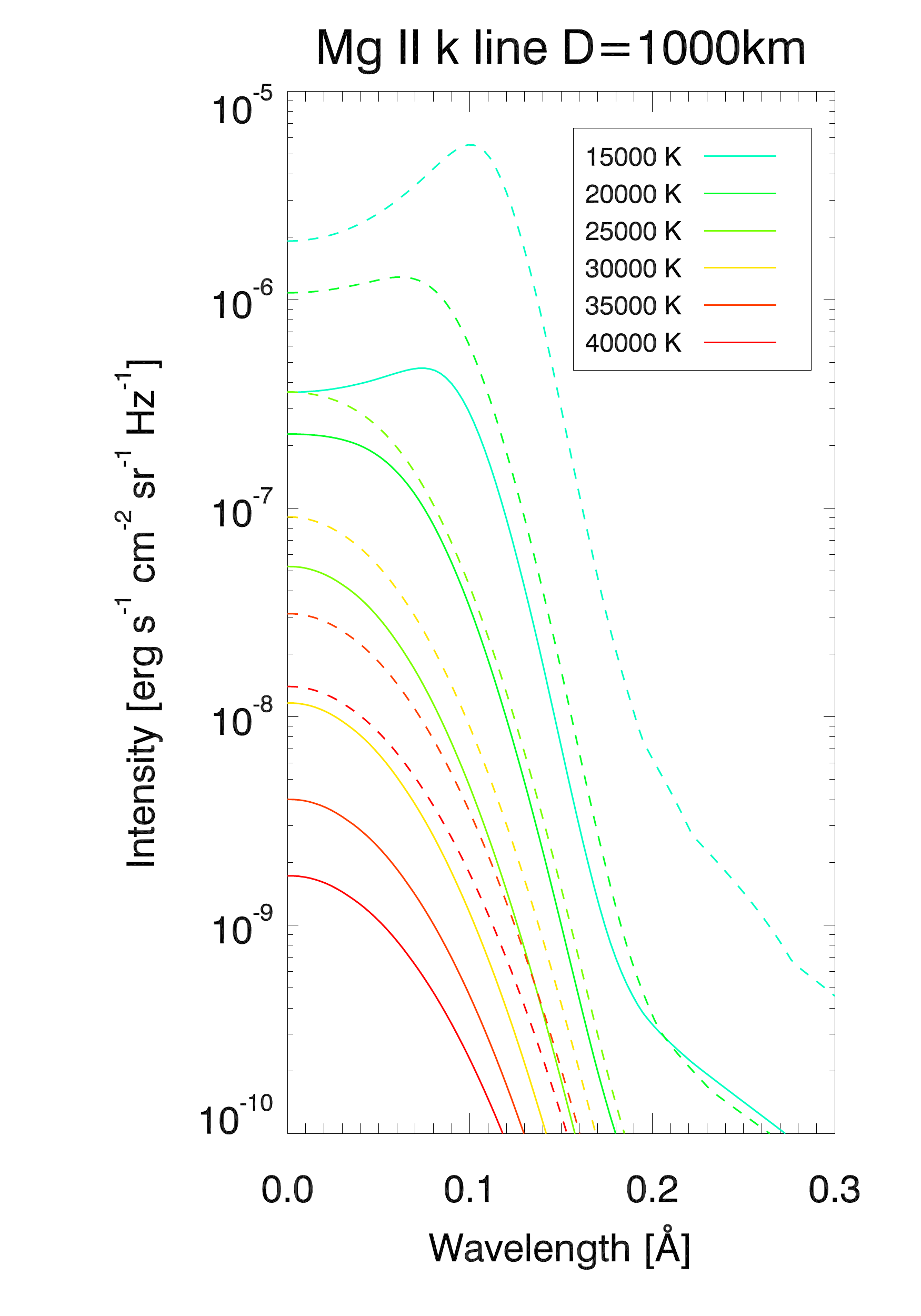}
                \caption{Half-line emergent profiles for the \ion{Mg}{ii} h (\textit{left}) and k (\textit{right}) lines from {isothermal isobaric} models with $T \geq 15000$~K. All profiles are from models with $D = 1000$~km. Profiles for two pressures are shown, with $P = 0.1$~dyne~cm$^{-2}$ (solid lines) and $P = 0.5$~dyne~cm$^{-2}$ (dashed lines). Temperatures are colour-coded: $15000$~K (black), $20000$~K (maroon), $25000$~K (red), $30000$~K (pink), $35000$~K (orange), and $40000$~K (yellow).}
                \label{fig:high_t}
        \end{center}
\end{figure}

\begin{figure}
        \begin{center}
                \includegraphics[width=0.49\hsize,clip=true,trim=1cm 0.7cm 0.5cm 0cm]{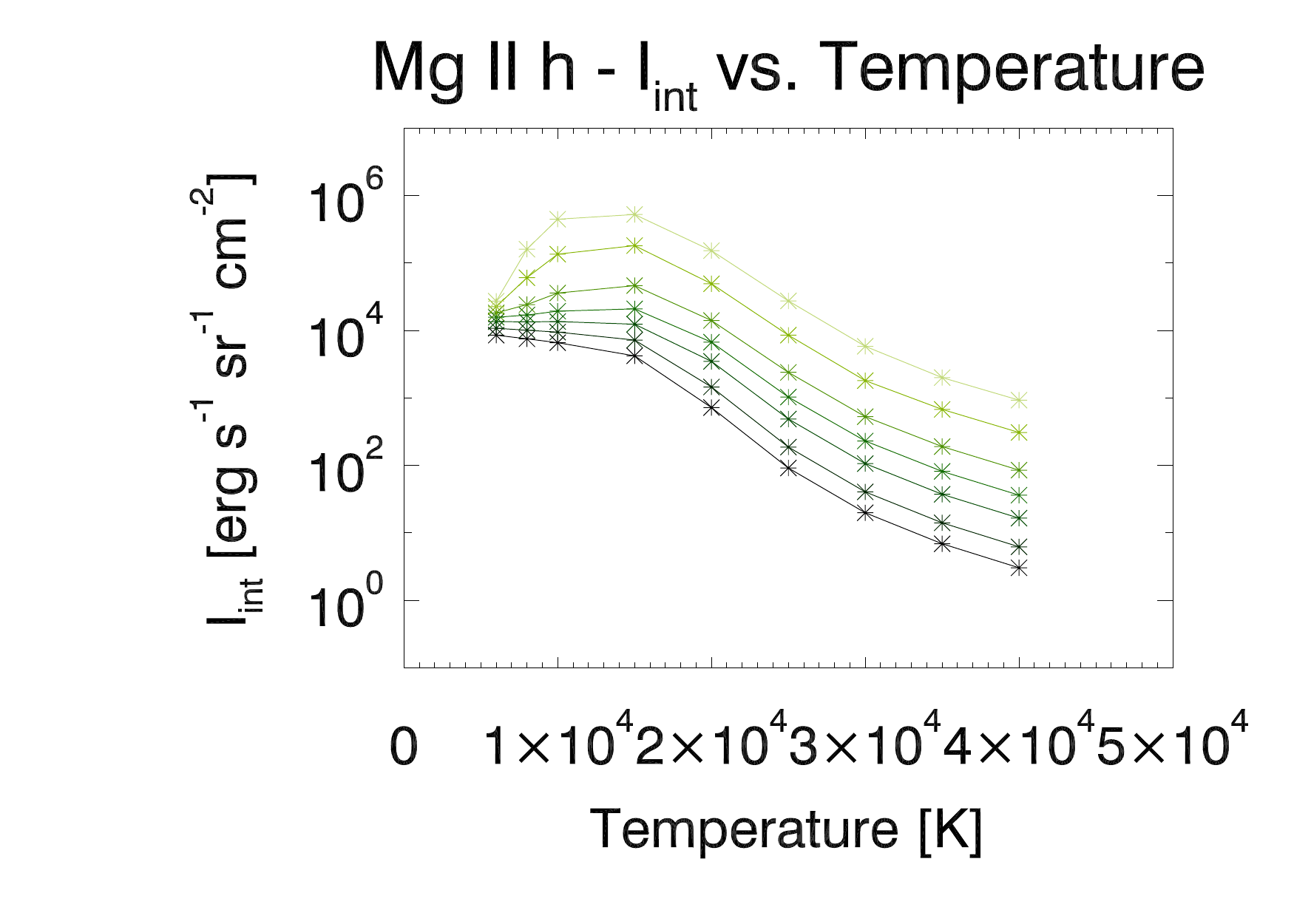}
                \includegraphics[width=0.49\hsize,clip=true,trim=1cm 0.7cm 0.5cm 0cm]{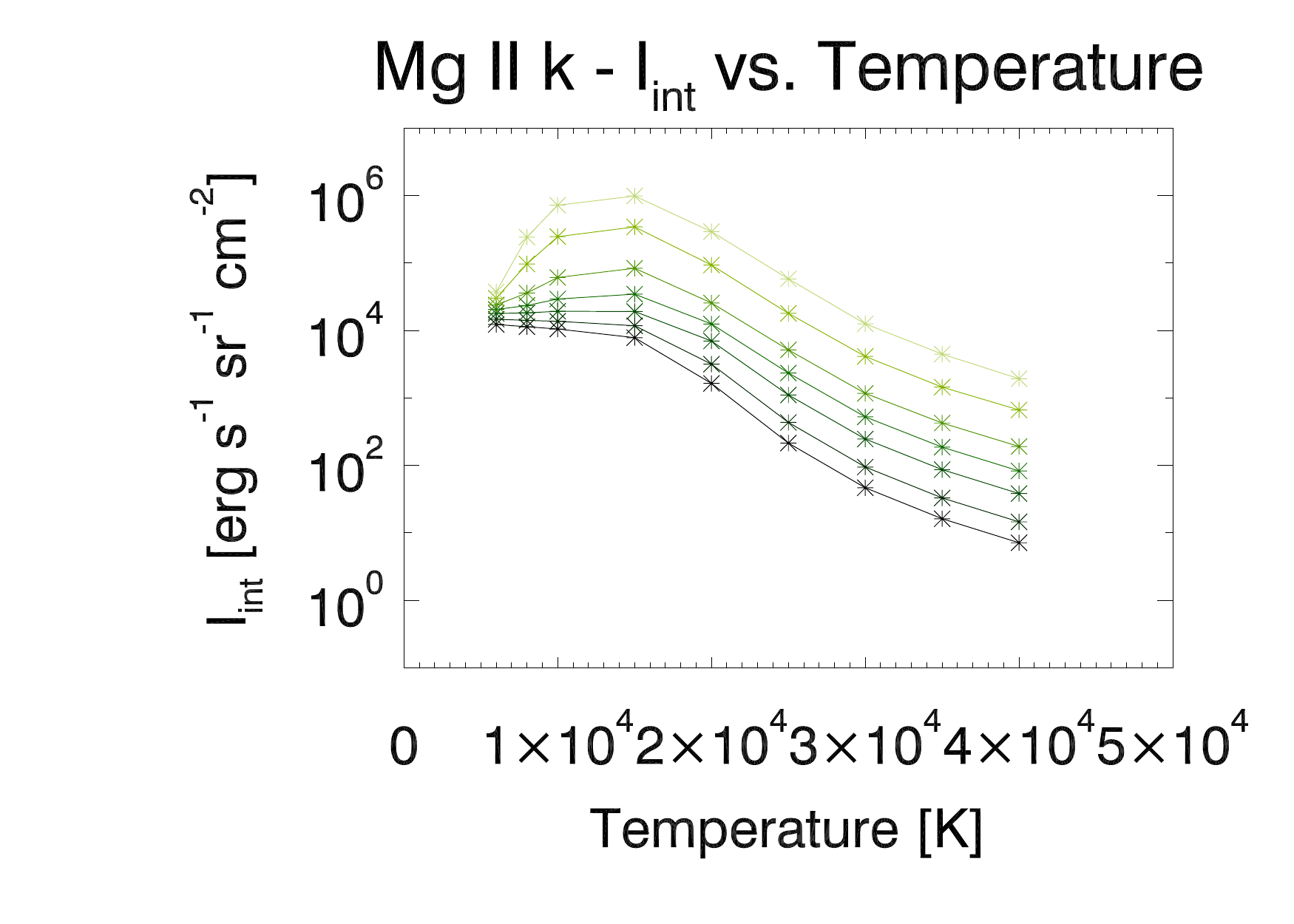}
                \caption{Plots of {frequency-}integrated intensity against temperature for {isothermal isobaric} models with $D = 1000$~km. Colour indicates pressures of (from dark to light) $0.01$, $0.02$, $0.05$, $0.1$, $0.2$, $0.5$, and $1$~dyne~cm$^{-2}$. \textit{Left:} \ion{Mg}{ii} h. \textit{Right:} \ion{Mg}{ii} k.}
                \label{fig:inti_t}
        \end{center}
\end{figure}

{To explore the relevance of  high-temperature models to observations in more detail, we present in} Figure~\ref{fig:high_t} the emergent half h and k profiles for {isothermal isobaric} models with $T \geq 15000$~K at one slab thickness, $D = 1000$~km, and two gas pressures, $P = 0.1$~dyne~cm$^{-2}$ (solid lines), and $P = 0.5$~dyne~cm$^{-2}$ (dashed lines). 
Each temperature is given a different colour to distinguish them. 
The profiles are reversed for the $15000$~K models, becoming flat-topped at $20000$~K, and are not reversed above $25000$~K;  this can already be seen in Figure~\ref{fig:252_rev_tau_comb} where all high-temperature models do not show central reversals. 
Also, as the temperature increases the line intensity decreases. 
This can be seen for line centre intensity in Figure~\ref{fig:high_t}, and also for the {frequency-}integrated line intensity which is plotted as a function of temperature in Figure~\ref{fig:inti_t}. 
Figure~\ref{fig:inti_t} shows plots for only {isothermal isobaric} models with $D = 1000$~km, but includes all the pressures  indicated by the colour gradient. 
These plots mostly have peaks at a temperature of around $10000 - 15000$~K for higher pressure models, around the formation temperature of \ion{Mg}{ii} \citep{Leenaarts2013,Leenaarts2013b}. 
In these cases, the peak occurs where collisional processes would be expected to begin to dominate over radiative ones.
At low pressures the emission seen will mostly be the result of resonant scattering of photons from the solar disc, causing radiative excitation of the lower levels of the h and k lines. 
As the pressure increases, the collisional processes become more important so the line intensity generally increases with respect to lower pressure models. 
Above $15000$~K a larger proportion of \ion{Mg}{ii} is ionised to \ion{Mg}{iii} so there is  less \ion{Mg}{ii} in the relevant energy levels for the h and k transitions, and  the intensity drops off as temperature increases. 

\begin{figure}
        \begin{center}
                \includegraphics[width=0.49\hsize,clip=true,trim=1cm 0.7cm 0.5cm 0cm]{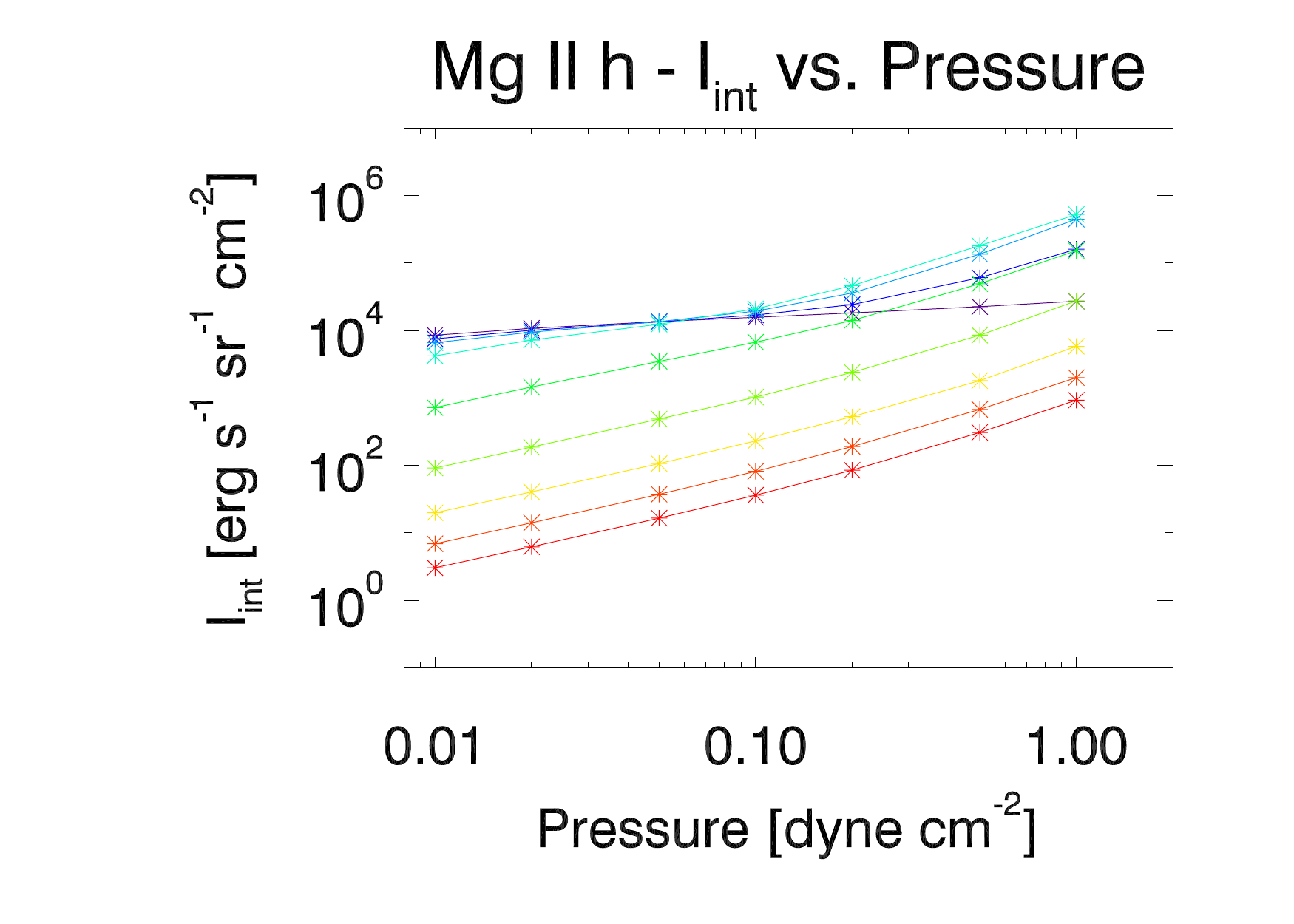}
                \includegraphics[width=0.49\hsize,clip=true,trim=1cm 0.7cm 0.5cm 0cm]{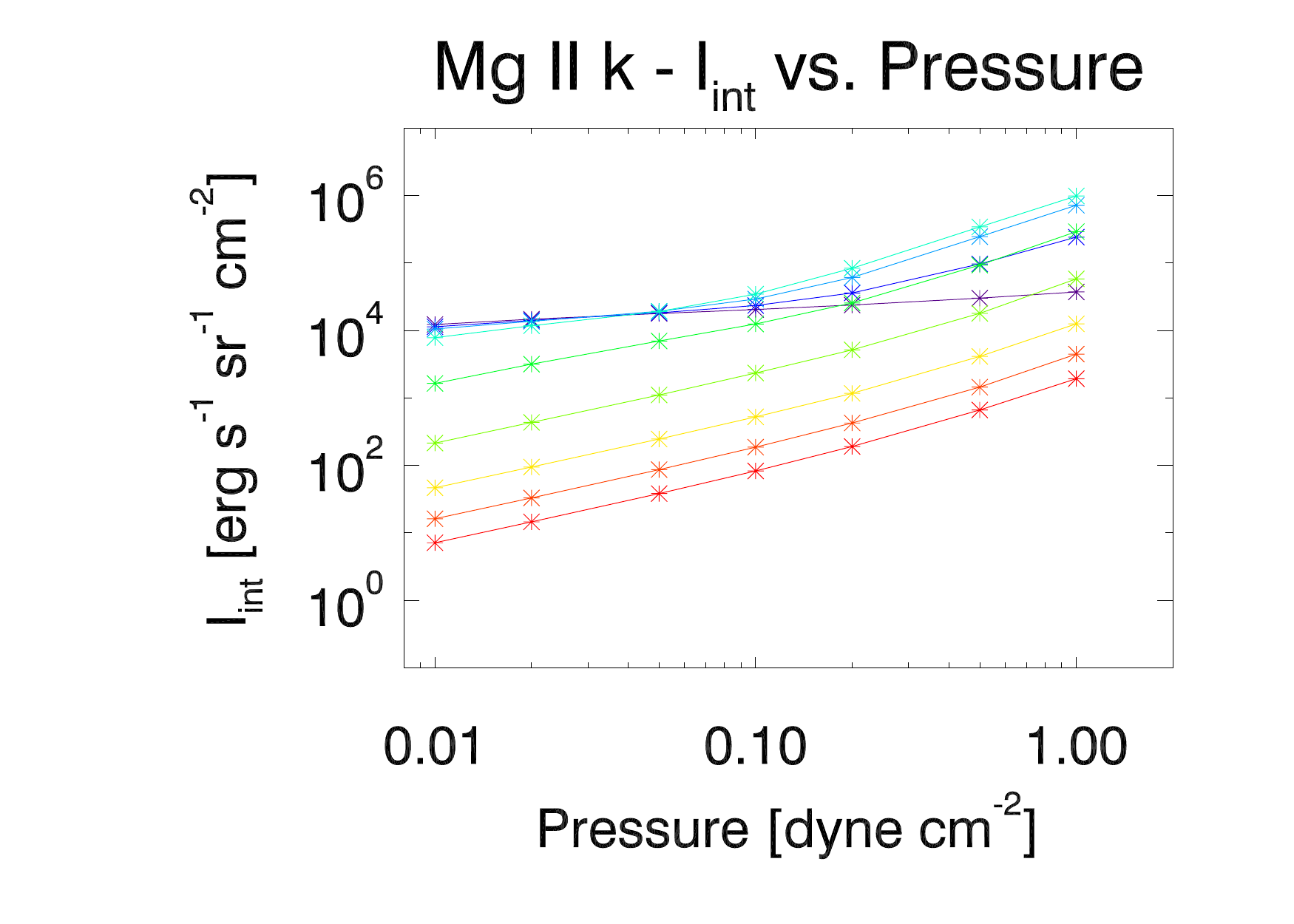}\\
                \includegraphics[width=0.49\hsize,clip=true,trim=1cm 0.7cm 0.5cm 0cm]{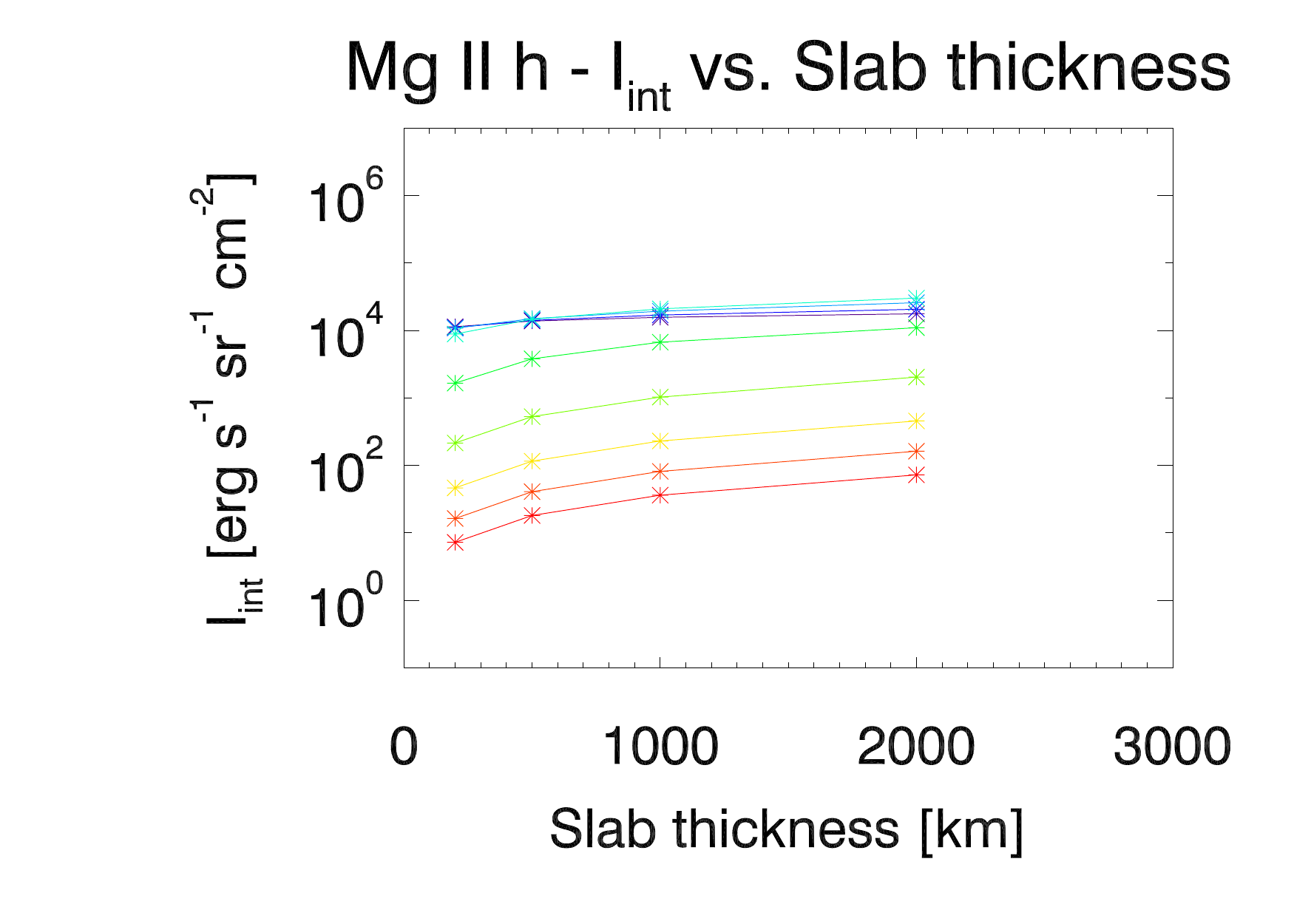}
                \includegraphics[width=0.49\hsize,clip=true,trim=1cm 0.7cm 0.5cm 0cm]{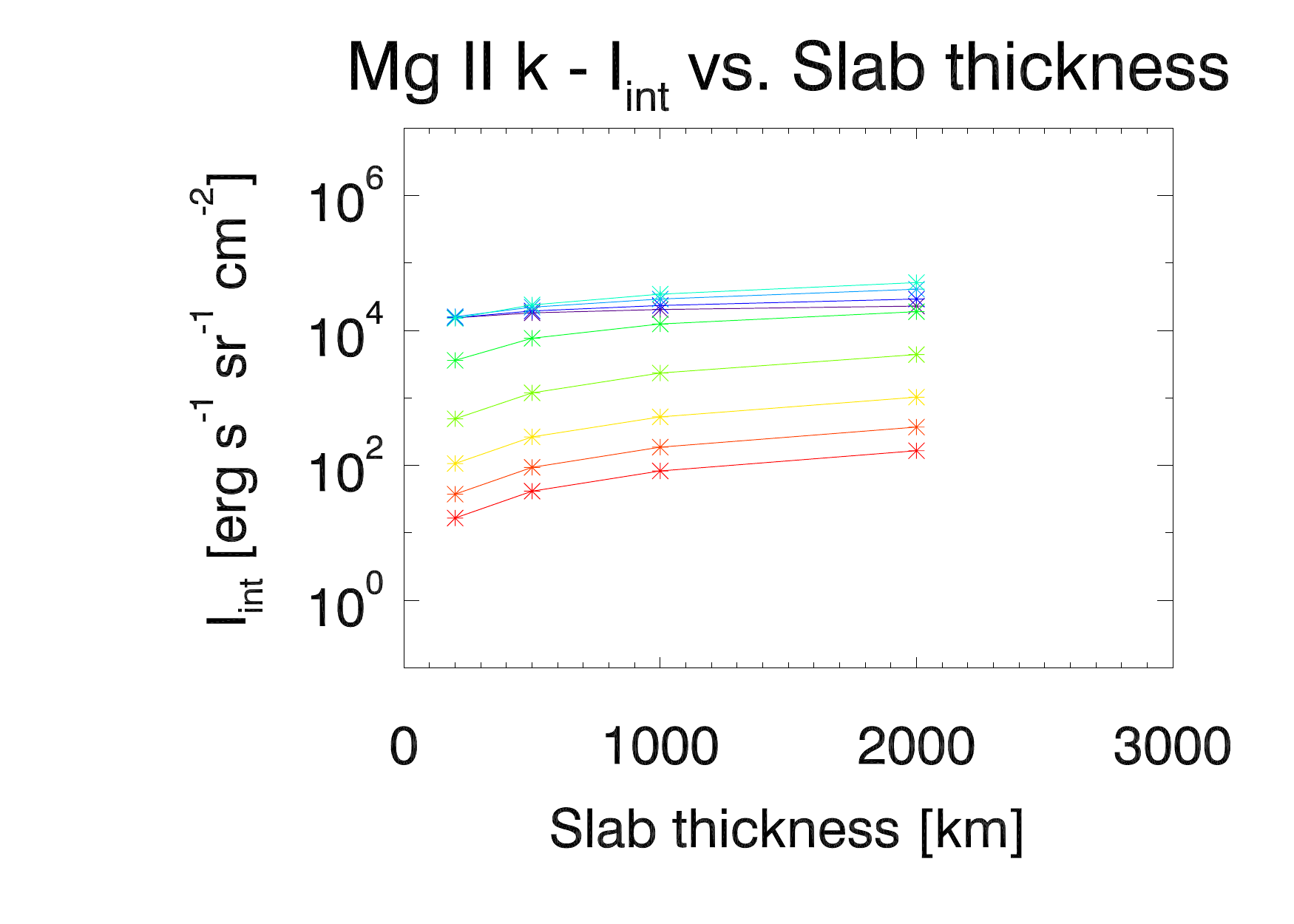}
                \caption{Plots of {frequency-}integrated intensity {for \ion{Mg}{ii} h (\textit{left}) and \ion{Mg}{ii} k (\textit{right})} against gas pressure (\textit{top}) and slab thickness (\textit{bottom}) {for isothermal isobaric models}. Pressure plots are for models with $D = 1000$~km, slab thickness plots are for models with $P = 0.1$~dyne~cm$^{-2}$. {Colour indicates temperature, increasing from purple to red.}}
                \label{fig:inti_pd}
        \end{center}
\end{figure}
Figure~\ref{fig:inti_pd} shows plots of {frequency-}integrated intensity against gas pressure  and slab thickness  for the \ion{Mg}{ii} h  and k  lines {in the case of isothermal isobaric models}. 
The top panels are shown for one slab thickness ($D = 1000$~km) and the bottom panels are shown for one gas pressure ($P = 0.1$~dyne~cm$^{-2}$). 
Temperatures are  colour-coded, with low temperatures having darker colours and high temperatures having lighter colours. 
{The frequency-integrated intensities of the h and k lines increase with gas pressure and with slab thickness. These plots clearly show, at the temperatures most representative of prominence conditions, that these dependencies are weak, which is due to the relatively high optical thickness of the plasma.}
The temperature structure seen in Figure~\ref{fig:inti_t} (peaked around $10000 - 15000$~K) can also be identified in these plots. 

The results shown here indicate that the formation temperature of the \ion{Mg}{ii} lines estimated from purely collisional excitation, $30000$~K \citep{Sigut1995,Dere1997,DelZanna2015}, is an overestimation for the case of solar prominences. 
This  was also found for the chromosphere \citep{Leenaarts2013,Leenaarts2013b}, where the importance of radiative excitation means that photons of \ion{Mg}{ii} h and k are created much more frequently at lower temperatures, around $10000$~K. 
The effects of ionisation from \ion{Mg}{ii} to \ion{Mg}{iii} at $30000$~K are also considerable, with the population of the \ion{Mg}{ii} ground level being reduced to the stage that the plasma becomes optically thin at h and k wavelengths, and no central reversals are seen.

{Our two grids} of models provide a large extension to the $27$ {isothermal isobaric} models presented by \citet{Heinzel2014}. 
However, models with higher {central} temperatures ($> 20000$~K) present non-reversed profiles with extremely low peak intensity values, below $10^{-7}$~erg s$^{-1}$ cm$^{-2}$ sr$^{-1}$ Hz$^{-1}$ and as low as $5 \times 10^{-11}$~erg s$^{-1}$ cm$^{-2}$ sr$^{-1}$ Hz$^{-1}$ for some of the highest temperature slabs. 
They also show no central reversal above $20000$~K.
This does not seem to match observations, where central reversals are observed, and for non-reversed profiles line-centre intensities of more than $10^{-7}$~erg s$^{-1}$ cm$^{-2}$ sr$^{-1}$ Hz$^{-1}$ are recovered \citep[e.g.][]{Levens2017}.
It is therefore not necessary to consider models with extremely high temperatures in comparison to observations, but it appears to be reasonable to consider models with temperatures up to $\sim 20000$~K.

\begin{figure}
        \begin{center}
                \includegraphics[width=0.49\hsize,clip=true,trim=1cm 0.7cm 0.5cm 0cm]{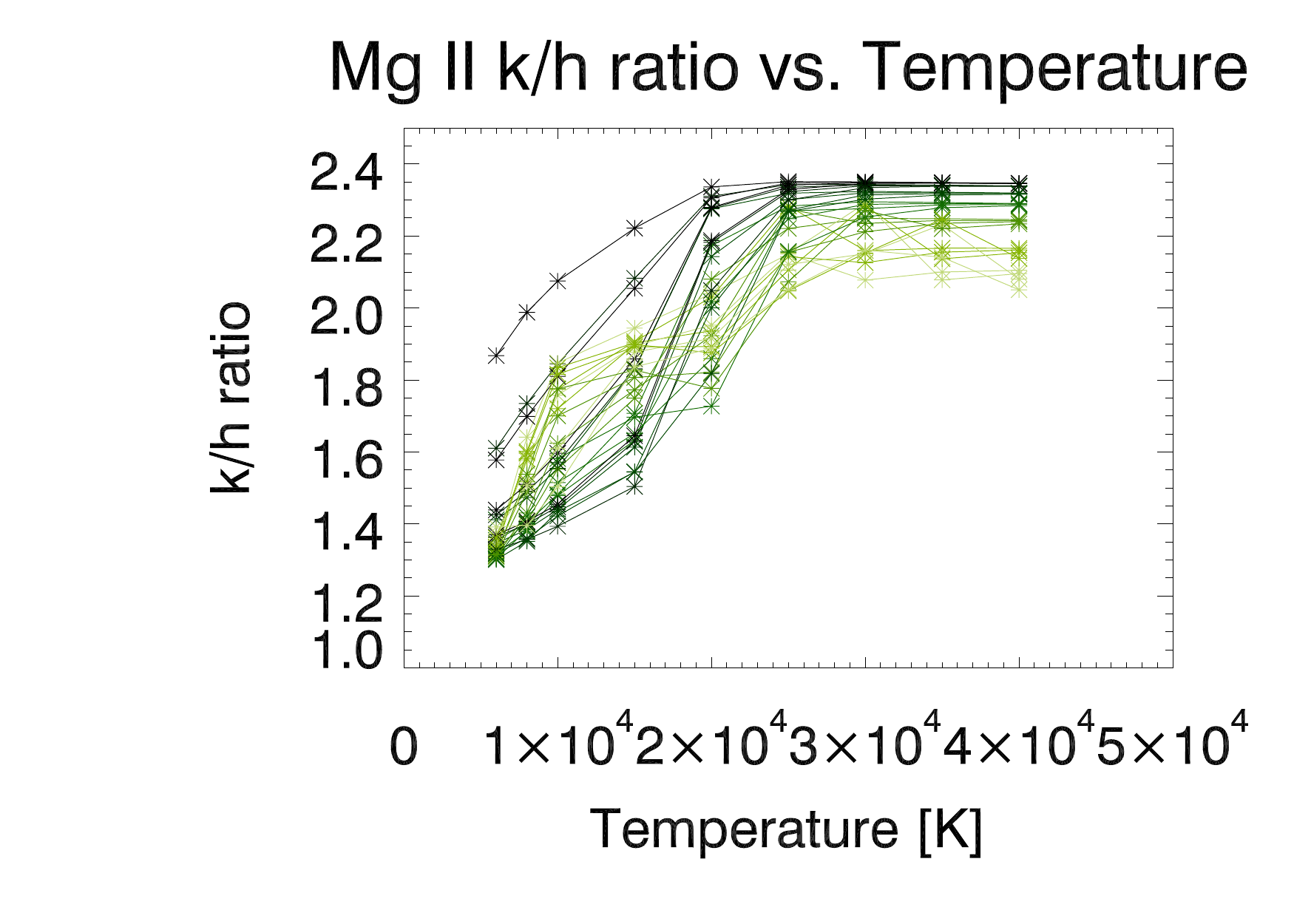}
                \includegraphics[width=0.49\hsize,clip=true,trim=1cm 0.7cm 0.5cm 0cm]{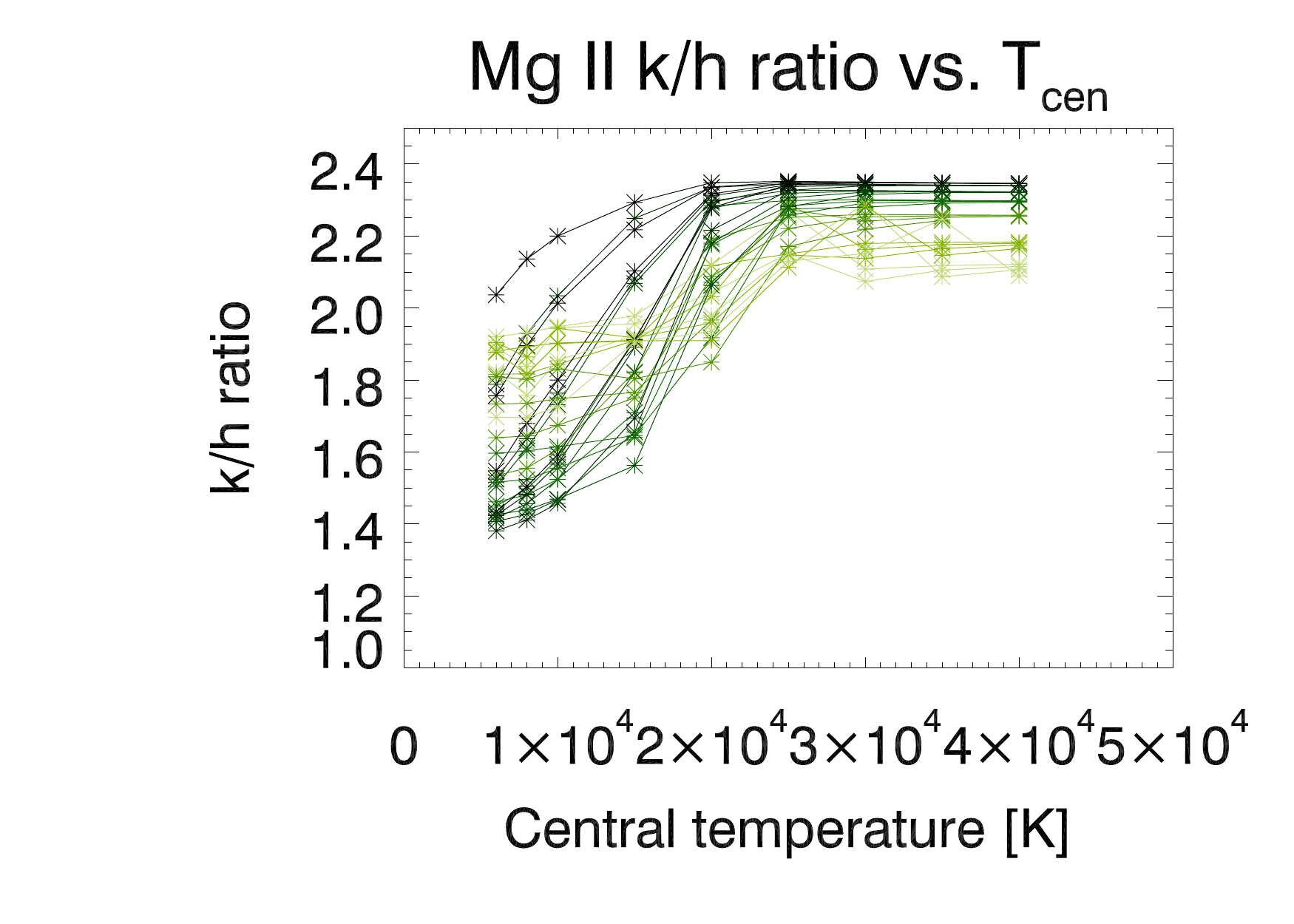}\\
                \includegraphics[width=0.49\hsize,clip=true,trim=1cm 0.7cm 0.5cm 0cm]{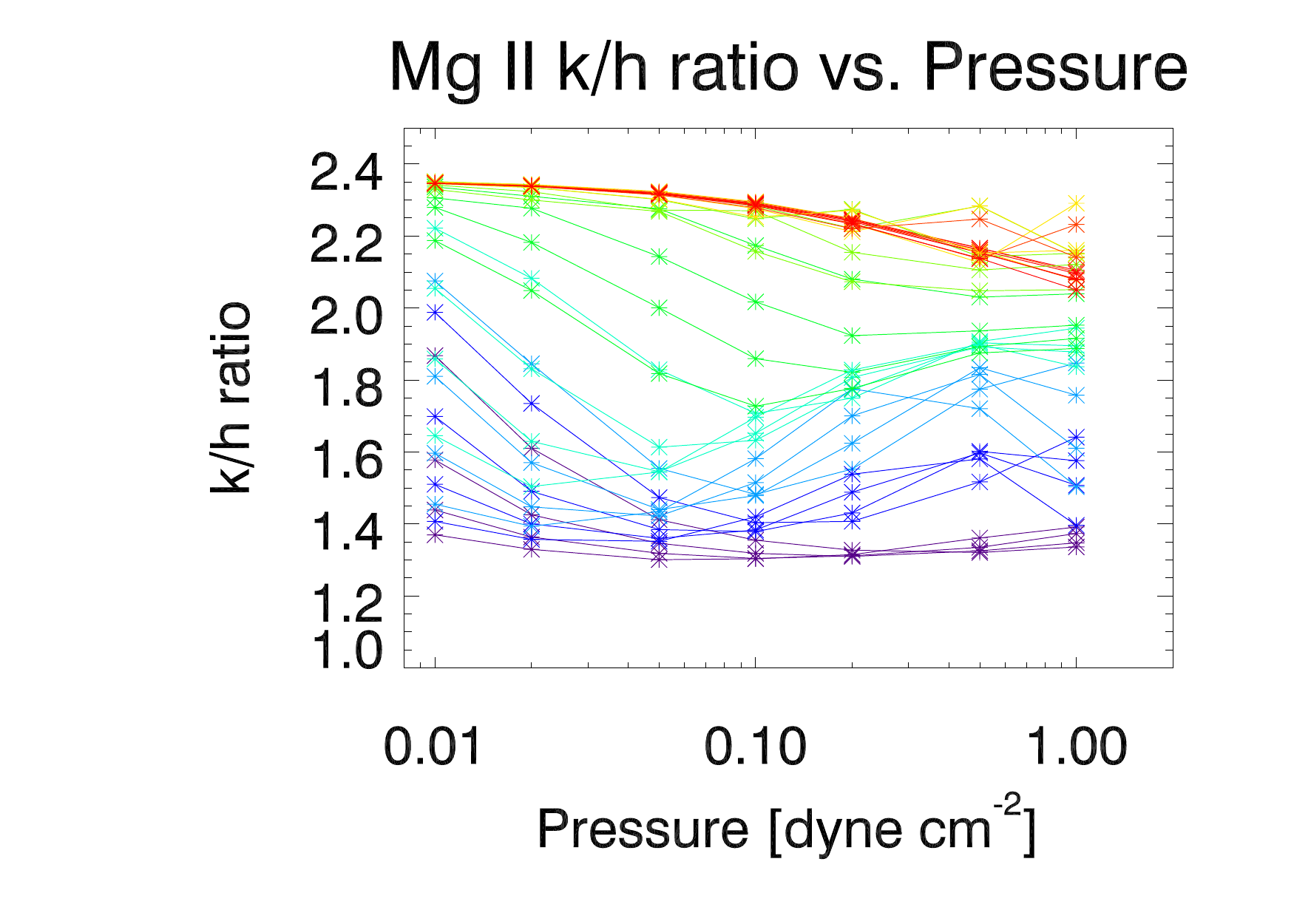}
                \includegraphics[width=0.49\hsize,clip=true,trim=1cm 0.7cm 0.5cm 0cm]{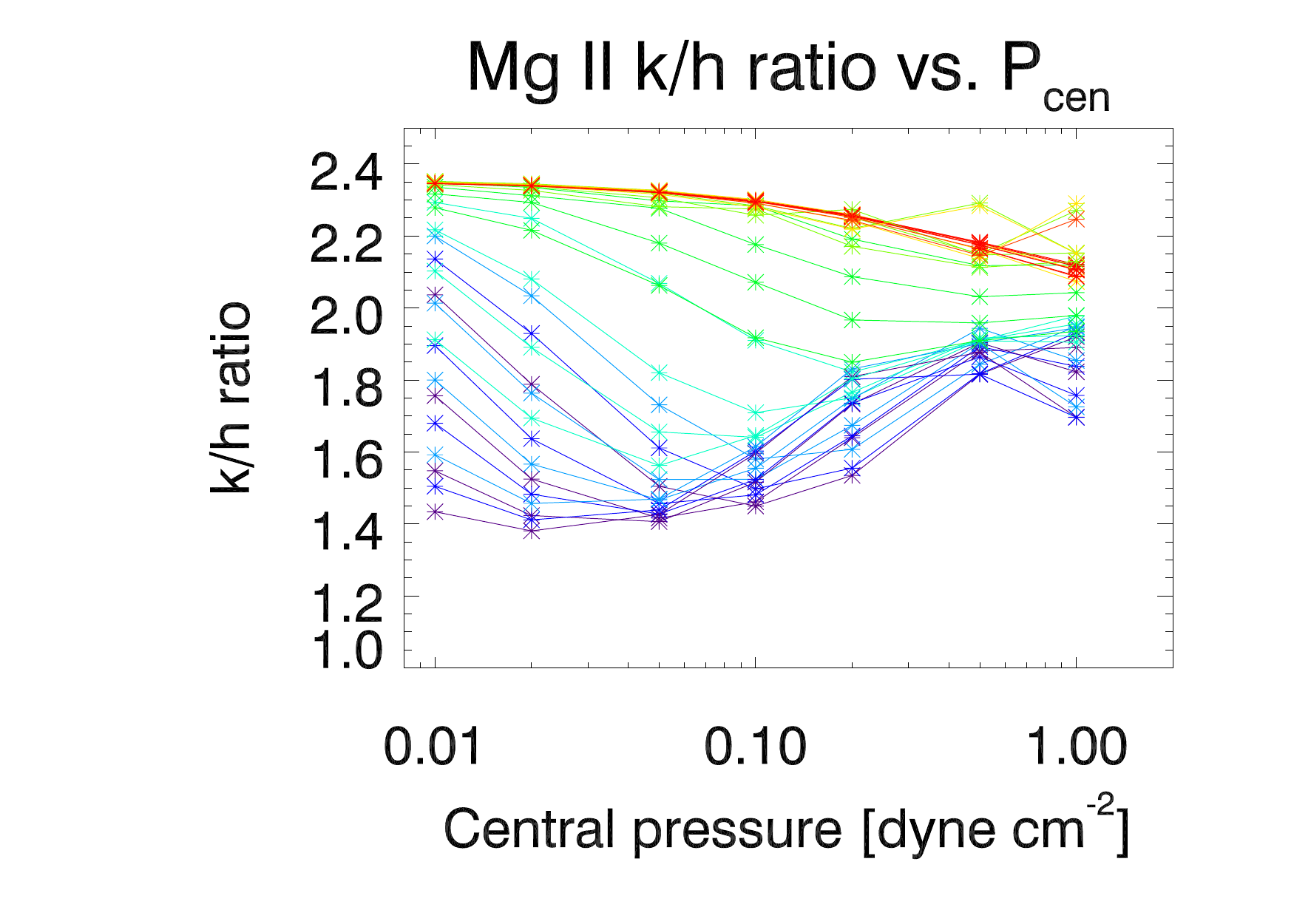}\\
                \includegraphics[width=0.49\hsize,clip=true,trim=1cm 0.7cm 0.5cm 0cm]{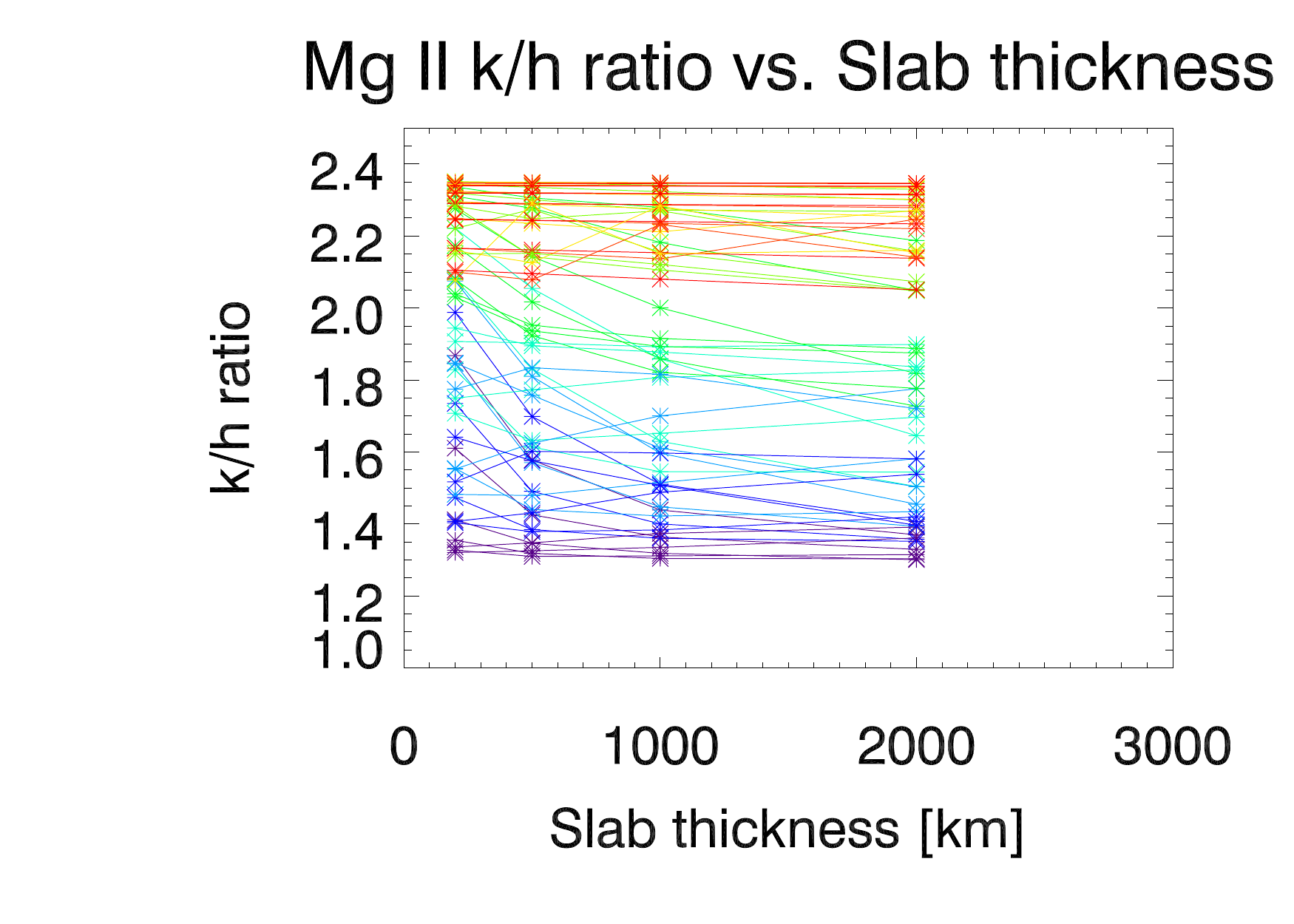}
                \includegraphics[width=0.49\hsize,clip=true,trim=1cm 0.7cm 0.5cm 0cm]{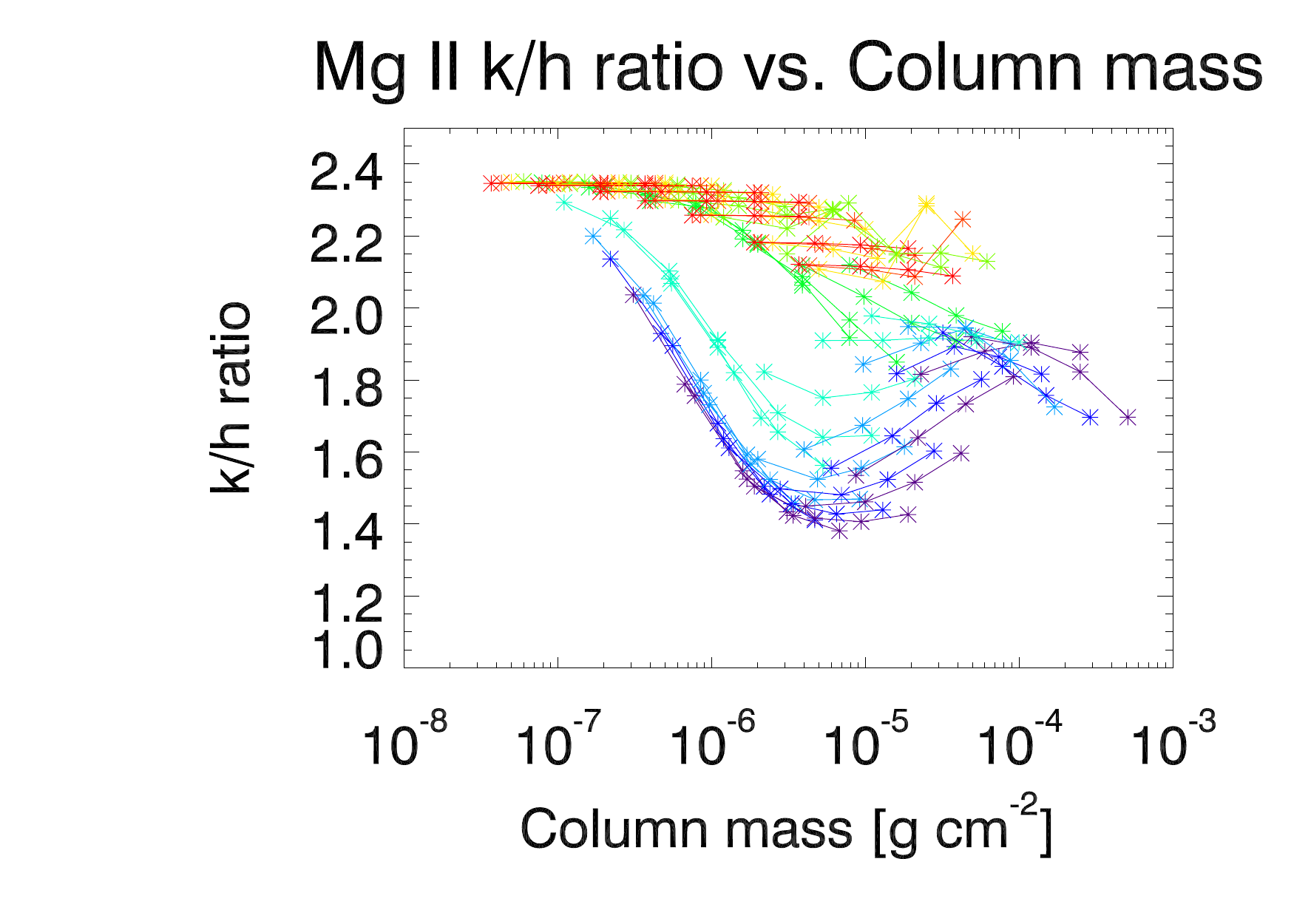}\\
                \includegraphics[width=0.49\hsize,clip=true,trim=1cm 0.7cm 0.5cm 0cm]{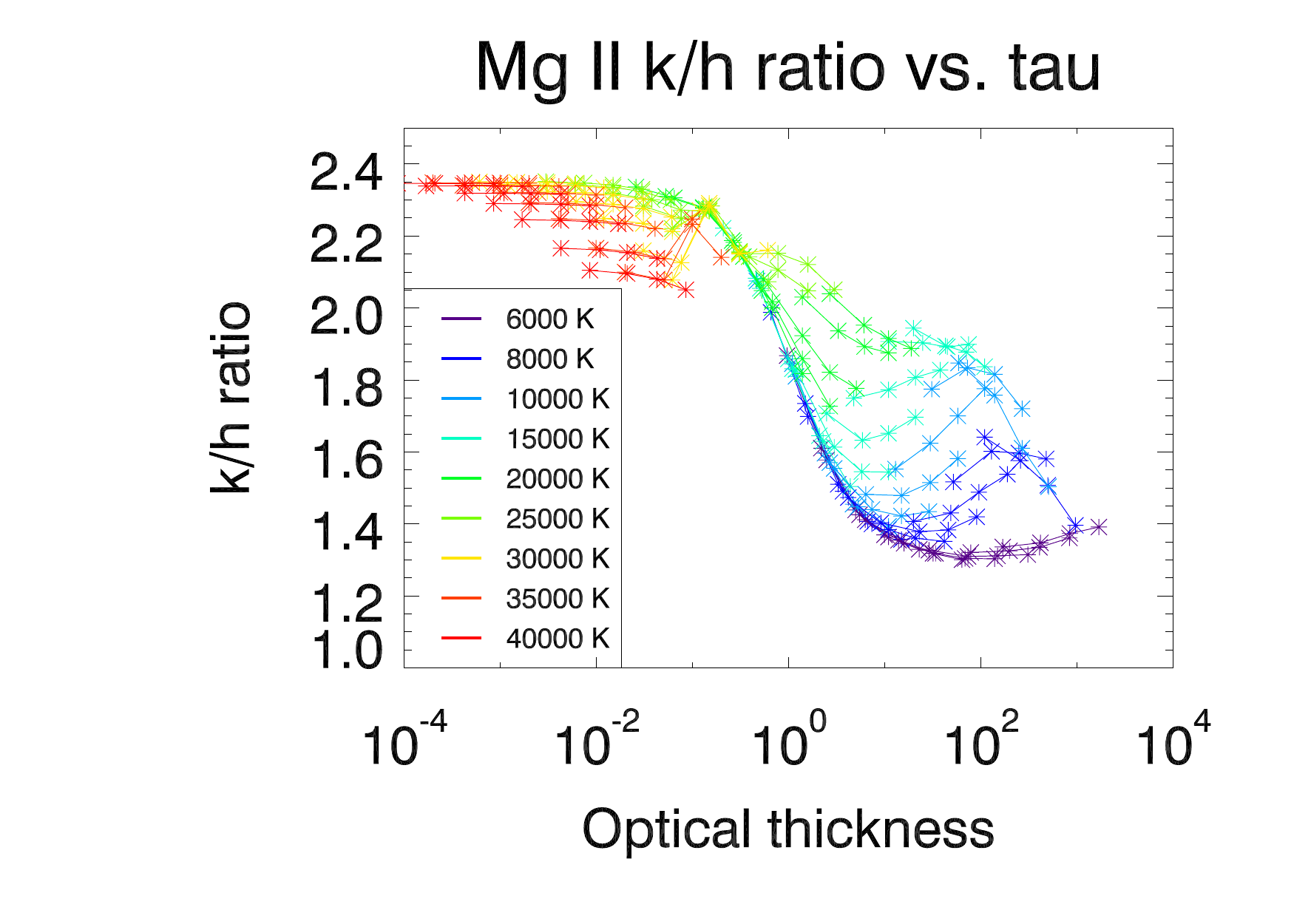}
                \includegraphics[width=0.49\hsize,clip=true,trim=1cm 0.7cm 0.5cm 0cm]{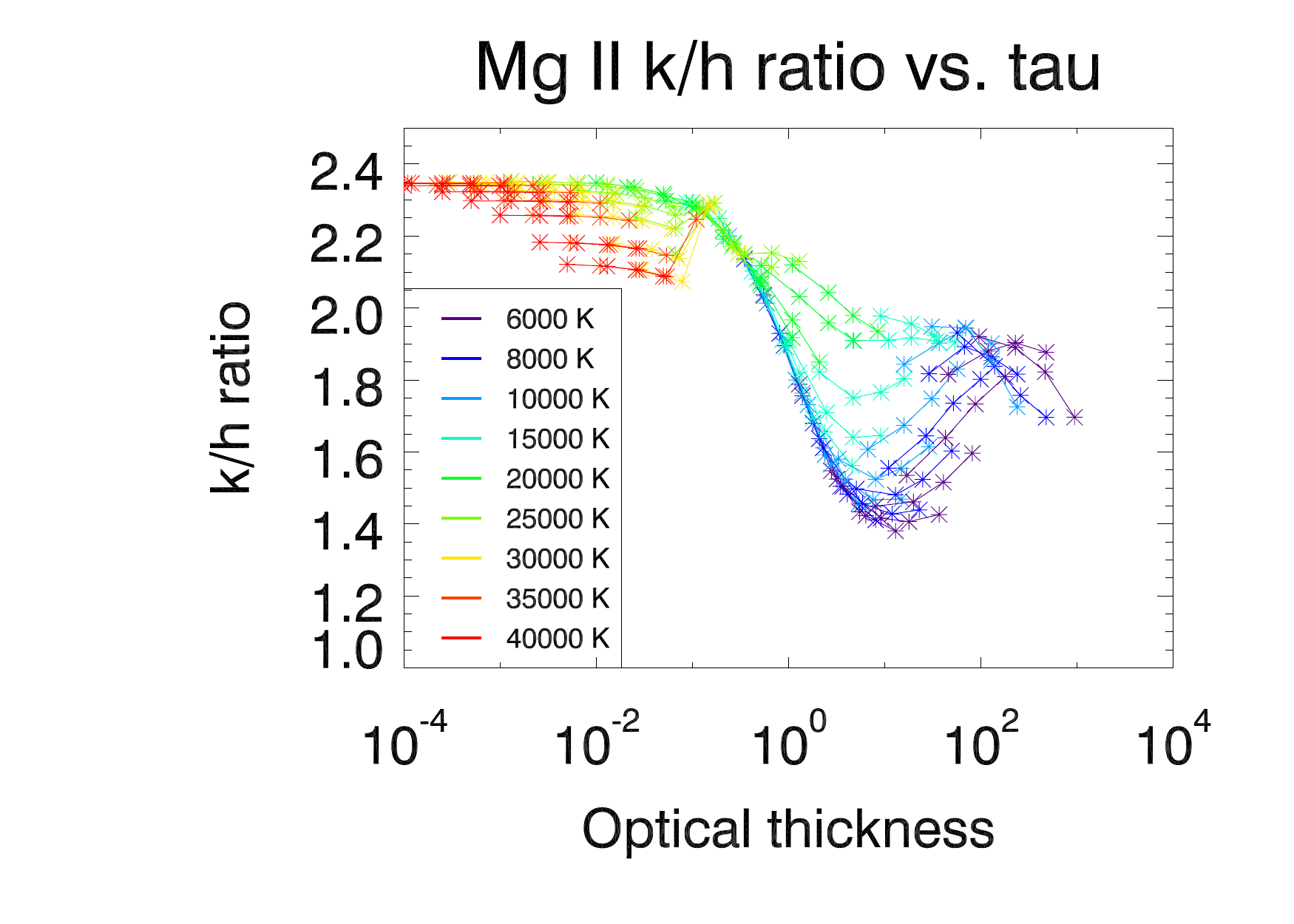}
                \caption{Plots of the \ion{Mg}{ii} k/h {line} ratio against physical parameters for the isothermal isobaric {(left column) and PCTR (with $\gamma = 2$, right column)} models. 
                        {\textit{First row:}} k/h {line ratio}  vs. $T$ (\textit{left}) {or $T_{\mathrm{cen}}$ (\textit{right})}, colour scale indicates pressure. 
                        {\textit{Second row:}} k/h {line ratio}  vs. $P$ (\textit{left}) {or $P_{\mathrm{cen}}$ (\textit{right})}, colour scale indicates temperature. 
                        {\textit{Third row:}} k/h {line ratio}  vs. $D$ (\textit{left}) {or $M$ (\textit{right})}, colour scale indicates temperature. 
                        \textit{Bottom row:} k/h {line ratio}  vs. $\tau_\mathrm{h}$, colour scale indicates temperature.}
                \label{fig:kh_vs_params_comb}
        \end{center}
\end{figure}

The k/h {line} ratio is one of the simplest parameters to calculate from observations, so understanding how it is related to the physical parameters of the plasma could potentially be very powerful for analysis. 
Figure~\ref{fig:kh_vs_params_comb} shows the k/h {line} ratio against temperature, pressure, slab thickness, and optical thickness for isothermal isobaric models {and for PCTR models with $\gamma=2$}. 
Plotting the k/h {line ratio}  against temperature {(Fig.~\ref{fig:kh_vs_params_comb}, first row)} reveals that, for all pressures, the k/h {line} ratio increases approximately linearly with temperature up to around $25000$~K, where the k/h {line} ratio saturates between $2$ and $2.4$. 
This suggests that for lower (more realistic) temperatures the k/h {line} ratio is a close analogy to temperature, regardless of slab thickness or pressure. 
The plots of k/h {line ratio}  against pressure ({second row}) do not show any significant correlations {in the isothermal isobaric case}. The central pressure has a larger effect on the k/h {line} ratio for PCTR models than in the isothermal isobaric case. 
At low temperatures, the k/h {line} ratio {may} increase with central pressure, whereas for high-temperature models the opposite is true. 
It is also interesting to note that  observed values of k/h {line} ratios around $1.4$ are only seen for the lowest pressure PCTR models, where $P_{\mathrm{cen}} = 0.01, 0.02$~dyne~cm$^{-2}$. 
The plots of k/h {line ratio}  against slab thickness {(third row, left)} do not show any significant correlations {in the isothermal isobaric case}. 
The relationship between k/h {line ratio} and column mass {for PCTR models gives a totally different picture and} is similar to that of the k/h {line} ratio and optical thickness. 
The bottom {row} of Figure~\ref{fig:kh_vs_params_comb} shows the k/h {line} ratio against optical thickness of the h line;  the k/h {line} ratio against k line optical thickness is not shown as it is almost identical. 
There is a general trend that shows lower optical thickness has a larger k/h {line} ratio, with higher optical thicknesses corresponding to lower k/h {line} ratios. 
The transition between the two cases is at $\tau$ values around the transition between optically thin and optically thick. 
There are some diversions from this trend (so there is some model dependence), but generally the k/h {line} ratio does appear to depend on the optical thickness, and when the lines are optically thin ($\tau < 1$) the k/h {line} ratio is higher than $2$.
{In the PCTR case}, there are {fewer} models at  k/h {line} ratios of $1.4$ {than in the isothermal isobaric case} ; that ratio appears to correlate only to a few models where the optical thickness is around $10$, corresponding to those models with low $P_{\mathrm{cen}}$, mentioned previously.

\begin{figure*}
        \begin{center}
                \includegraphics[width=0.24\hsize,clip=true,trim=1cm 0.7cm 0.5cm 0cm]{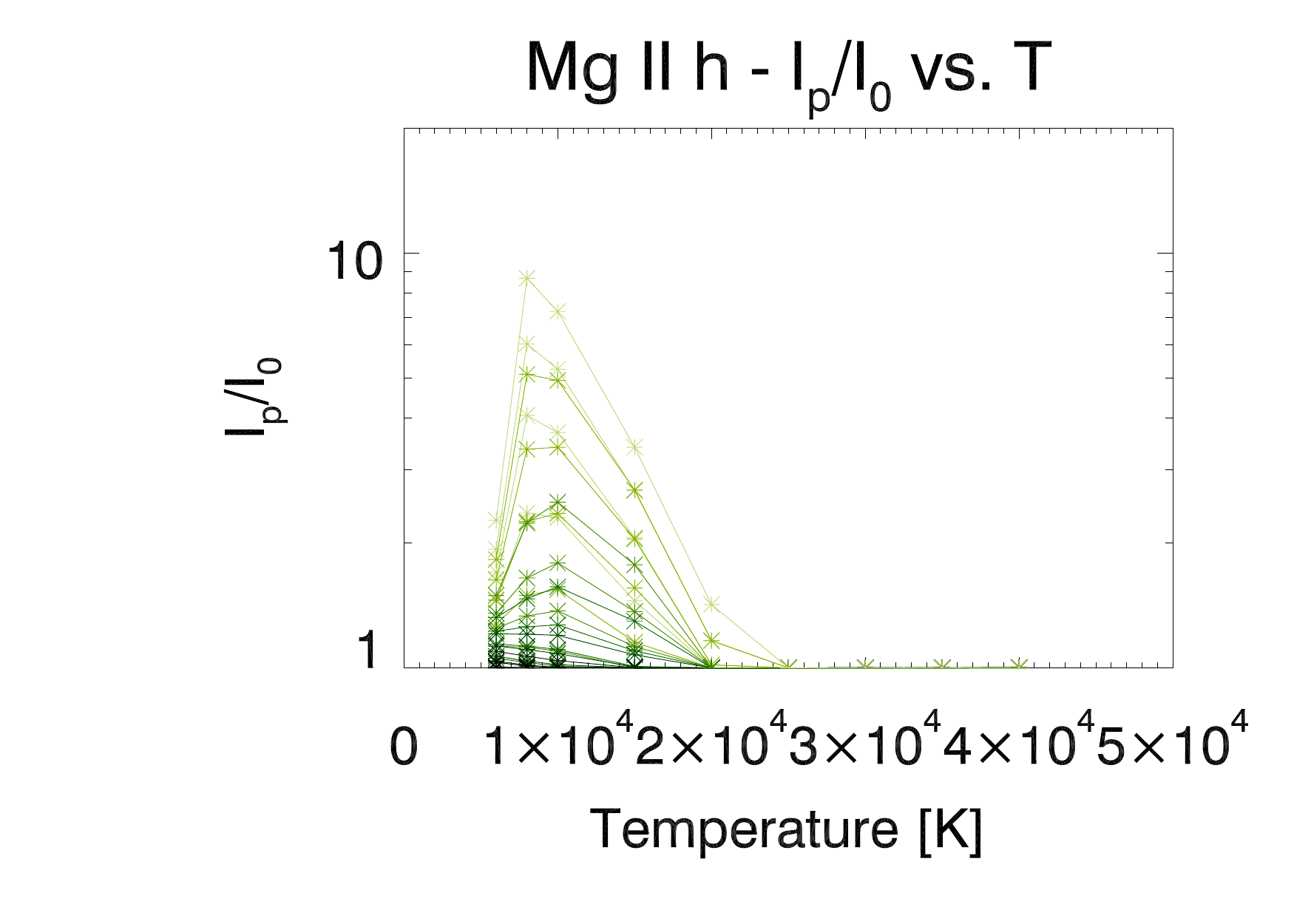}
                \includegraphics[width=0.24\hsize,clip=true,trim=1cm 0.7cm 0.5cm 0cm]{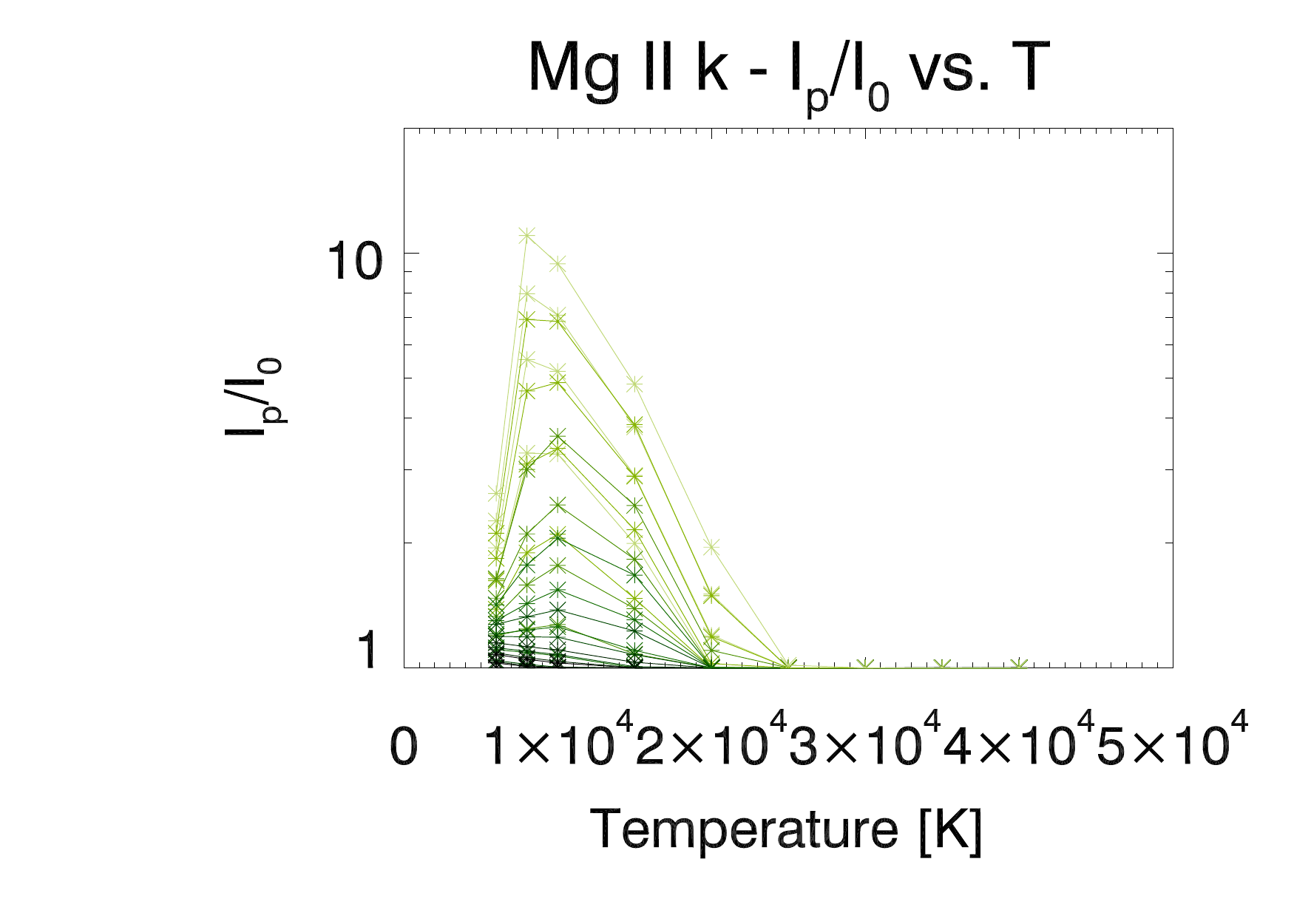}
                \includegraphics[width=0.24\hsize,clip=true,trim=1cm 0.7cm 0.5cm 0cm]{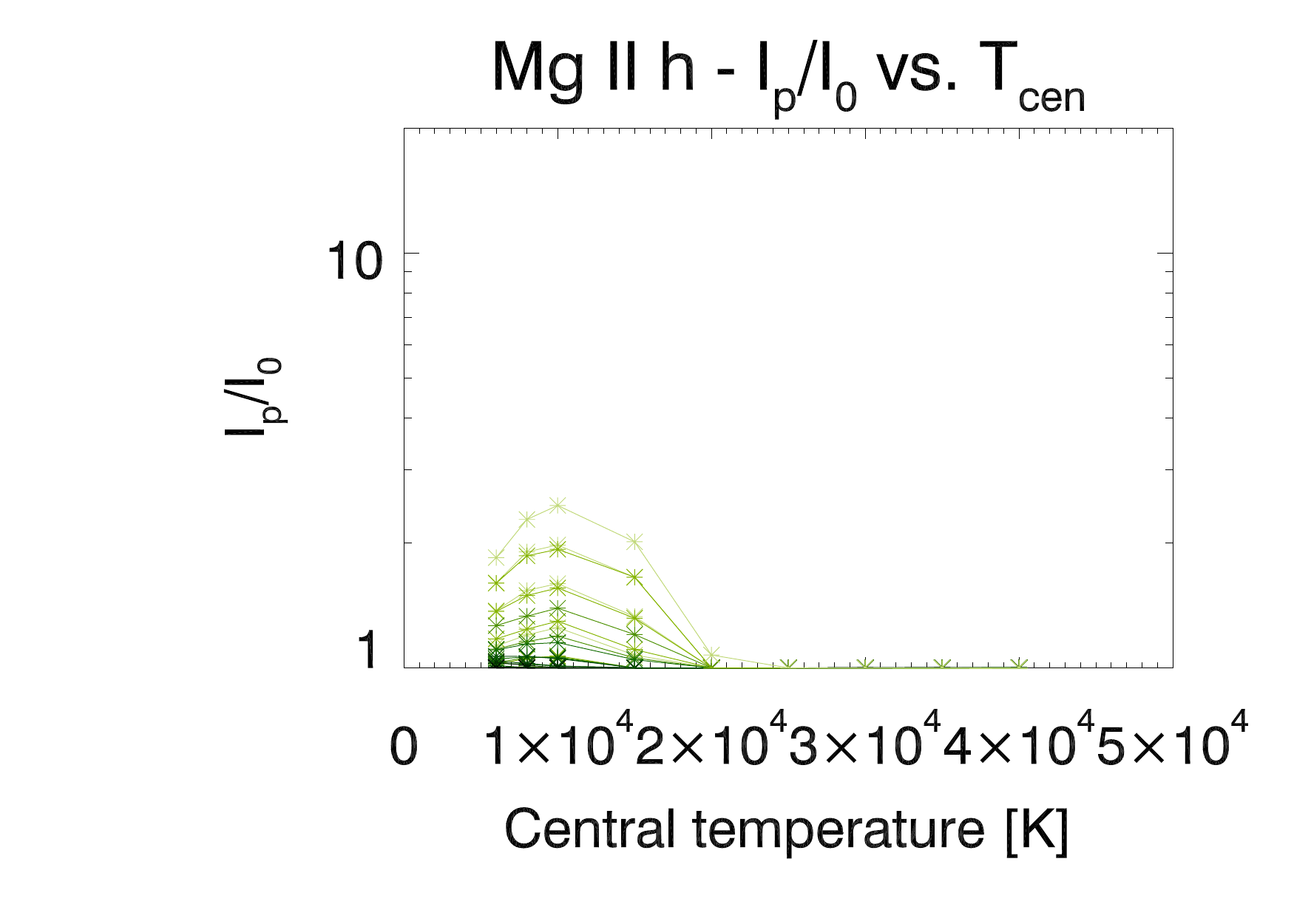}
                \includegraphics[width=0.24\hsize,clip=true,trim=1cm 0.7cm 0.5cm 0cm]{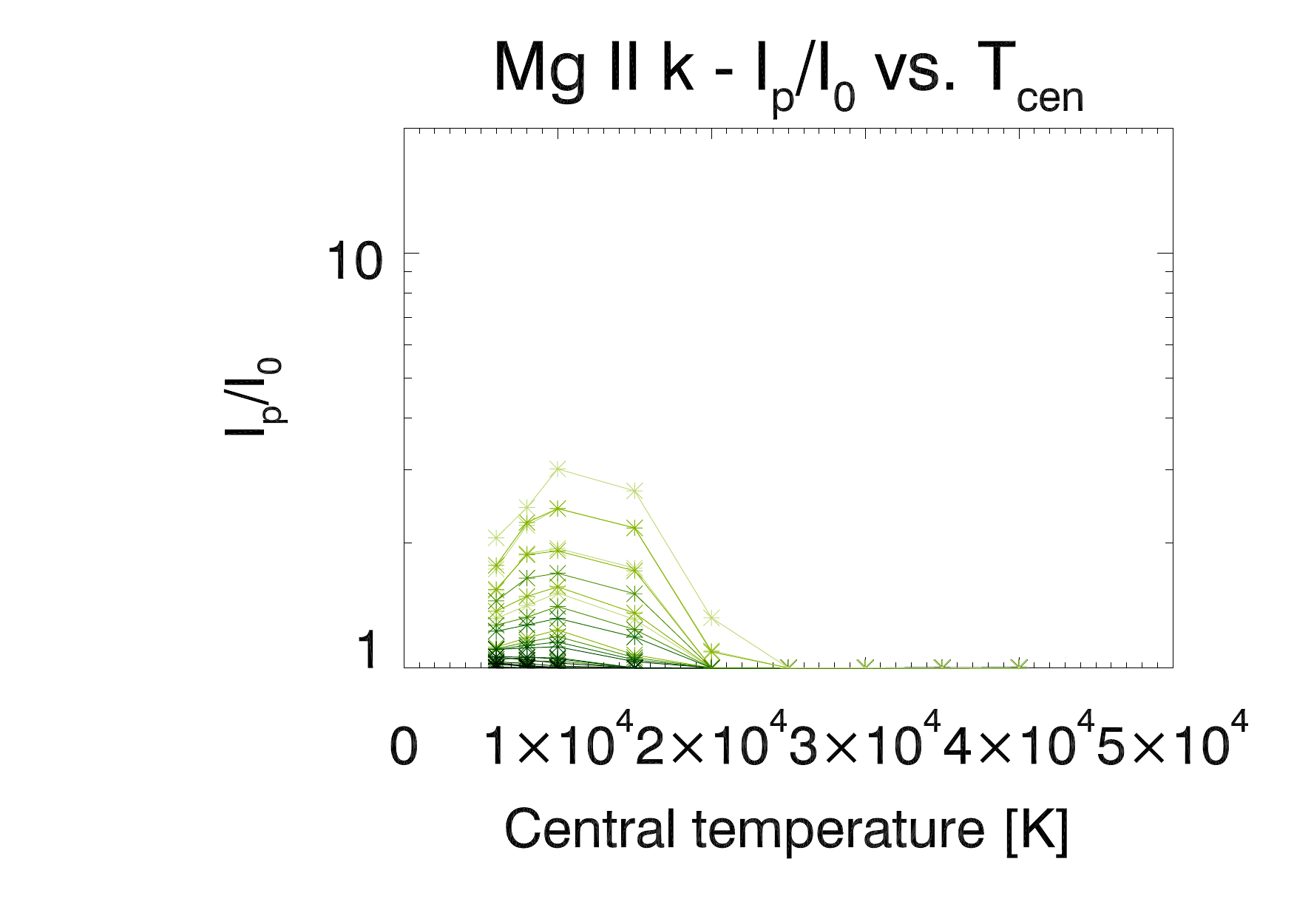}\\
                \includegraphics[width=0.24\hsize,clip=true,trim=1cm 0.7cm 0.5cm 0cm]{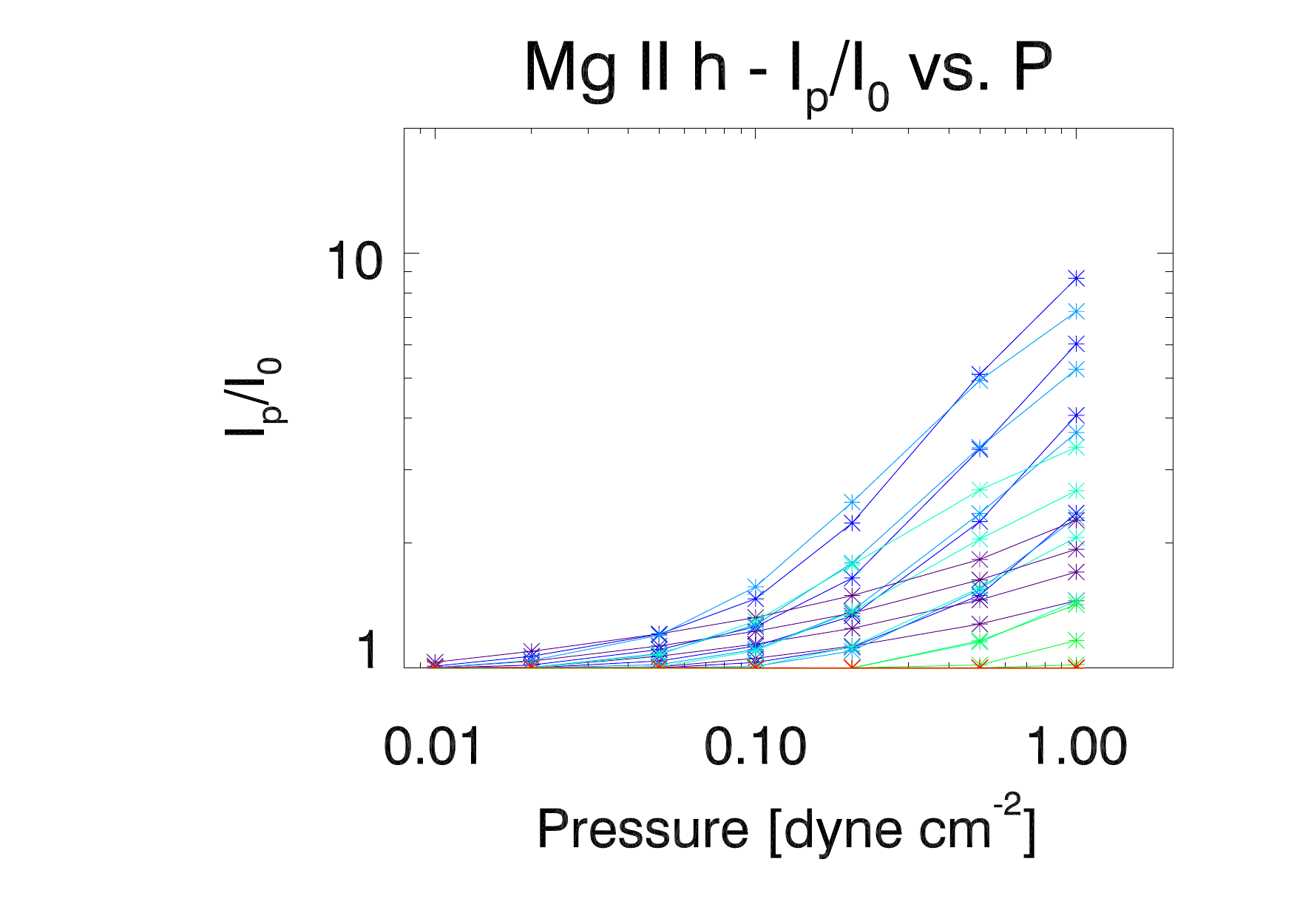}
                \includegraphics[width=0.24\hsize,clip=true,trim=1cm 0.7cm 0.5cm 0cm]{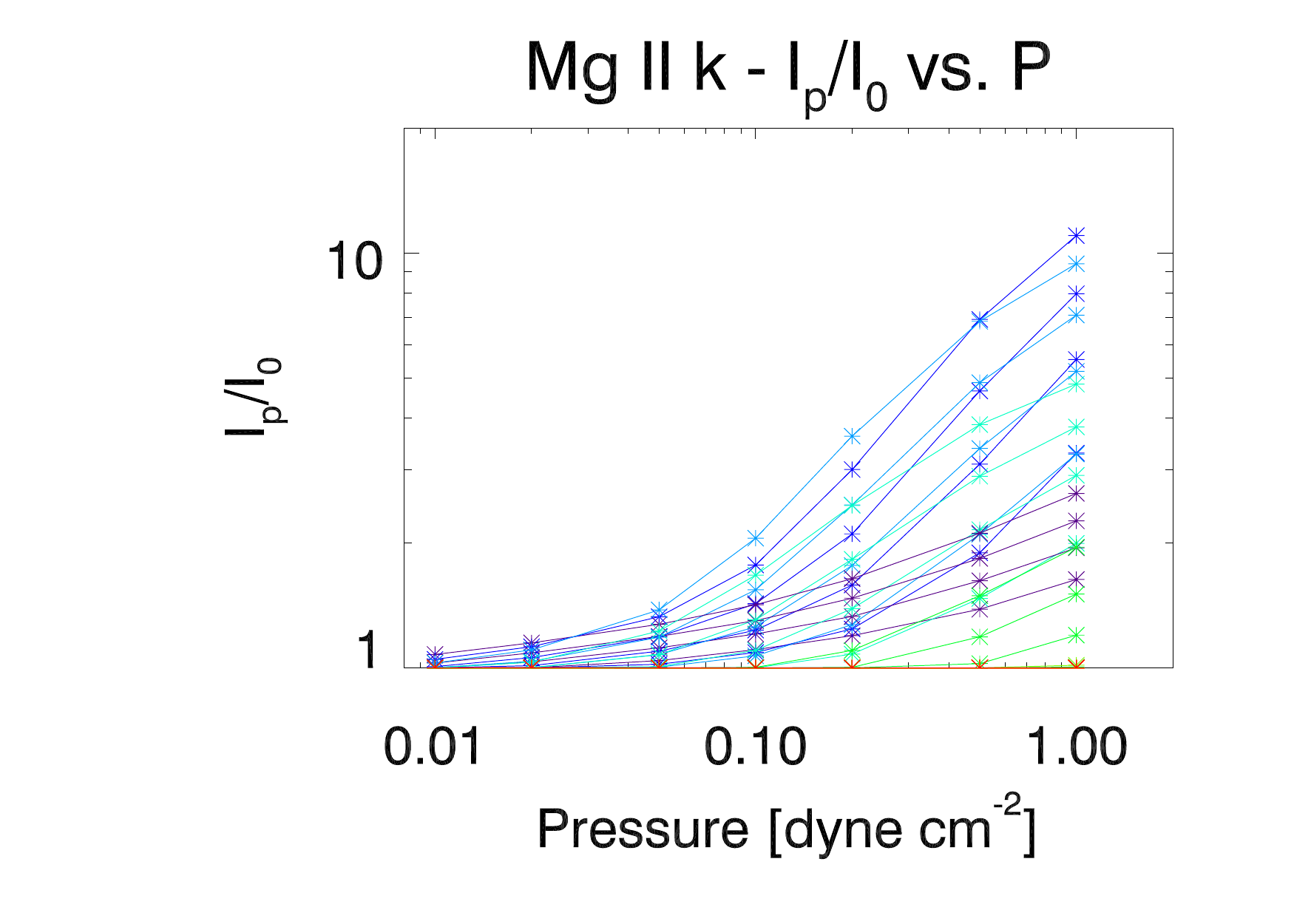}
                \includegraphics[width=0.24\hsize,clip=true,trim=1cm 0.7cm 0.5cm 0cm]{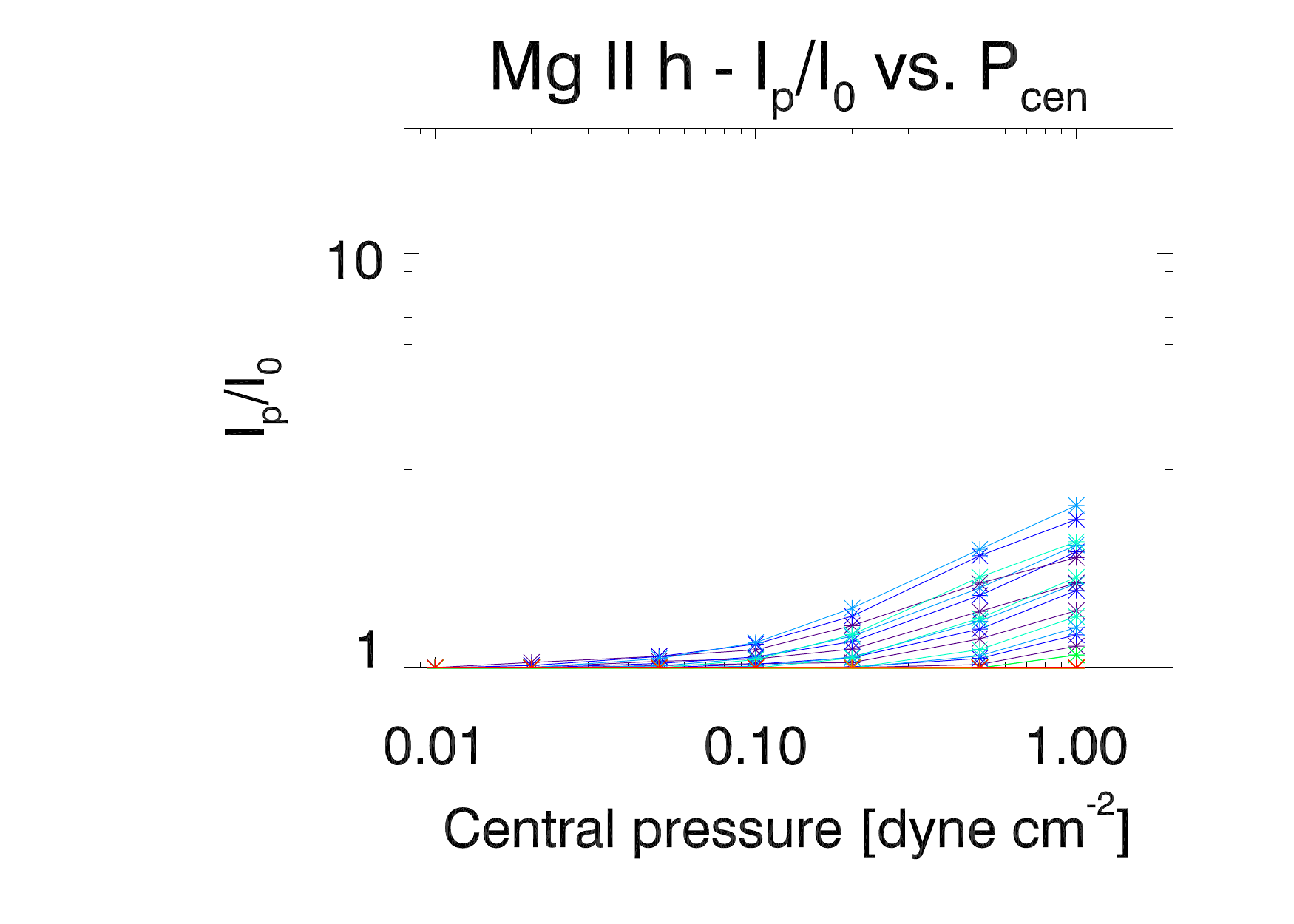}
                \includegraphics[width=0.24\hsize,clip=true,trim=1cm 0.7cm 0.5cm 0cm]{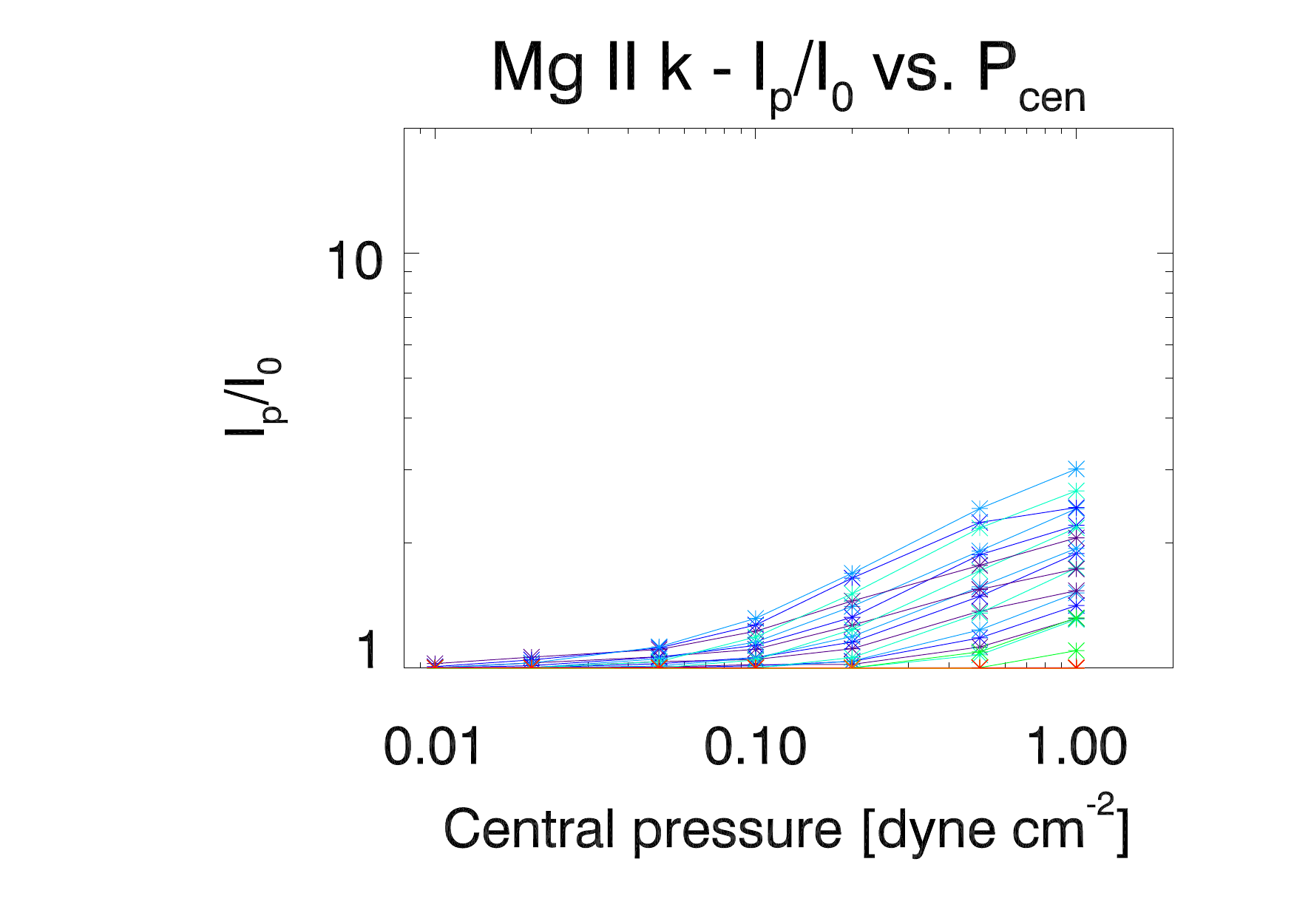}\\
                \includegraphics[width=0.24\hsize,clip=true,trim=1cm 0.7cm 0.5cm 0cm]{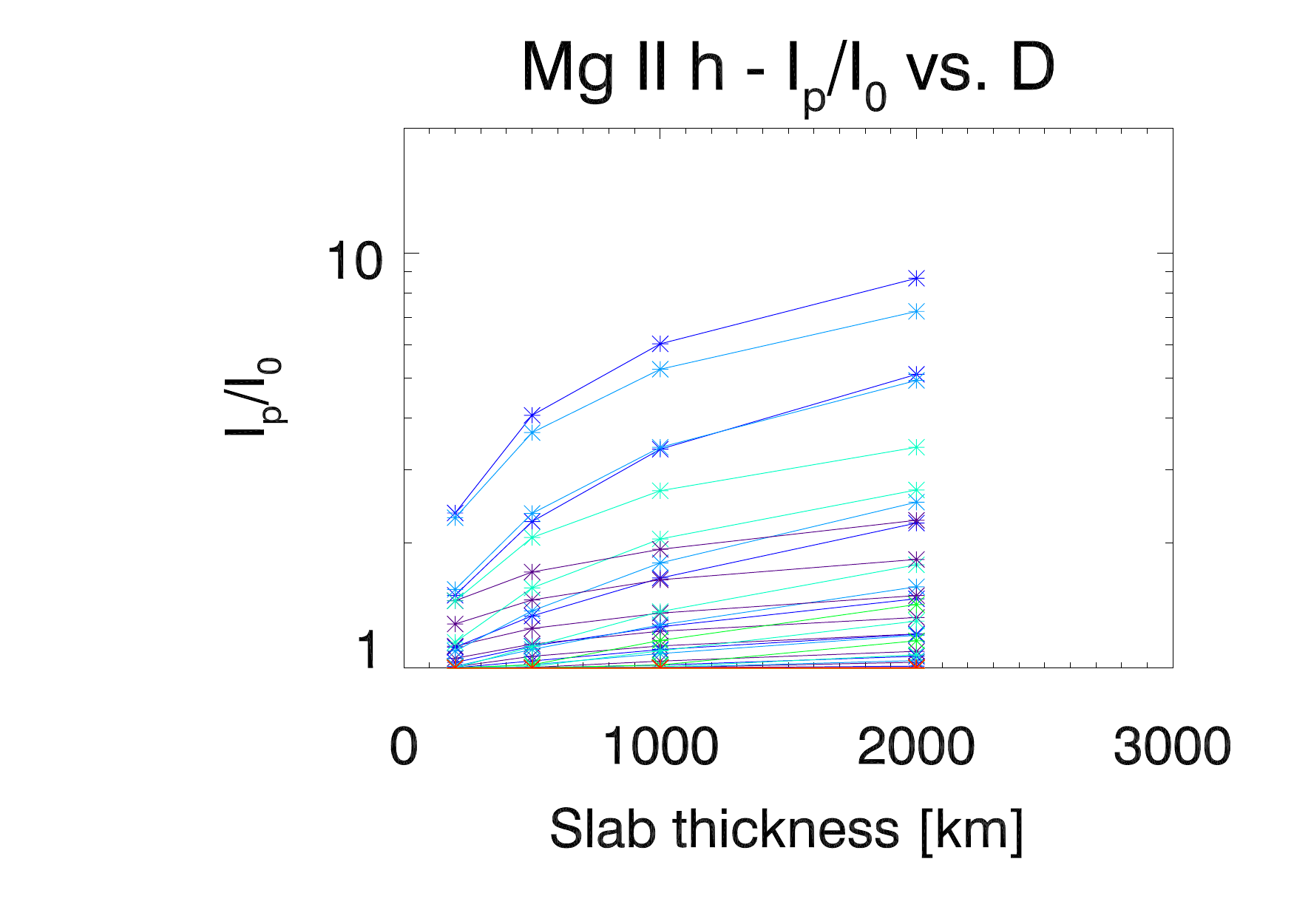}
                \includegraphics[width=0.24\hsize,clip=true,trim=1cm 0.7cm 0.5cm 0cm]{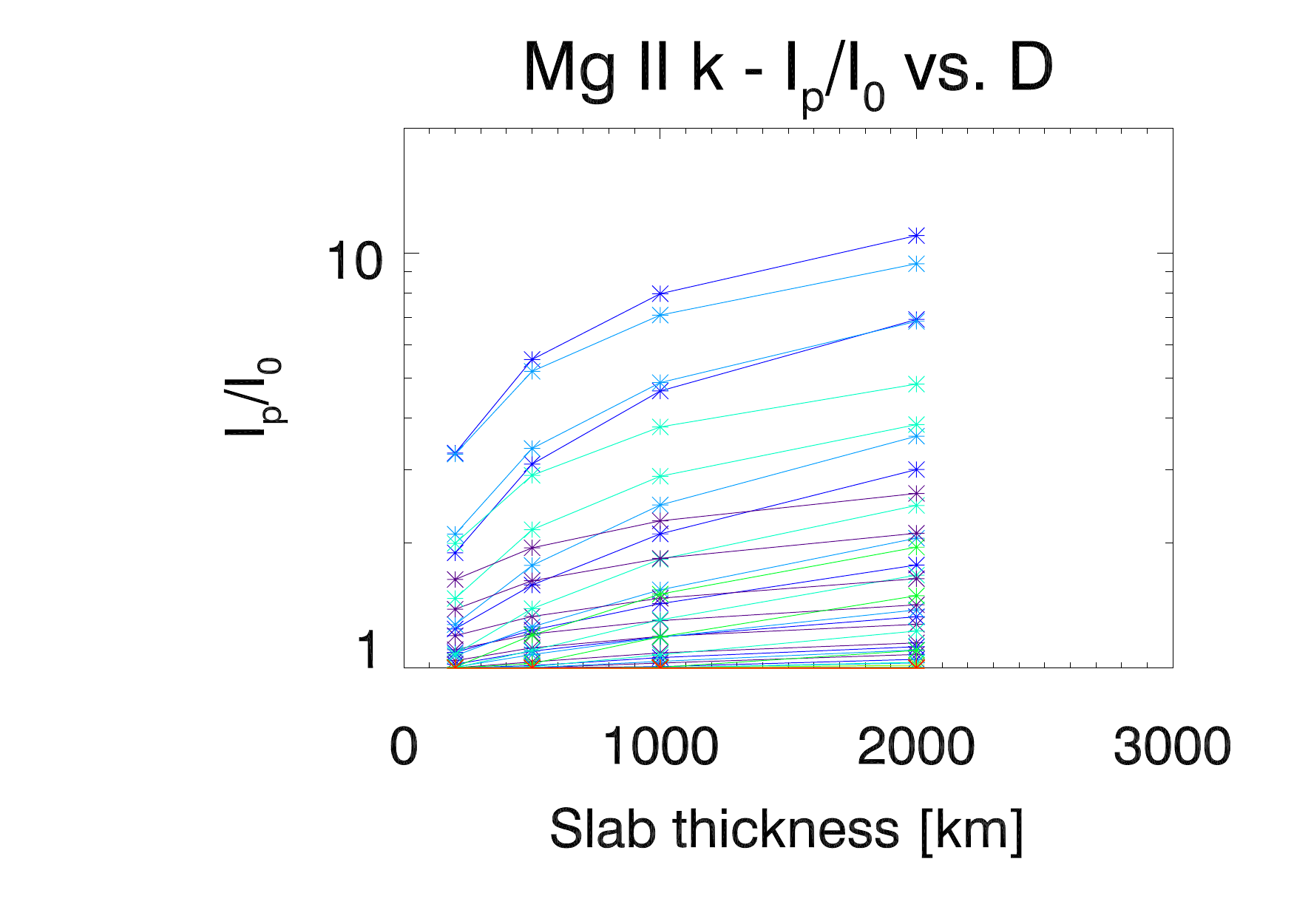}
                \includegraphics[width=0.24\hsize,clip=true,trim=1cm 0.7cm 0.5cm 0cm]{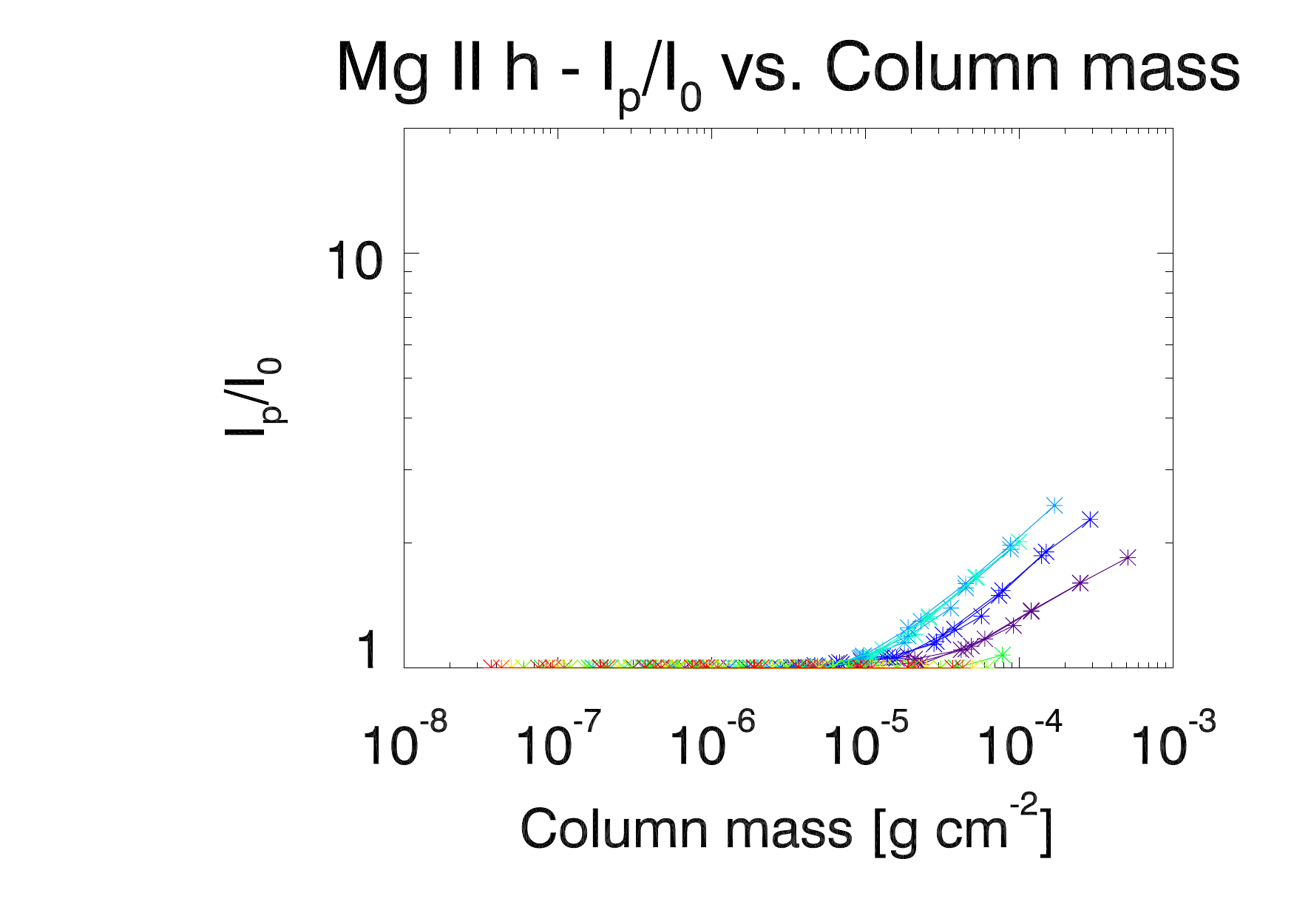}
                \includegraphics[width=0.24\hsize,clip=true,trim=1cm 0.7cm 0.5cm 0cm]{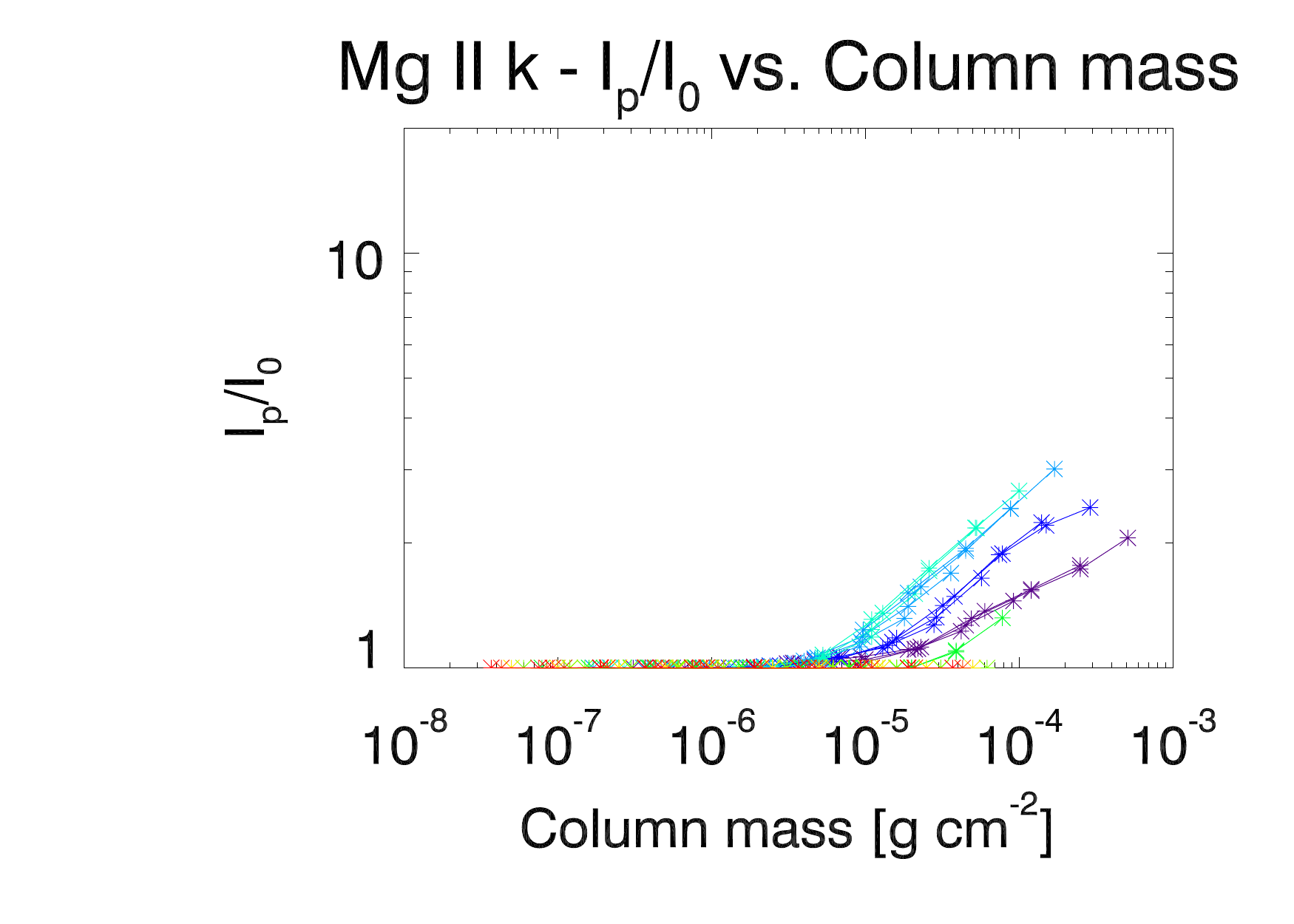}
                \caption{Plots of reversal level of the \ion{Mg}{ii} h line {(first and third columns)} and k line {(second and fourth columns)} against physical parameters for the isothermal isobaric models {(first two columns) and the PCTR models with $\gamma = 2$ (last two columns)}. \textit{Top:} Reversal {level} vs. {central temperature}, colour scale indicates pressure. \textit{Middle:} Reversal {level} vs. {central pressure}, colour scale indicates temperature. \textit{Bottom:} Reversal {level} vs. {slab thickness (for isothermal isobaric models) or column mass (for PCTR models)}, colour scale indicates temperature.}
                \label{fig:rev_vs_params_comb}
        \end{center}
\end{figure*}

As mentioned {earlier}, the reversal level of the h and k lines is heavily dependent on the optical thickness of the lines, as seen in Figure~\ref{fig:252_rev_tau_comb}. 
Exploring how the  reversal {level} is related to other parameters is useful for narrowing down the physical prominence conditions that give rise to the observed profiles. 
Figure~\ref{fig:rev_vs_params_comb} shows plots of h and k line reversal {levels} against {central} temperature, {central} gas pressure, and slab thickness {or column width} {for isothermal isobaric models (first two columns) and PCTR models with $\gamma=2$ (last two columns)}. 
The lower maximum reversal level {for PCTR models} as compared to isothermal isobaric models {is clearly seen in all plots}.
The reversal level vs. temperature plots (\textit{top panels}) show that the reversal {level} of both h and k lines is heavily dependent on temperature, but also on gas pressure (shown by the green colour gradient). 
In terms of temperature, the reversal level is greatest at around $8000 - 10000$~K, dependent on the model. 
These plots also generally show that the level of reversal increases with pressure {(see middle row)}. 
The bottom panels of Figure~\ref{fig:rev_vs_params_comb} show that for reversed profiles the slab thickness does have an effect on the {level}  of reversal seen in the emergent profile {from isothermal isobaric models}, with the largest line reversal {levels} only being found for the thickest prominence slabs. 
Reversal level as a function of optical thickness was already discussed in Fig.~\ref{fig:252_rev_tau_comb}, where only models with optical thickness higher than $\sim 10$ show a central reversal. 
Column mass {(used in PCTR models)} is related to reversal level in a similar way to optical thickness. 
Only column masses above $5 \times 10^{-6}$~g {cm$^{-2}$} have a central reversal, and again it is only lower temperature models that show any reversal.

\subsection{Correlations between \ion{Mg}{ii} and \ion{H}{i} intensities}
\begin{figure}
        \begin{center}
                \includegraphics[width=\hsize]{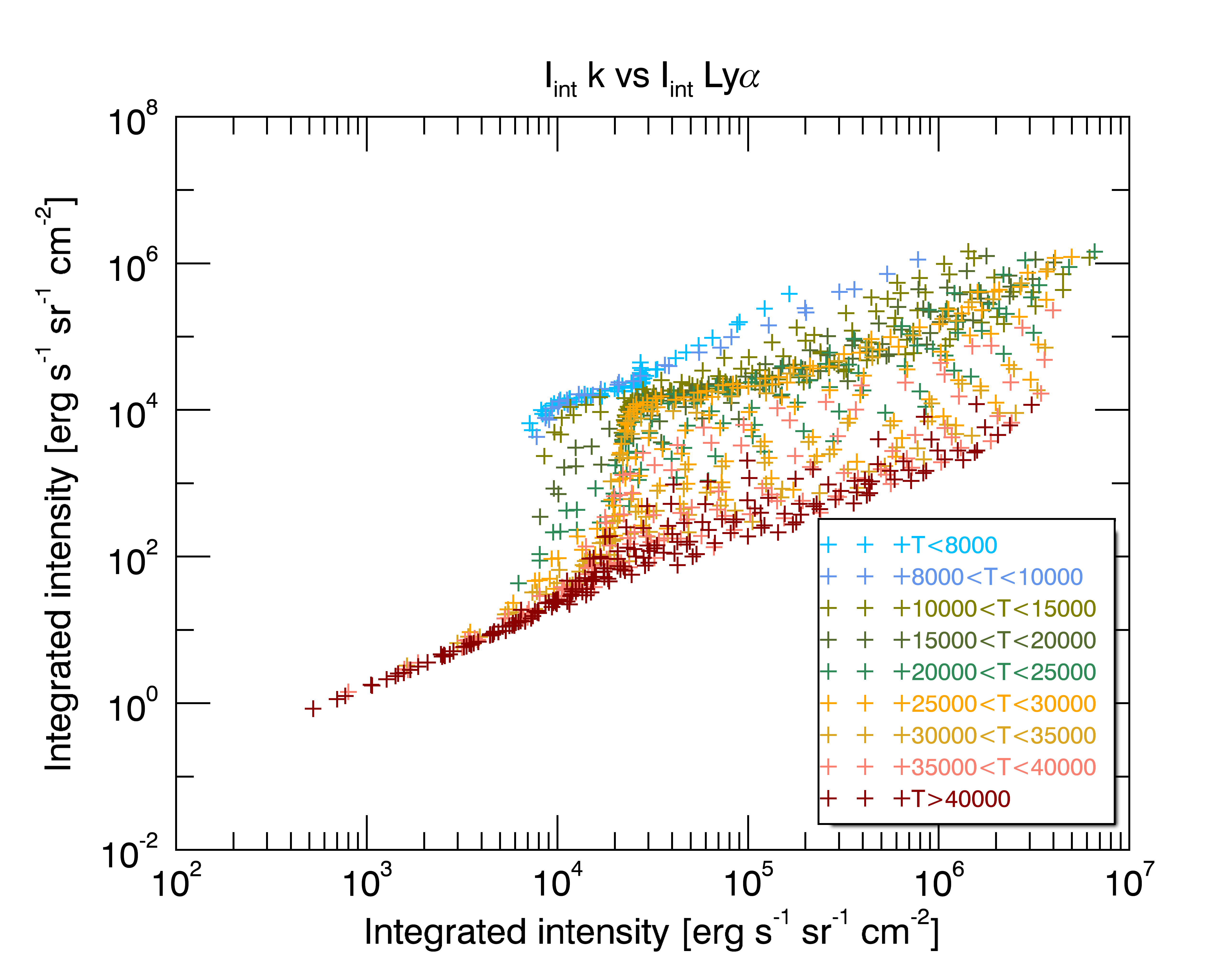}
                \includegraphics[width=\hsize]{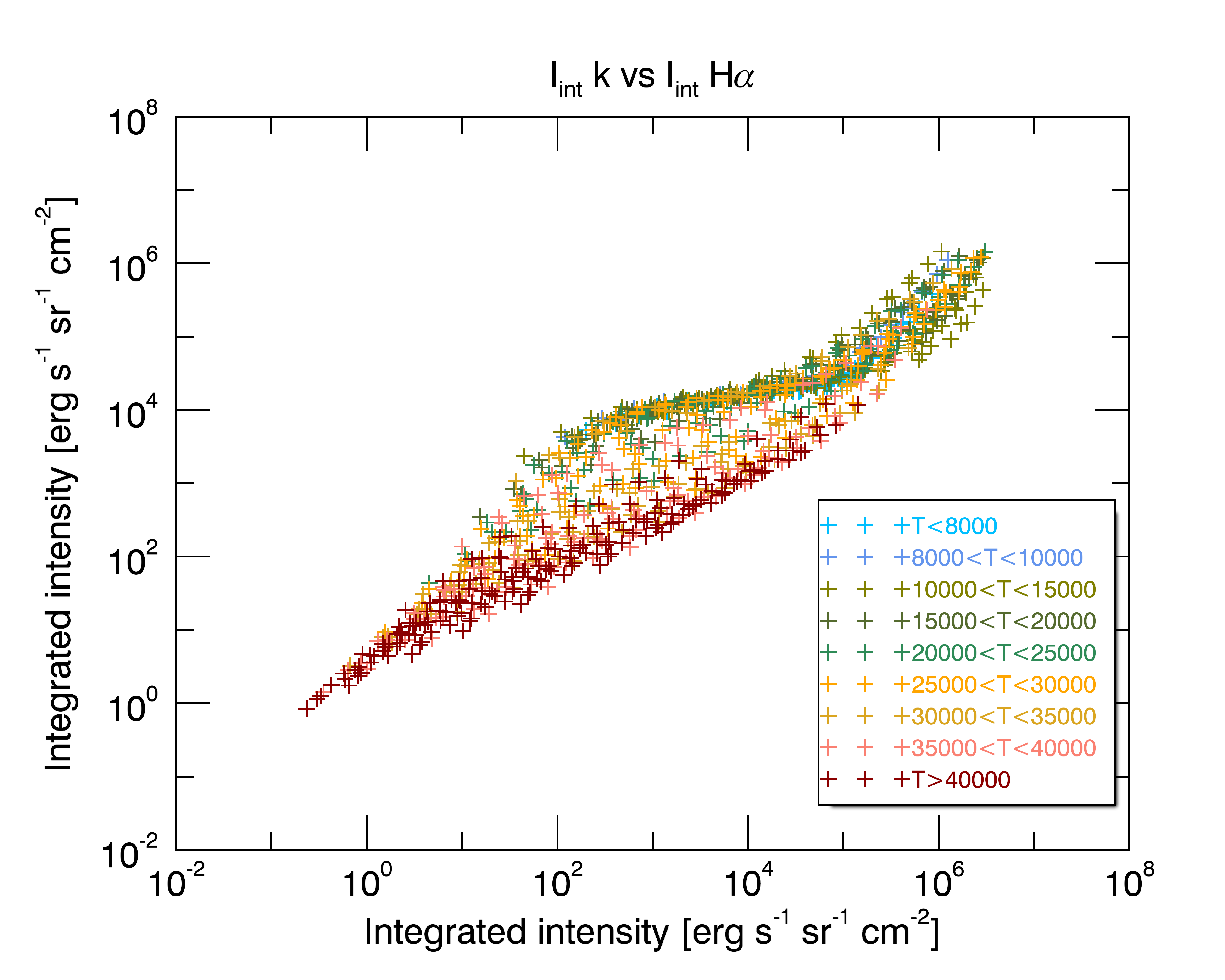}
                \caption{{Frequency-}integrated intensity in the k  line as a function of the Ly$\alpha$ (top) and H$\alpha$ (bottom) {frequency-}integrated intensities. The results are plotted for  isothermal and for isobaric models, and for models with a PCTR. Each colour corresponds to a given range of mean temperatures.}
                \label{f:mgvshlines}
        \end{center}
\end{figure}

It is also interesting to note how the \ion{Mg}{ii} lines correlate with intensities emitted in hydrogen lines. Figure~\ref{f:mgvshlines} shows how the intensity in the k  lines depends on the Ly$\alpha$  and H$\alpha$  intensities. The corresponding diagram for the Ly$\beta$ line is very similar to that of Ly$\alpha$ and is therefore not shown here. In Fig.~\ref{f:mgvshlines}, each colour represents a different range of mean temperatures, where the mean temperature is  the weighted-mean of the temperature along the line of sight as in \cite{Labrosse2004}. Of course, the mean temperature has the same value as the central temperature for isothermal models. In the case of PCTR models, however, the mean temperature can be significantly higher than the temperature at {the} slab centre. In the PCTR models used here (Table~\ref{tab:pctr}), the lowest mean temperature is nearly 10500\,K.

\begin{figure}
        \begin{center}
                \includegraphics[width=\hsize]{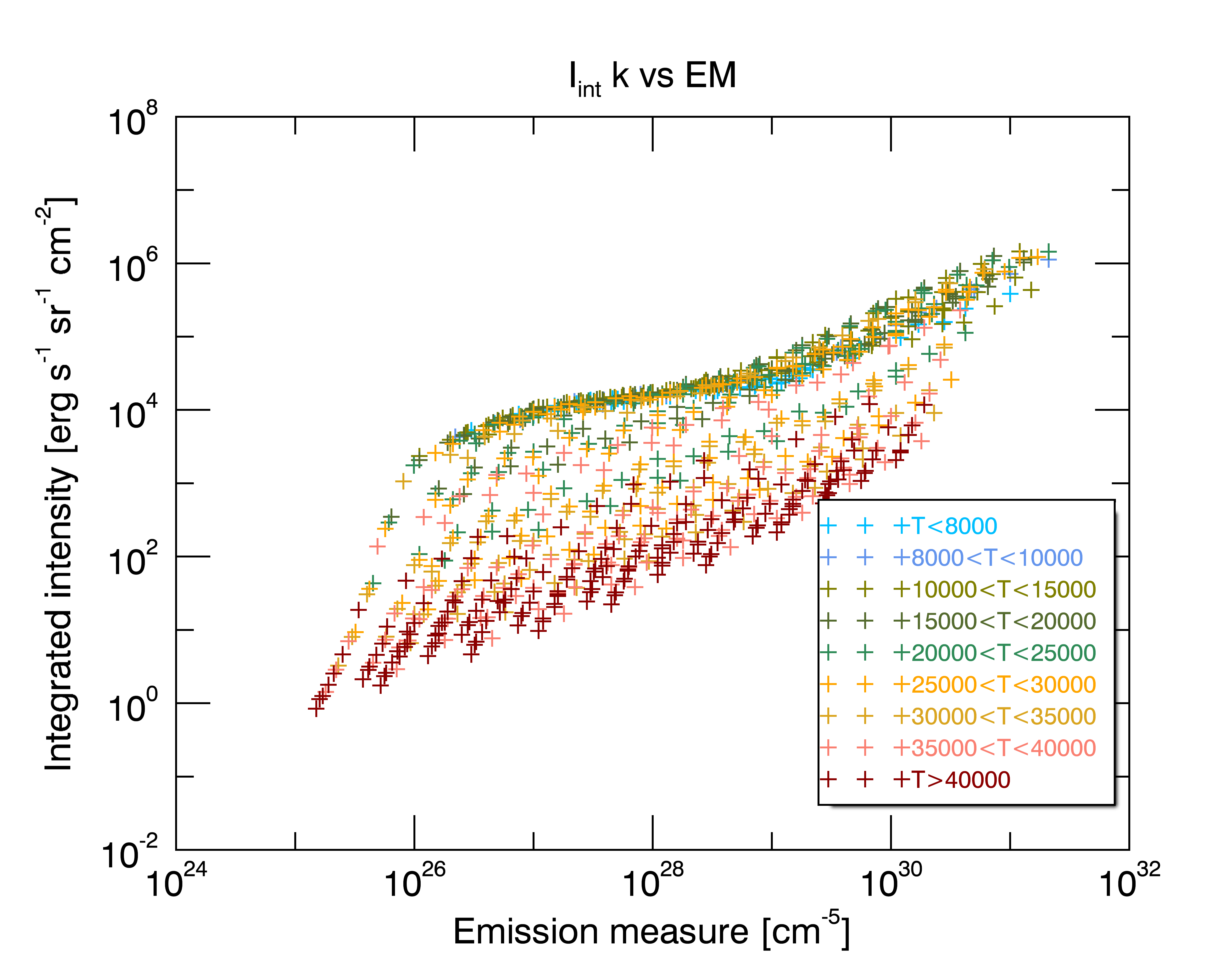}
                \includegraphics[width=\hsize]{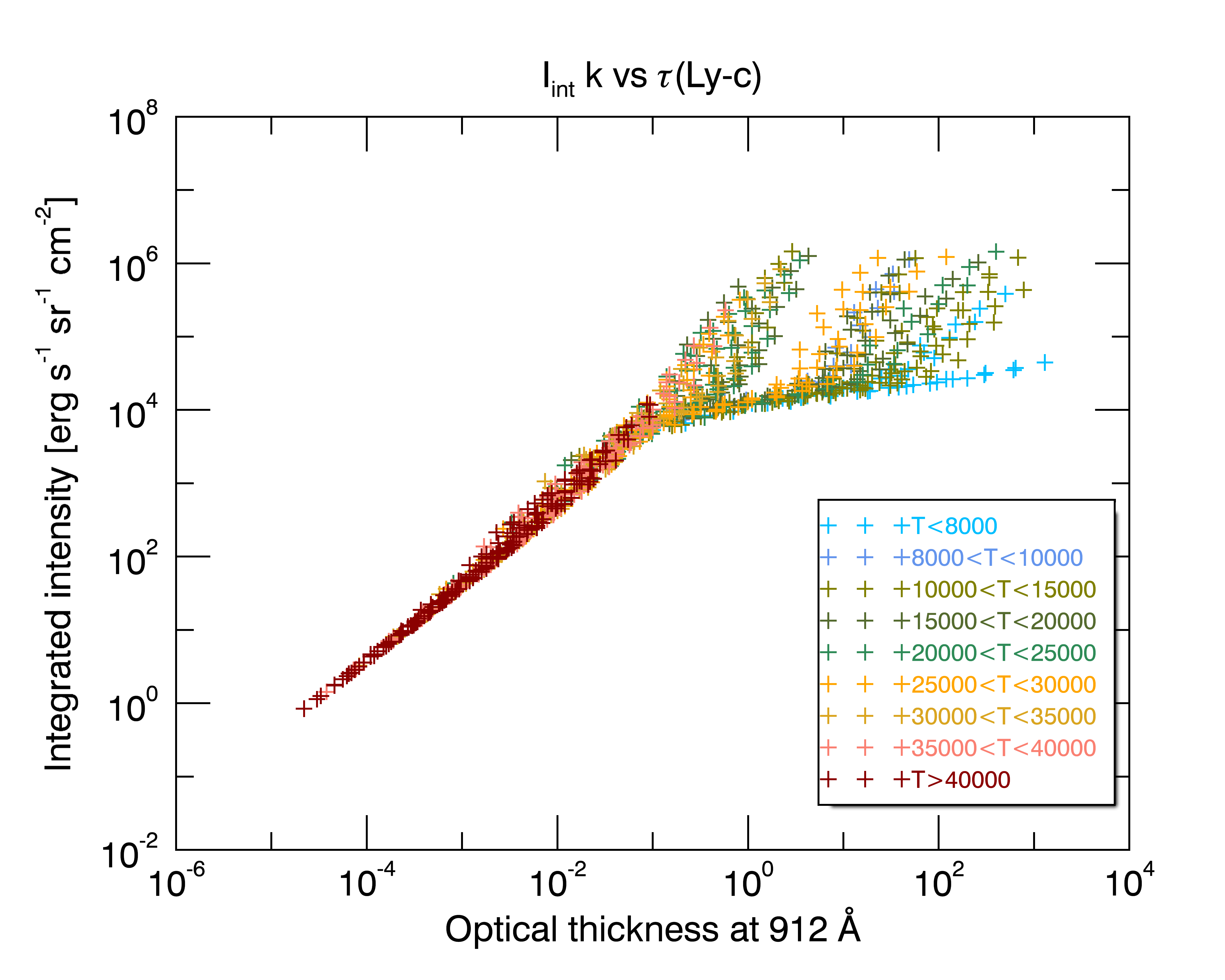}
                \caption{{Frequency-}integrated intensity in the k  line as a function of emission measure (top) and the optical thickness at the head of the Lyman continuum (912\,\AA). The results are plotted for both isothermal and isobaric models, and models with a PCTR. Each colour corresponds to a given range of mean temperatures.}
                \label{f:kvsEM}
        \end{center}
\end{figure}

In the top panel of Fig.~\ref{f:kvsEM}, we show the relation between the k line {frequency-}integrated intensity and the emission measure, defined here as $\mathrm{EM}=n_e^2 D$. If the thickness of the prominence slab (in other words the extent of the observed prominence along the line of sight) can be estimated, a good correlation between the two quantities means that the k line intensity  provides a way to measure the electron density in the prominence.
We immediately note that the trend shown in the top panel of  Fig.~\ref{f:kvsEM}  is similar to that seen in the bottom panel of Fig.~\ref{f:mgvshlines}, namely the correlation between the k and H$\alpha$ line intensities. This is not surprising as we know that a strong relation between H$\alpha$ intensities and emission measure was found in 1D isothermal isobaric models \citep{Gouttebroze1993}. This correlation still exists for PCTR models. Here we consider a broader grid of models than in GHV, and H$\alpha$ intensities generally become lower as the mean temperature of the model increases. This is what produces the apparent spread in these plots.

We also show the relation between the k line intensity and the optical thickness at the head of the Lyman continuum (912\,\AA) in the bottom panel of Fig.~\ref{f:kvsEM}. The latter gives us an idea of the penetration of the incident radiation in the slab,  in particular with  less ionising radiation penetrating deep in the slab for models where the plasma is optically thick at 912\,\AA.
We see in this plot that the intensities in the \ion{Mg}{ii} k line (and of the h line, too) will be first determined by the amount of external radiation received by the ions in the prominence for the coolest models. As the mean temperature increases, the collisions become more important in the formation of the lines, while at the same time letting more radiation penetrate (as the temperature increase results in a lower optical thickness at 912~\AA). For very high temperatures, higher  than 30000\,K, all our models are optically thin in the Lyman continuum, while \ion{Mg}{ii} is  depleted and populates \ion{Mg}{iii}.

As noted earlier, we expect the most representative prominence models to have temperatures lower than 25000\,K, and these results indicate that we can expect good correlations between the \ion{Mg}{ii} k line intensities and the intensities of hydrogen lines, as well as the emission measure.

\section{Summary and conclusion}
\label{s:concl}
This study is dedicated to developing prominence models of the \ion{Mg}{ii} h and k resonance lines, and using them to investigate how these lines behave under a wide range of physical conditions.
Since the launch of \iris, the \ion{Mg}{ii} h and k lines have been well studied and modelled in the chromosphere \citep{Leenaarts2013,Leenaarts2013b}, in flares \citep[e.g.][]{Kerr2016}, and in prominences \citep{Heinzel2014,Heinzel2015}. 
The \citet[][HVA]{Heinzel2014} grid of prominence \ion{Mg}{ii} models explored $27$ isothermal isobaric models and investigated the inclusion of a PCTR to some of their models as well. 
The goal of the present work  is to extend that grid of models in order to more fully understand the behaviour of the \ion{Mg}{ii} h and k lines. 

The atomic model of magnesium used here was constructed using standard atomic data for a five-level plus continuum \ion{Mg}{ii} atom.
Incident radiation was determined using \iris\ quiet-sun profiles for the \ion{Mg}{ii} h and k lines, as well as continuum-level intensities for the three \ion{Mg}{ii} subordinate lines. 
Our results are qualitatively similar to those from HVA. 

An extended grid of 1D models was designed in order to investigate the effects of higher temperatures and pressures on the emergent h and k line profiles. 
We then looked for correlations between observable line parameters (line {frequency-}integrated intensity, k/h {line} ratio, reversal level) and  model parameters (temperature, pressure, and slab thickness or total column mass), as well as optical thickness. 
The {frequency-}integrated intensity of the h and k lines is closely related for all models, displaying a power-law relationship with a small `bump' which is caused by the differences in the optical thickness of the h and k lines. 
Line reversal {levels} of both h and k are found to be closely related to optical thickness;  only models where $\tau > 10$ show any central reversal of the lines. 
{The frequency-}integrated intensity is also related to the optical thickness of the lines for both h and k, with the relationship changing at the transition between optically thin and optically thick. 

Investigating higher temperature models ($T \geq 15000$~K) reveals firstly that there are no central reversals found for models above $20000$~K. 
Secondly, the line centre intensity decreases with temperature, down to {values} that are not observed in real prominences (within the instrumental sensitivity of \iris). 
The cause of this decreased intensity is the increased amount of ionisation to \ion{Mg}{iii} at these temperatures. 
Therefore, there is  less \ion{Mg}{ii} in the relevant energy states to produce photons that contribute to the h and k lines, causing a lower intensity in those lines. 
This indicates that extending the grid of isothermal isobaric models much past $\sim 20000$~K is not necessary, as the computed emergent line profiles do not reflect those observed. 
However, it will be useful to have a finer grid of models between $\sim 6000 - 15000$~K to properly investigate the differences in low-temperature cases. 

Models with a prominence-corona transition region (PCTR) represent a significant improvement on isothermal isobaric slab models due to their temperature and pressure gradients between the cool prominence core and the hot surrounding atmosphere. 
Temperature and pressure gradients from \citet{Anzer1999} are used here to represent the PCTR. 
The \ion{Mg}{ii} h and k lines from PCTR models show similar inter-line relationships, and also {a} similar correlation with the line optical thickness to the isothermal isobaric cases. 
Considering the emergent line profiles themselves there are some notable changes. 
Primarily at low central temperatures ($< 10000$~K) there is a much broader range of line-centre intensities recovered than in the isothermal case with the same central temperature. 
This is caused by the presence of higher temperature plasma in the PCTR, which causes more collisional excitation within the slab.

Some correlations are found between observable parameters of the \ion{Mg}{ii} h and k lines (namely the k/h {line} ratio and the reversal level) and the temperature of the plasma slab, for both isothermal isobaric and PCTR models. 
For all pressures and slab thicknesses, the k/h {line} ratio increases approximately linearly with temperature, up to a point where the k/h {line} ratio saturates somewhere between $2$ and $2.4$, depending on the model. 
There is also some correlation between the k/h {line} ratio and the line optical thickness. 
In the isothermal isobaric case, for optically thin slabs (at higher $T$) the k/h {line} ratio is high, around $2$ to $2.4$, switching around $\tau = 1$ to lower values of around $1.4$, mostly for low $T$ models. 
There are models that deviate from this trend, but generally the correlation seen is good. 
This is interesting in comparison with the \iris\ observations presented in \cite{Levens2017}, where k/h {line ratio} values closely scattered around a value of $1.4$ were found. 
From the model results shown here this indicates that the observed profiles are all optically thick, and the prominence temperatures are low ($\sim 6000 - 8000$~K). 
A similar trend is seen for models with a PCTR, but there is some deviation for models with high optical thickness. 
The observed trend indicates that only a small number of PCTR models can explain the observed k/h {line} ratio of $1.4$ -- those with low pressure ($0.01$ or $0.02$~dyne~cm$^{-2}$) and an optical thickness of around $10$. 
It remains to be seen which of these scenarios is more likely. However, the PCTR models are generally better at explaining observed line profiles \citep{Schmieder2014,Heinzel2015}. 
The reversal level of the h and k lines can also be used as an indicator of both the plasma temperature and pressure, with the largest peak-to-core ratios coming from models (isothermal isobaric and PCTR) with temperatures of $8000 - 10000$~K. 
Higher pressure models generally show higher reversal levels too; however, there is some model dependence of this. 
We find good correlations between the \ion{Mg}{ii} k line intensities and the intensities of hydrogen lines, as well as with the emission measure.

These models represent a  simplification of the physical conditions of prominence plasma, but are nonetheless a useful tool for investigating the prominence conditions with respect to the observed \ion{Mg}{ii} h and k profiles. 
Although there is some improvement in moving from isothermal isobaric models to those with a PCTR, 
it will  be useful to move to more complex multi-thread models for detailed comparisons to observations. 
The 1D isothermal isobaric and PCTR models presented here can be used as a base for these more complex prominence models, including ones that include 1D/2D multi-thread structures to attempt to replicate the fine structure and small-scale prominence dynamics that are observed \citep{Gunar2007,Gunar2008,Labrosse2016}.

\begin{acknowledgements}
{We are grateful to the referee, Andrii Sukhorukov, for his constructive and insightful comments which improved the quality of this paper.}
        The authors would like to thank L.~Green and E.P. Kontar for comments on earlier versions of this work presented in \cite{levens2018diagnostics}. We are grateful to the International Space Science Institute for their support during two meetings of the International Team 374 led by N.L.
        P.J.L. acknowledges support from an STFC Research Studentship ST/K502005/1. N.L. acknowledges support from STFC through grant ST/P000533/1.
        IRIS is a NASA small explorer mission developed and operated by LMSAL with mission operations executed at the NASA Ames Research Center and major contributions to downlink communications funded by the Norwegian Space Center (NSC, Norway) through an ESA PRODEX contract.
\end{acknowledgements}

\bibliographystyle{aa} 
\bibliography{mgii_prom}

\end{document}